\documentclass[reprint,superscriptaddress,amsmath,amssymb,aps,prx,onecolumn]{revtex4-2}

\usepackage{graphicx}
\usepackage[hyperfootnotes=false]{hyperref}
\usepackage{dcolumn}
\usepackage{bm}
\usepackage{array}

\newtheorem{theorem}{Theorem}
\newtheorem{corollary}{Corollary}[theorem]
\newtheorem{lemma}[theorem]{Lemma}

\begin{document}

\preprint{APS/123-QED}

\title{Out-of-distribution generalisation for learning quantum channels with low energy coherent states}

\author{Jason L. Pereira}
\email{Jason.Pereira@lip6.fr}
\affiliation{INFN Sezione di Firenze, via G. Sansone 1, Sesto Fiorentino (FI) I-50019, Italy}
\affiliation{Ming Hsieh Department of Electrical and Computer Engineering, University of Southern California, Los Angeles, California 90089, USA}
\affiliation{Sorbonne Universit\'{e}, CNRS, LIP6, 4 Place Jussieu, Paris F-75005, France}
\author{Quntao Zhuang}
\affiliation{Ming Hsieh Department of Electrical and Computer Engineering, University of Southern California, Los Angeles, California 90089, USA}
\affiliation{Department of Physics and Astronomy, University of Southern California, Los Angeles, California 90089, USA}
\author{Leonardo Banchi}
\affiliation{Department of Physics and Astronomy, University of Florence, via G. Sansone 1, Sesto Fiorentino (FI) I-50019, Italy}
\affiliation{INFN Sezione di Firenze, via G. Sansone 1, Sesto Fiorentino (FI) I-50019, Italy}

\date{\today}

\begin{abstract}
    When experimentally learning the action of a continuous variable quantum process by probing it with inputs, there will often be some restriction on the input states used. One experimentally simple way to probe a quantum channel is using low energy coherent states. Learning a quantum channel in this way presents difficulties, due to the fact that two channels may act similarly on low energy inputs but very differently for high energy inputs. They may also act similarly on coherent state inputs but differently on non-classical inputs. Extrapolating the behaviour of a channel for more general input states from its action on the far more limited set of low energy coherent states is a case of out-of-distribution generalisation. To be sure that such generalisation gives meaningful results, one needs to relate error bounds for the training set to bounds that are valid for all inputs. We show that for any pair of channels that act sufficiently similarly on low energy coherent state inputs, one can bound how different the input-output relations are for any (high energy or highly non-classical) input. This proves out-of-distribution generalisation is always possible for learning quantum channels using low energy coherent states, as long as enough samples are used.
\end{abstract}

\maketitle

\tableofcontents

\section{Introduction}

Many physical systems can be modelled as quantum channels that map one (input) quantum state into another (the output). Learning about a physical process can therefore be regarded as the task of finding the input-output relations of the enacted quantum channel~\cite{gebhart2023learning}. Experimentally, we would do this by sending probe states through the channel and characterising the outputs. However, a full characterisation can be very ``expensive" in terms of the number of experiments required. This is especially true for continuous variable (CV) systems~\cite{braunstein2005quantum,weedbrook_gaussian_2012}. Since the dimension is infinite, it is not possible to, for example, probe a channel with every basis state individually.

Instead, we must probe the channel with some finite set of input states and then extrapolate the results to all possible inputs. One particularly simple set of input states is the coherent states. These are classical states that can easily be generated experimentally. However, they form an overcomplete basis, so are sufficient to characterise a CV channel~\cite{lobino_complete_2008,rahimi-keshari_quantum_2011,cooper_characterization_2015,kervinen_extended_2024}. In any real experiment, we will also have a maximum energy for our probe states, so the inputs will be restricted to coherent states of bounded energy.

Whenever we extrapolate from limited data, we need to understand how reliable our learned answer is outside of our limited dataset, i.e., whether it generalises. The problem is compounded if we want our solution to hold even outside of the parameter region in which our dataset lies (i.e., outside the space spanned by inputs in the training distribution). If our dataset consists of only the outputs for low energy, classical inputs, we may wonder how reliable our input-output relations are for higher energy or non-classical systems. This is called the out-of-distribution generalisation problem.

The problem of out-of-distribution generalisation for learning quantum channels has been investigated in the discrete variable (DV), unitary case~\cite{caro_out--distribution_2023,zhang_scalable_2024}. In~\cite{caro_out--distribution_2023}, it is shown that by probing a unitary with a restricted set of input states, one can learn how it acts on a very different set of inputs.
The authors also state that their proofs can be extended to a class of channels called doubly stochastic channels. Whilst many follow-on works about out-of-distribution generalisation in the DV setting largely focused on the unitary setting, the noisy versions of certain unitary learning problems can also be considered as specific instances of channel learning~\cite{gibbs_dynamical_2024}. Other works about incoherent learning of quantum dynamics can be seen as out-of-distribution generalisation results for specific classes of quantum to classical channels~\cite{jerbi_power_2023,du_efficient_2025}. Similarly, in Ref.~\cite{huang_learning_2023}, the authors show that observables learned on one distribution can be generalised to other distributions, as long as both distributions have a particular property called local flatness.

In the CV case, it is known that we can extrapolate the input-output relations for coherent states to general states~\cite{lobino_complete_2008,rahimi-keshari_quantum_2011,volkoff_universal_2021}, but not how an uncertainty in those relations for coherent states propagates when applied out-of-distribution to general states. Since we will never learn the action of a process perfectly (with finite samples), even on a limited set of inputs, an understanding of out-of-distribution error propagation is crucial.

In particular, understanding of out-of-distribution generalisation for CV channels impacts the study of many different quantum systems. In terms of quantum sensing and communication systems, CV channels are important for optomechanical systems~\cite{li2021cavity,xia2023entanglement,brady2023entanglement}, microwave cavity quantum electrodynamics (cQED) systems~\cite{eickbusch2022fast}, radiofrequency-photonic sensors~\cite{xia2020demonstration}, radar and lidar detection~\cite{zhuang2022ultimate}, optical spectroscopy processes~\cite{coddington2016dual,shi2023entanglement,hariri_entangled_2025,herman_squeezed_2025}, and optical communication and quantum networks~\cite{kozlowski2019towards,shi2020practical}. At the same time, CV channels also model nanophotonic chips for machine learning~\cite{shen2017deep} and optical quantum computers~\cite{kok2007linear}.

In this work, we present a general formalism for understanding out-of-distribution generalisation in CV channels. We show that if any pair of CV channels has sufficiently close outputs for low energy coherent state inputs, it is possible to construct a bound on the output distance for any input state. Hence, we prove that out-of-distribution generalisation is always possible for learning quantum channels using low energy coherent states.

We start, in Section~\ref{sec: overview}, by introducing the problem and giving an overview of our main results. In Section~\ref{sec: coherent state}, we go into more detail about the generalisation from low energy coherent states to higher energy coherent states. In Section~\ref{sec: coherent examples}, we investigate some examples of this type of generalisation. In Section~\ref{sec: general state}, we show the extension from coherent states to general states, and in Section~\ref{sec: general examples}, we give examples of this extension. In Section~\ref{sec: metrology}, we discuss how our work can be framed in terms of quantum process tomography, quantum metrology, and quantum channel discrimination. In Section~\ref{sec: conclusion}, we present our conclusions and discuss possible future research directions.

\section{Overview}\label{sec: overview}

Learning a quantum channel means understanding how any input state will be mapped to an output state by that channel. We can express our knowledge of the target channel as a ``learned" channel that mimics as closely as possible the input-output relations of the target.
This knowledge could (non-exhaustively) take the form of a classical description of the input-output relations, the parameter values for a specific form of parametrised channel, or some set of settings for a physical device that lets us enact the learned channel.
The better we learn the target channel, the closer the two channels will be.

Our aim is to bound the error in learning the action of a CV channel on an arbitrary state after learning on samples from a restricted set of input coherent states. Specifically, we want to bound the distance (trace norm) between the output of a target channel, $\Psi$, and our learned channel, $\Phi$, for test state $\rho$, with average photon number $\bar{n}$. To be useful (non-trivial), the bound on $\|\Psi[\rho]-\Phi[\rho]\|$ should approach $0$ as we learn $\Psi$ better over our restricted set of inputs.
We will occasionally refer to the ``trace distance", which is just half of the trace norm of the state difference.

\begin{figure}[t]
	\centering
	\includegraphics[width=0.9\textwidth]{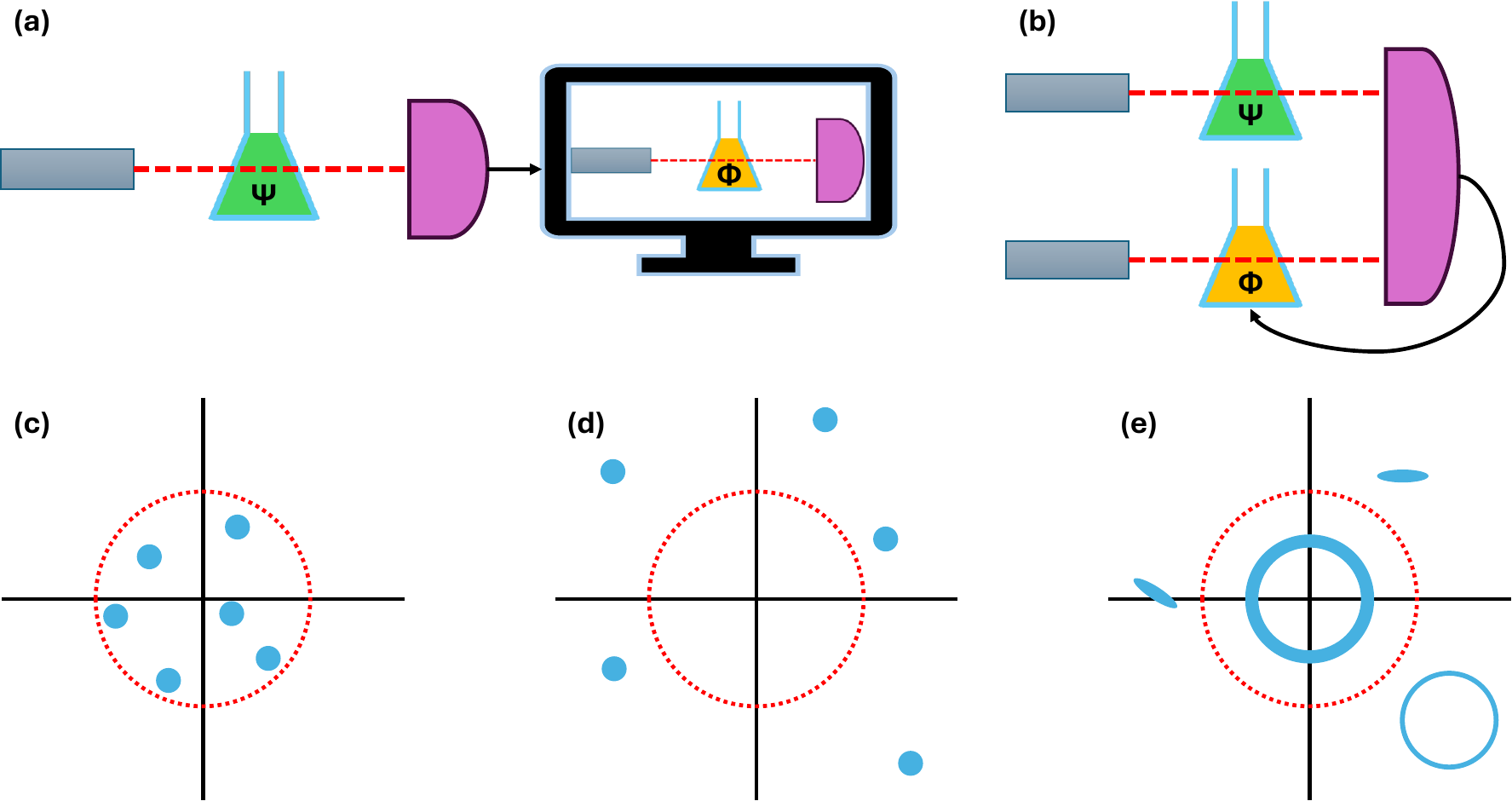}
	\caption{Parts (a) and (b) show two different ways we could learn a channel, $\Phi$, that mimics the action of a target process, $\Psi$, for low energy coherent states.
    (a) is ``classical" learning of a quantum channel, whilst (b) is a form of quantum machine learning.
    In this illustration, we depict probing the optical properties of an unknown (green) sample, although the process could be a magnetic field, a non-linear medium, a reflective cavity, or any other transformation that could be applied to a state of light. We search through a set of possible (yellow) optical mediums to find the one that best imitates the green substance.
    In (a), we use a (finite energy) laser to send low energy coherent states through the target and characterise the outputs with measurements. The results are post-processed (e.g., by feeding them into a classical computer, per the diagram) to obtain classical knowledge of the input-output relations. We can then ``simulate" the action of $\Psi$ on an unknown state, using our classical computer and the transformation $\Phi$.
    In (b), we have a tuneable quantum device (such as a different substance with tuneable optical properties, a tuneable magnetic field, or some parametrised optical circuit) that we want to use to imitate the target process. We probe both processes with the same coherent states (we show two lasers, but a real implementation could use a single laser and a balanced beamsplitter, to ensure the probing states are identical). Instead of characterising both outputs, we measure them jointly, to determine their trace distance. We could then use the measurement results to tune the quantum device and so reduce the distance. Once $\Phi$ is tuned, we hope that an unknown state sent through the quantum device will be transformed similarly to if it were sent through $\Psi$.
    Parts (c-e) illustrate the three types of input for which we can bound the closeness of two channels.
    (c) corresponds to the input states for parts (a) and (b); the probes used in our physical measurements are low energy coherent states. The dotted line shows the maximum average energy of the inputs. We can construct a bound on the output distance between our target and learned processes, $\Psi$ and $\Phi$, based on our measurement results.
    In (d), we consider higher energy coherent state inputs. These are outside of the class of inputs for which we have actual measurement results; rather, we want to be able to trust that our simulation, $\Phi$, of $\Psi$ is still accurate for these inputs (in the scenario of part (a)) or that our tuned quantum device still mimics our target well (in the scenario of part (b)). We use our bound for the inputs shown in (c) to construct a bound on the inputs shown in (d). The scaling of this bound depends on the class of channels to which the target and learned channel belong.
    In (e), we consider other types of input states, including non-classical states such as squeezed vacuums and Fock states. We again want to trust that $\Phi$ is a good simulation of $\Psi$ for these types of input, and so we extend our bounds for the inputs in (d) to the inputs in (e).}
	\label{fig: three_steps}
\end{figure}

If such a bound exists, this tells us that \textit{out-of-distribution generalisation is possible}, in the sense that learning the input-output relations for a very restricted set of classical state inputs is sufficient to mimic them for completely general (including highly non-classical) input states.
In some cases, this may require us to learn the channel very well for our restricted set of input states.
The tightness of the bound depends on the class of channels to which $\Psi$ and $\Phi$ belong, e.g. Gaussian~\cite{weedbrook_gaussian_2012,serafini2023quantum}, unitary operation, etc. If $\rho$ has a known, finite negativity (of its P-representation) $\mathcal{N}$~\cite{tan_negativity_2020}, we may also wish to take this into account to obtain a tighter bound.
A bound of this type can be constructed in three stages (as illustrated in Fig.~\ref{fig: three_steps}).

\begin{enumerate}
  \item Given $N$ samples of low energy coherent states sent through a channel $\Psi$, bound the error in learning the action of this channel on the same distribution of low energy coherent states. As sketched in Fig.~\ref{fig: three_steps}~(c), we have $N$ samples of the form $\Psi[| r e^{i\phi} \rangle \langle r e^{i\phi} |_{\mathrm{coh}}]$, for $r\leq\tau$. Find a protocol that uses them to learn a channel $\Phi$, such that (with high probability) $\|(\Psi-\Phi)[| r e^{i\phi} \rangle \langle r e^{i\phi} |_{\mathrm{coh}}]\|\leq \epsilon_0$ for all $r\leq \tau$, where $\epsilon_0$ is a decreasing function of $N$ (so that more samples let us learn the channel better). In other words, we bound the ``in-distribution" error for learning the action of channel $\Psi$ on coherent states with an average photon number of $\leq \tau^2$. Decreasing $\epsilon_0$ means using more samples to find a new channel $\Phi$, such that the output distance (for our subset of coherent states) is smaller. We require that $\epsilon_0 \to 0$ as $N \to \infty$.
    
  \item Assume stage 1 was successful. In stage 2, depicted in Fig.~\ref{fig: three_steps}~(d), we aim to solve the restricted case of out-of-distribution generalisation in which we only consider coherent states beyond the energy of the actual probe states from stage 1. More precisely, we need to find a function $\epsilon(\epsilon_0,r^2)$ such that $\|(\Psi-\Phi)[| re^{i\phi} \rangle \langle re^{i\phi} |_{\mathrm{coh}}]\|\leq \epsilon(\epsilon_0,r^2)$ for all $r$. We also require that $\epsilon(\epsilon_0,r^2)$ is a decreasing function of $\epsilon_0$ that reaches $0$ for $\epsilon_0=0$, ensuring that our out-of-distribution error reaches $0$ if our in-distribution error does.
    
  \item Assume stage 2 was successful. In the final stage, depicted in Fig.~\ref{fig: three_steps}~(e), we generalise to arbitrary states: for an arbitrary input $\rho$, we need to bound $\|(\Psi-\Phi)[\rho]\|$ in terms of $\bar{n}(\rho)$, $\epsilon(\epsilon_0,r)$, and possibly $\mathcal{N}(\rho)$ (if finite). This gives us a complete bound for out-of-distribution generalisation in the general case, and we require that it reaches $0$ for $\epsilon=0$.
\end{enumerate}

The first stage involves using a limited set of inputs to probe a quantum process and applying in-distribution generalisation to bound the error over this set.
Learning the input-output relations for a channel over a limited set of states is a somewhat underspecified problem, since any concrete statements about it would strongly depend on the learning process and the classes of channel under consideration. This is intentional, since the aim of this paper is to be as general as possible and to give a formalism for extending a bound for low energy coherent state inputs to general states, regardless of how the initial bound was obtained. However, in Section~\ref{sec: metrology}, we discuss briefly how this task has been addressed in the frameworks of quantum process tomography, quantum metrology and quantum machine learning.
Elsewhere, we do not focus on this step, but rather assume we already have the bound $\|(\Psi-\Phi)[| r e^{i\phi} \rangle \langle r e^{i\phi} |_{\mathrm{coh}}]\|\leq \epsilon_0$ (for $r\leq \tau$) and want to extend it to more general inputs. $\epsilon_0$ lies in the range $[0,2]$ (with a value of $2$ being trivial).

We focus on the remaining two stages. Our goals are twofold: we want to make some general statements that hold for all classes of channels and all input states and we also want to give tighter bounds for cases in which we have more information about the channels and/or input states.
Our main results can be informally summarised as follows:
\begin{enumerate}
    \item If two channels have exactly the same output for low energy coherent state inputs, they also have the same output for high energy coherent state inputs. If two channels have sufficiently similar outputs (i.e., a tightly bounded output distance) for low energy coherent state inputs, they also have a bounded output distance for high energy coherent state inputs. As the channel outputs become more similar for low energy inputs, they also become more similar for high energy inputs (as one bound converges to $0$, the other does too).
    For particular example classes of channels, we derive explicit bounds.
    \item If two channels have similar outputs for coherent state inputs, they also have a bounded output distance for any input state. As the channel outputs become more similar for coherent states, they also become more similar for general inputs.
    We give explicit expressions for particular examples of non-classical states.
\end{enumerate}
Together, these results show that \textit{out-of-distribution generalisation is possible for every input state and all target channels}, in the sense that a sufficiently small in-distribution error bound can always be translated to an out-of-distribution error bound. It is therefore possible to achieve an arbitrarily small error in reproducing the input-output relations for any input state, as long as sufficient samples are taken from the set of low energy coherent states.

\subsection{Main results: Out of distribution generalisation for coherent states}

We start by investigating how the output distance between a pair of channels, for a coherent state input, scales with the energy of the input. The question we want to answer is: \textit{if the outputs of two channels are close for low energy, pure coherent state inputs, how quickly do they become far apart as the input energy increases?} The first observation we make (proven in Appendix~\ref{app: oodg for coherent exact}) is:
\begin{lemma}\label{th: eps convergence}
    Suppose $\|(\Psi-\Phi)[| r e^{i\phi} \rangle \langle r e^{i\phi} |_{\mathrm{coh}}]\| = 0$ for all $\phi$ and all $r\leq \tau$, for some $\tau>0$.
    Then, for all $\phi$ and all $r$, $\|(\Psi-\Phi)[| r e^{i\phi} \rangle \langle r e^{i\phi} |_{\mathrm{coh}}]\| = 0$.
\end{lemma}

In other words, if we learn the action of a channel exactly on an (infinite but compact) subset of the pure coherent states, we also learn it exactly on all coherent states. Whilst intuitive, this is important to state, as otherwise one might think there are pairs of channels that exactly coincide for low energy coherent states but that diverge elsewhere in phase space.
This follows from the overcompleteness of the coherent states, or alternatively from the fact that every differential of every matrix element (in any basis) of $(\Psi-\Phi)[| r e^{i\phi} \rangle \langle r e^{i\phi} |_{\mathrm{coh}}]$ is identically zero, and is in line with what we might expect from Refs.~\cite{rahimi-keshari_quantum_2011} and \cite{volkoff_universal_2021}.

On the other hand, even if we can only bound the error on the subset with some $\epsilon_0 > 0$, we can still bound the error for all coherent states. In Appendix~\ref{app: oodg for coherent bound}, we show:
\begin{theorem}\label{th: eps existence}
    Suppose $\|(\Psi-\Phi)[| r e^{i\phi} \rangle \langle r e^{i\phi} |_{\mathrm{coh}}]\| \leq \epsilon_0$ for all $\phi$ and all $r\leq \tau$, for some $\tau>0$.
    Then, there exists a concave (in $r^2$) function $\epsilon(\epsilon_0,r^2)$ such that $\|(\Psi-\Phi)[| r e^{i\phi} \rangle \langle r e^{i\phi} |_{\mathrm{coh}}]\| \leq \epsilon(\epsilon_0,r^2)$ for all $\phi$ and $r$, with the following properties:
    \begin{enumerate}
        \item For any $r$ and for $\epsilon_0'<\epsilon_0$, $\epsilon(\epsilon_0',r^2) \leq \epsilon(\epsilon_0,r^2)$.
        \item For any $r$, $\epsilon(\epsilon_0,r^2) \to 0$ as $\epsilon_0 \to 0$.
    \end{enumerate}
\end{theorem}

We express $\epsilon$ as a function of $r^2$, rather than $r$, since $r^2$ is the mean photon number of the coherent state input, and so is a more natural variable to use. This theorem shows that if we learn a channel sufficiently well on a subset of the coherent states, our target and learned channel will also have similar outputs for coherent states outside of that subset.

Although Theorem~\ref{th: eps existence} shows the existence of a bounding function $\epsilon(\epsilon_0,r^2)$ that is non-trivial for sufficiently small $\epsilon_0$ (and, in fact, the proof is constructive), the function found in Appendix~\ref{app: oodg for coherent bound} requires very small $\epsilon_0$ to be non-trivial for large $r$. It is therefore better to find channel-specific $\epsilon$-functions. In Section~\ref{sec: coherent examples}, we use specific examples to demonstrate how we can obtain analytical expressions for the function $\epsilon(\epsilon_0,r^2)$ in certain cases, e.g., if our target channel is in a known class.

\subsection{Main results: Out of distribution generalisation for general states}

Next, we address more general states. The question now is: \textit{if the outputs of two channels are within a bounded distance of each other for coherent state inputs, how far apart are they for other types of input states?} We now assume we have a function $\epsilon(\epsilon_0,r^2)$ that upper bounds the trace norm between channel outputs for a coherent state input. Starting from this function, the goal is to prove a bound on the trace norm between channel outputs for any bounded energy input state $\rho$.

Such a bound can be obtained using the P-representation of CV states and the convexity of the trace norm. The calculations are outlined in Section~\ref{sec: general state} and presented in greater detail in the Appendices. In the most general case, the output distance for any non-classical input can be bounded with the following theorem:
\begin{theorem}\label{th: general states}
    Let $\Psi$ and $\Phi$ be a pair of quantum channels for which $\|(\Psi-\Phi)[| re^{i\phi} \rangle \langle re^{i\phi} |_{\mathrm{coh}}]\|\leq \epsilon(\epsilon_0,r^2)$ for all $r$ and for some concave (in $r^2$) function $\epsilon$. Let $\rho$ be any input state with average photon number $\bar{n}$. Then, for any positive integer $M$ and any $0\leq s < \frac{1}{2}$ obeying $\frac{(1-s)(1-2s)}{s(M-1)}>1$,
    \begin{equation}
        \|(\Psi-\Phi)[\rho]\| \leq \frac{2(1-s)^{M}M}{s^{M-1}(1-2s)} \epsilon\left(\epsilon_0, \frac{s(1-s)(M+1)}{1-2s} \right)  + 4\sqrt{s(1+ 2\bar{n})} + 4\sqrt{\frac{\bar{n}}{M}}.
        \label{eq: generic UB}
    \end{equation}
\end{theorem}

This theorem has two free parameters, $s$ and $M$, which must be tuned to attain the tightest bound.
It was obtained by bounding the output distance for an input state close to $\rho$, denoted $\sigma_{s,M}$, and applying a continuity bound. The physical meaning of the free parameters is as follows: to obtain $\sigma_{s,M}$ from $\rho$, apply a hard energy cut-off of $M-1$ photons, then pass the resulting state through a Gaussian additive noise channel, with a noise parametrised by $s$.
We can fix particular relationships between $s$, $M$, and $\bar{n}$ to study specific cases, though any such choice may not be optimal (in terms of giving the tightest bound).
For instance, we can study the scaling with the input energy, $\bar{n}$, by making $s$ proportional to $M^{-1}$, with some constant $\kappa$, obtaining the corollary:
\begin{corollary}\label{cor: energy scaling}
    Let $\rho$ be any input state with average photon number $\bar{n}$. For any $\kappa>1$ and positive integer $M$,
    \begin{equation}
        \|(\Psi-\Phi)[\rho]\| \leq 
        \mathcal{O}[e^{M\log(M\kappa)}\kappa^{-1} \epsilon (\epsilon_0, \kappa^{-1} ) ]
        + \mathcal{O}[\bar{n}^{\frac{1}{2}} M^{-\frac{1}{2}} ]
        \label{eq: worst case scaling}
    \end{equation}
\end{corollary}

Eq.~(\ref{eq: worst case scaling}) allows us to make a few observations. Firstly, if we fix $\kappa$, then as $\epsilon_0$ approaches $0$, $\epsilon (\epsilon_0, \kappa^{-1} )$ will also approach zero, and we will be able to send $M\to\infty$, so that Eq.~(\ref{eq: generic UB}) approaches $0$ too. Hence, as long as we are able to keep decreasing $\epsilon_0$, by taking more samples, we are also able to achieve an arbitrarily small output trace distance for any input state.
Secondly, fixing $M\sim \bar{n}$, we see that the leading order term scales at most poly-exponentially in the input energy ($\sim\mathcal{O}[e^{\bar{n}\log(\bar{n})}\epsilon (\epsilon_0, \kappa^{-1} )]$). A linear increase in the allowed $\bar{n}$ (for fixed error) can therefore always be achieved with an (at most) exponential decrease in $\epsilon (\epsilon_0, \kappa^{-1} )$.

In the specific case of a classical (but not necessarily coherent state) input, we get the corollary (see Appendix~\ref{app: classical}):
\begin{corollary}\label{cor: classical}
  Let $\rho_{\mathrm{class}}$ be a classical input state with average photon number $\bar{n}$. Then,
    \begin{equation}
        \|(\Psi-\Phi)[\rho_{\mathrm{class}}]\| \leq \epsilon(\epsilon_0,\bar{n}).\label{eq: classical bound}
    \end{equation}
\end{corollary}
In the more general, but still not universal, case in which the input state has a finite negativity, $\mathcal{N}$, and a known P-representation, $P(re^{i\phi})$, we define the quantities:
\begin{equation}
    \mu_P = 1+ 2\mathcal{N} = \iint \big|P(re^{i\phi})\big| r d\phi dr,
    \quad \nu_P = \iint \big|P(re^{i\phi})\big| r^3 d\phi dr,
    \quad \bar{n}_{\pm} = \frac{\iint P_{\pm}(re^{i\phi}) r^3 d\phi dr}{\iint P_{\pm}(re^{i\phi}) r d\phi dr},
\end{equation}
where $P_{\pm}$ means we integrate over the positive/negative domain of $P$. Then, we have the corollary
\begin{corollary}\label{cor: finite negativity}
    Let $\rho_{P}$ be an input state with P-representation $P(re^{i\phi})$ and finite negativity $\mathcal{N}$. Then,
    \begin{equation}
        \|(\Psi-\Phi)[\rho_{P}]\|
        \leq (1+\mathcal{N})\epsilon(\epsilon_0,\bar{n}_{+}) + \mathcal{N}\epsilon(\epsilon_0,\bar{n}_{-})
        \leq \mu_P\epsilon\bigg(\epsilon_0,\frac{\nu_P}{\mu_P}\bigg).
        \label{eq: bound finite negativity}
    \end{equation}
\end{corollary}

\begin{figure}[t]
	\centering
	\includegraphics[width=0.9\textwidth]{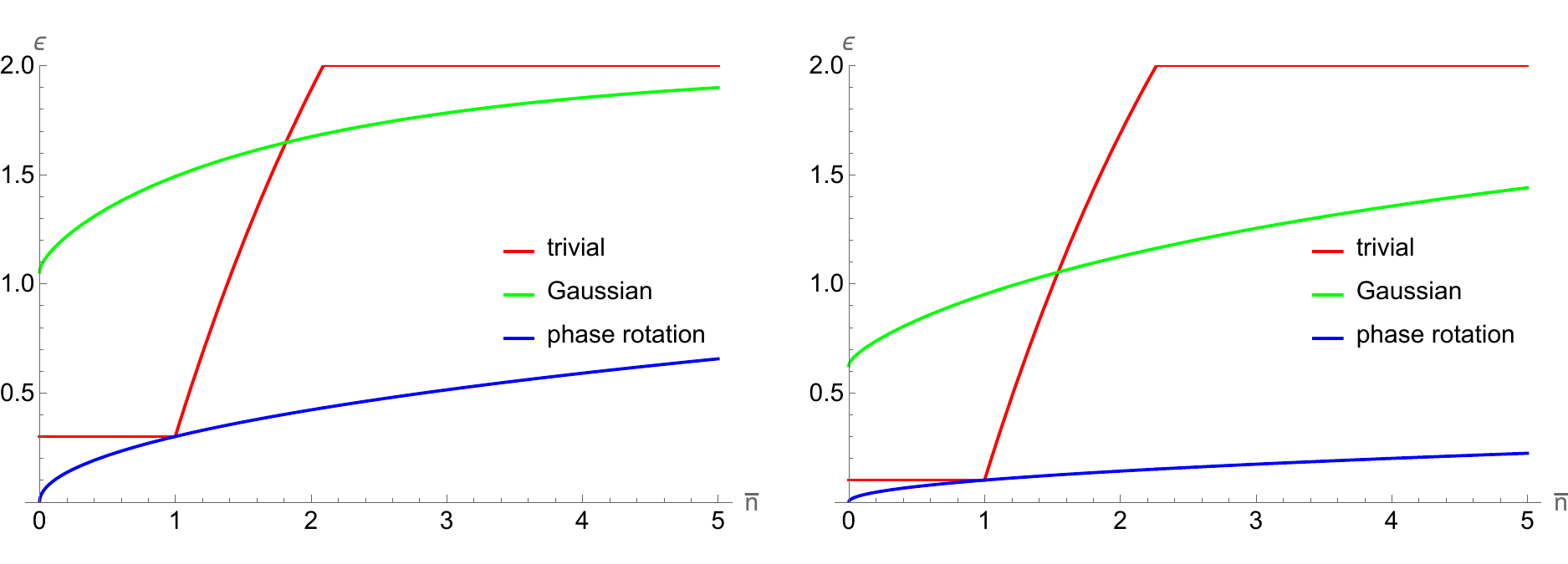}
	\caption{Trace norm bounds as a function of the energy of the input coherent state using a variety of techniques. The red lines show the trivial, piecewise step function bound, the green lines show the bound for any pair of Gaussian channels (from Eq.~(\ref{eq: gaussian epsilon})), and the blue lines show the bound when both channels are phase rotations (from Eq.~(\ref{eq: PR epsilon})). In the plot on the left, we set $\epsilon_0=0.3$, whilst on the plot on the right, we set $\epsilon_0=0.1$. In both cases, $\tau^2=1$.}
	\label{fig: coherent_bounds}
\end{figure}

\section{Out-of-distribution generalisation for coherent states}\label{sec: coherent state}

Suppose two channels, $\Psi$ and $\Phi$, output similar output states for input coherent states in a given region. This similarity is quantified by $\|(\Psi-\Phi)[| r e^{i\phi} \rangle \langle r e^{i\phi} |_{\mathrm{coh}}]\| \leq \epsilon_0$ for $re^{i\phi}$ lying in some closed subset. We want to generalise this bound to every complex value of $re^{i\phi}$. For simplicity (and physical relevance), we assume the subset of coherent states over which we know $\Psi$ and $\Phi$ are similar is a circle centred on the origin, so that the radius of the subset corresponds to the maximum average energy of the coherent states lying in it. Hence, our starting point is $\|(\Psi-\Phi)[| r e^{i\phi} \rangle \langle r e^{i\phi} |_{\mathrm{coh}}]\| \leq \epsilon_0$ for $r^2 \leq \tau^2$, and our goal is to find $\epsilon(\epsilon_0,r^2)$ such that $\|(\Psi-\Phi)[| r e^{i\phi} \rangle \langle r e^{i\phi} |_{\mathrm{coh}}]\|\leq \epsilon(\epsilon_0,r^2)$ for all $r$ (with the inequality as tight as possible).

It is not necessarily the case that the channel outputs grow further apart as $r$ increases. For instance, if $\Psi$ and $\Phi$ are both replacement channels (i.e., channels that ignore the input and prepare some fixed output), the output distance will be constant for any input. For some channels, the channel outputs could even converge for large $r$. However, in either of these scenarios, out-of-distribution generalisation is trivial, as we can simply define $\epsilon(\epsilon_0,r^2)=\epsilon_0$. We therefore focus on scenarios in which we expect $\epsilon(\epsilon_0,r^2)$ to grow with $r$. For instance, if $\Psi$ and $\Phi$ are lossy channels, a small difference in the loss parameters will still result in similar outputs for low energy inputs, but the difference will grow with the energy of the inputs.

Trivially, we could choose the step function, $\epsilon_{\mathrm{step}}(\epsilon_0,r^2)$, defined by
\begin{equation}
    \epsilon_{\mathrm{step}}(\epsilon_0,r^2) =
    \begin{cases}
        \epsilon_0 & r \leq \tau\\
        2 & r > \tau
    \end{cases},
    \label{eq: step function}
\end{equation}
however this clearly gives us no out-of-distribution information. A slightly better choice would be to
interpolate between the two regions, using the data processing inequality, which tells us that changing the input state by a distance of $\delta$ cannot increase the output distance by more than $2\delta$.
Nonetheless, this improved form still lacks an important property: it is trivial for large $r$, even when $\epsilon_0\to 0$. Hence, even if we gain perfect information about the subset of coherent states, the step function tells us nothing about 
large $r$. Lemma~\ref{th: eps convergence}, which is a simple consequence of expressing the input state in a fixed basis (see Appendix~\ref{app: oodg for coherent exact}), tells us that at $\epsilon_0=0$, we should have $\epsilon(0,r^2)=0$.

To be useful, the function $\epsilon(\epsilon_0,r^2)$ should be a decreasing function of $\epsilon_0$ for every value of $r^2$ (i.e., we require that for any $r$ and for $\epsilon_0'<\epsilon_0$, $\epsilon(\epsilon_0',r^2) \leq \epsilon(\epsilon_0,r^2)$). We also require it to approach $0$ as $\epsilon_0$ does so (i.e., for any $r$, $\epsilon(\epsilon_0,r^2) \to 0$ as $\epsilon_0 \to 0$). In theory, $\epsilon(\epsilon_0,r^2)$ could have an arbitrary dependence on $r$, since we only required it to be a decreasing function of $\epsilon_0$. However, given such a function with an arbitrary $r$-dependence, we can always construct $\epsilon'(\epsilon_0,r^2)$ as an upper concave hull of $\epsilon(\epsilon_0,r^2)$, i.e., a concave function (in terms of $r^2$) such that $\epsilon'(\epsilon_0,r^2)\geq\epsilon(\epsilon_0,r^2)$ for every value of $r^2$.
We use the concavity of $\epsilon(\epsilon_0,r^2)$ in Appendix~\ref{app: classical}.
Thus, we will henceforth assume that our upper bound on the output trace norm for coherent states is an increasing and concave function of $r^2$. Theorem~\ref{th: eps existence} (proven in Appendix~\ref{app: oodg for coherent bound}, using various concepts from later sections) tells us that it is always possible for us to find such a function.

If we know the classes of channel to which $\Psi$ and $\Phi$ belong (for instance, if we know the general form of $\Psi$ but are trying to learn specific channel parameters), we can find a function $\epsilon(\epsilon_0,r^2)$ by finding the worst case scenario set of errors in the parameters such that we still meet the constraint $\|(\Psi-\Phi)[| r e^{i\phi} \rangle \langle r e^{i\phi} |_{\mathrm{coh}}]\| \leq \epsilon_0$ for $r^2 \leq \tau^2$. In Section~\ref{sec: coherent examples}, we demonstrate this for three example classes of channel.

In Fig.~\ref{fig: coherent_bounds}, we illustrate how we can construct various bounding functions based on the degree of knowledge we have about the channels. The trivial bound holds for every channel, but does not approach $0$ for large $\bar{n}$, no matter how small we make $\epsilon_0$. If our target and learned channels are Gaussian, we can apply a tighter and always non-trivial bound (see Section~\ref{sec: gaussian channel example}). If we know that we are looking for a phase rotation channel (i.e., the task is to learn the rotation parameter), then the bound becomes exact. Note that the non-trivial, Gaussian bound is looser than the trivial bound in some regions; this is a consequence of the way it has been constructed (the looseness of the Fuchs-van de Graaf inequalities and taking multiple worst case values that cannot all be attained simultaneously). Comparing the plots for $\epsilon_0=0.3$ and $\epsilon_0=0.1$, we see that improving the in-distribution error can greatly improve the out-of-distribution error. Whilst Fig.~\ref{fig: coherent_bounds} refers to a specific scenario in which the channel we want to learn is a phase rotation, Theorem~\ref{th: eps existence} tells us that a non-trivial bounding function always exists for small enough $\epsilon_0$.

Without knowing the specific form of $\epsilon (\epsilon_0, r^2 )$, it is difficult to say anything general about its scaling with $\epsilon_0$ and $\tau$, however we can calculate the scaling for particular channel classes.
For instance (as we will see in Section~\ref{sec: gaussian channel example}), for Gaussian channels it is approximately given by $\epsilon (\epsilon_0, r^{2} ) \sim \mathcal{O}[\sqrt{\epsilon_0} r \tau^{-1}]$. Even if we know the scaling with $\epsilon_0$ and $\tau$, it is not possible to say anything definitive about the scaling with the number of samples, $N$, since this depends on the learning method employed in stage 1 of the process detailed in Section~\ref{sec: overview}, about which we are intentionally agnostic. Hence, without specifying the learning method, we cannot say whether it is ``better" (in terms of sample efficiency) to increase $\tau$ or to decrease $\epsilon_0$.

As an example, if we were to employ a tomography based method (discussed in Appendix~\ref{app: measurements}), we might expect each measurement of an output to require between $\mathcal{O}[E\epsilon_0^{-2}]$ and $\mathcal{O}[E^2\epsilon_0^{-3}]$ copies of the state, where $E$ is the average energy of the outputs. One method of guaranteeing the validity of the output distance bound over the entire $r\leq \tau$ region is to take evenly distributed samples, so that nowhere in the region is more than $\mathcal{O}[\epsilon_0]$ from a sampled point. In that case, the number of different coherent state inputs would be $\mathcal{O}[\epsilon_0^{-2}\tau^2]$, so the overall number of samples, $N$, would be between $\mathcal{O}[E\epsilon_0^{-4}\tau^2]$ and $\mathcal{O}[E^2\epsilon_0^{-5}\tau^2]$.
However, if we are dealing specifically with Gaussian channels (and perhaps other channel classes with a fixed, known form), the structure of the problem should allow us to sample significantly fewer (different) input states and still achieve the same guarantees over the $r\leq \tau$ region, improving the sample efficiency. In Ref.~\cite{volkoff_universal_2021}, it is shown that Gaussian unitaries can be learned by sampling only two distinct coherent state inputs, so for certain channel classes, the number of different input states required could even be independent of $\epsilon_0$ and $\tau$, meaning the required number of samples would depend only on the efficiency of state tomography.
One subtlety is that, in many physically relevant scenarios, the output energy, $E$, depends on the input energy, which is proportional to $\tau^2$, introducing a ``hidden" dependence on $\tau$.
We therefore emphasise that any statement about whether it is better to increase $\tau$ or decrease $\epsilon_0$ must be specific to one model of channel learning and one class of target channels.

\section{Examples: Out-of-distribution generalisation for coherent states}\label{sec: coherent examples}

We want to give analytical expressions for the function $\epsilon(\epsilon_0,r^2)$ for the case in which we have some prior knowledge about the target channel. The expressions necessarily depend on the specific forms of the target and learned channels, so to give more concrete statements, we focus on cases in which both belong to a particular, parametrised class of channels. We therefore have target parameters and learned parameters, and the closer the two sets of parameters are, the closer the channels (and hence the channel outputs) will be. As specific examples, we will consider Gaussian channels, the cubic phase unitary, and the Kerr interaction.

\subsection{Gaussian channels}\label{sec: gaussian channel example}

A Gaussian channel transforms a Gaussian input with first moment vector $q$ and covariance matrix $V$ according to
\begin{equation}
    q \to M q + d,
    \quad V \to M V M^T + N,
\end{equation}
where $d$ is a $2$-element real vector, $M$ and $N$ are $2$ by $2$ real matrices, $N=N^T \geq 0$, and $\det[N]\geq (\det[M]-1)^2$ (we have adopted the convention that $\hbar = 2$)~\cite{serafini2023quantum,weedbrook_gaussian_2012}. For the coherent state $|r e^{i\phi}\rangle$, $q$ has elements $2\mathrm{Re}[r e^{i\phi}]=2r\cos(\phi)$ and $2\mathrm{Im}[r e^{i\phi}]=2r\sin(\phi)$, and $V$ is the identity matrix. For Gaussian channels $\mathcal{G}_1$ and $\mathcal{G}_2$, characterised by $d_i$, $M_i$, and $N_i$, the output fidelity is~\cite{banchi_quantum_2015}
\begin{equation}
    F^2 = \frac{2\exp[-\frac{1}{2}\mu^T (V_1+V_2)^{-1} \mu]}{\sqrt{\Delta+\delta}-\sqrt{\delta}},
    \quad \mu = (M_2 - M_1)q + (d_2-d_1),
    \quad V_i = M_i M_i^{T} + N_i,
    \label{eq: output fidelity Gaussian}
\end{equation}
where $\delta=(\det[V_1]-1)(\det[V_2]-1)$ and $\Delta=\det[V_1+V_2]$. 
Examining the $r$-dependence, we can see $F^2$ is the exponential of a quadratic in $r$. Letting $x$, $y$, and $z$ be some unknown functions of $d_i$, $M_i$, $N_i$, and $\phi$, the output fidelity can be expressed as
\begin{equation}
    F^2 = x e^{-(yr^2+zr)}
    \leq xe^{-yr^2+|z|r},
    \label{eq: output fidelity Gaussian gen form}
\end{equation}
where we have folded the angular dependence of the outputs into the definitions of parameters $x$, $y$, and $z$. This form holds for every pair of Gaussian channels, so our goal is simply to bound $x$, $y$, and $z$.

If we know $\|(\mathcal{G}_1-\mathcal{G}_2)[| r e^{i\phi} \rangle \langle r e^{i\phi} |_{\mathrm{coh}}]\|\leq \epsilon_0$ for all $\phi$ and for $r\leq\tau$, we can lower bound $x$ and upper bound $y$ and $|z|$. Hence, we can lower bound $F^2$ and so, via the Fuchs-van de Graaf inequality~\cite{nielsen_quantum_2010}, upper bound the output distance. Specifically,
\begin{equation}
    x\geq \Big(\frac{2-\epsilon_0}{2}\Big)^2,
    \quad y\leq \frac{2}{\tau^2}\log\Big[\frac{2}{2-\epsilon_0}\Big],
    \quad |z|\leq \frac{1}{\tau}\log\Big[\frac{2}{2-\epsilon_0}\Big],
    \label{eq: xyz expressions}
\end{equation}
where we obtain the bounds by assessing Eq.~(\ref{eq: output fidelity Gaussian gen form}) at $r=0$ and $r=\pm \tau$ (strictly, $r\geq 0$ but for simplicity we treat $\phi\to\phi+\pi$ as $r\to-r$) and comparing the resulting expressions to our condition. Note that the three expressions in Eq.~(\ref{eq: xyz expressions}) cannot all be saturated simultaneously for our condition to hold, but they allow us to lower bound Eq.~(\ref{eq: output fidelity Gaussian gen form}) for every value of $r$. We get
\begin{equation}
    F^2 \leq \Big(\frac{2-\epsilon_0}{2}\Big)^{2\frac{r^2}{\tau^2}+\frac{r}{\tau}+2},
    \quad \|(\mathcal{G}_1-\mathcal{G}_2)[| r e^{i\phi} \rangle \langle r e^{i\phi} |_{\mathrm{coh}}]\|\leq
    \epsilon_{\mathrm{Gaussian}}(\epsilon_0,r^2) =
    2\sqrt{1-\Big(\frac{2-\epsilon_0}{2}\Big)^{2\frac{r^2}{\tau^2}+\frac{r}{\tau}+2}},
    \label{eq: gaussian epsilon}
\end{equation}
which we immediately see obeys the property of convergence to $0$ as $\epsilon_0\to 0$. Notably, it is never trivial (greater than $2$) for large $r$ and is concave in $r^2$ (though not in $r$). However, it can be significantly higher than $\epsilon_0$ even in the $r<\tau$ region (so is trivial in this region). This is the price we pay for bounding $x$, $y$, and $z$ individually; we could potentially get a tighter bound by working directly with Eq.~(\ref{eq: output fidelity Gaussian gen form}) to construct a (possibly piecewise) bound.
For small $\epsilon_0$, the leading order term is $\mathcal{O}[\sqrt{\epsilon_0} r \tau^{-1}]$.
Further calculation details are given in Appendix~\ref{app: gaussian}.

If we specify more details about the forms of the Gaussian channels under consideration, we may be able to do better. For instance, for an unknown displacement, the output fidelity is independent of the input coherent state, so $\epsilon_{\mathrm{dis}}(\epsilon_0,r^2) = \epsilon_0$. As another example, suppose the channels are rotationally symmetric (no angular dependence), as is the case for a range of channel classes, including lossy channels and quantum-limited amplifiers. Then we have $z=0$, so we get the tighter bound
\begin{equation}
    \epsilon_{\mathrm{sym}}(\epsilon_0,r^2) =
    2\sqrt{1-\Big(\frac{2-\epsilon_0}{2}\Big)^{2\big(\frac{r^2}{\tau^2}+1\big)}}.
\end{equation}

In the case of many single-parameter Gaussian unitaries, it is the case that a particular value of $\epsilon_0$, for a given value of $\tau$, is one-to-one with a specific value of the parameter difference, so that $\epsilon(\epsilon_0,r^2)$ can be tight.
For example, for phase rotation channels $\Psi_{\mathrm{PR}}$ and $\Phi_{\mathrm{PR}}$, with phases differing by $\theta$, the output trace norm for a coherent state input is
\begin{equation}
    \|(\Psi_{\mathrm{PR}}-\Phi_{\mathrm{PR}})[|re^{i\phi}\rangle \langle re^{i\phi}|]\|
    = 2\sqrt{1-\big|\langle re^{i\phi}|re^{i(\phi+\theta)}\rangle\big|^2}
    = 2\sqrt{1-e^{-2r^2(1-\cos(\theta))}},
    \label{eq: PR coherent distance}
\end{equation}
so we can write the exact expression
\begin{equation}
    \epsilon_{\mathrm{PR}}(\epsilon_0,r^2) =
        2\sqrt{1- \Big(1-\frac{1}{4}\epsilon_0^2\Big)^{\frac{r^2}{\tau^2}}}.
        \label{eq: PR epsilon}
\end{equation}
Similarly, for single-mode squeezing,
\begin{equation}
    \epsilon_{\mathrm{sq}}(\epsilon_0,r^2) =
        2\sqrt{1- \frac{1}{2\tau^2}W_0\big[ e^{2\tau^2} \tau^2 (2-\frac{1}{2}\epsilon_0^2) \big] \exp\bigg[r^2 \bigg( \frac{1}{\tau^2}W_0\big[ e^{2\tau^2} \tau^2 (2-\frac{1}{2}\epsilon_0^2) \big] - 2 \bigg) \bigg]},
        \label{eq: squeezing epsilon}
\end{equation}
where $W_0$ is the Lambert W function. We note that obtaining an exact expression for the output trace norm for coherent states, based on $\epsilon_0$ and $\tau$, is not possible for all single-parameter unitaries. The reason it is possible for the cases addressed here is because the expressions for the coherent state output trace norm satisfy the condition for a useful $\epsilon(\epsilon_0,r^2)$ stated in Section~\ref{sec: coherent state}: namely, that they are uniformly non-increasing in $\epsilon_0$. In other words, choosing closer parameter values for the target and learned channels, $\Psi$ and $\Phi$, and so decreasing $\epsilon_0$, always results in a smaller (or equal) value of $\epsilon(\epsilon_0,r^2)$ for every $r$. For non-Gaussian unitaries, this is not always the case (even when learning over a single parameter). We discuss an example of a unitary without this property in Section~\ref{sec: kerr}.

\subsection{Cubic phase unitaries}

The cubic phase gate is an important gate in CV quantum computing, because it is one option for the non-Gaussian operation required to make CV quantum computing universal~\cite{weedbrook_gaussian_2012}. It applies the unitary $V_{\gamma} = e^{i \gamma \hat{q}^3}$, where $q$ is one of the quadratures of a CV mode. To understand how it transforms a coherent state, it is easiest to work in the $\hat{q}$ basis. Specifically, we write
\begin{equation}
    |\alpha\rangle = \frac{1}{\sqrt[4]{2\pi}} \int_{-\infty}^{\infty} \exp\Big[-\frac{(q-2\mathrm{Re}[\alpha])^2}{4} + i(q-2\mathrm{Re}[\alpha])\mathrm{Im}[\alpha] \Big] |q\rangle dq,
\end{equation}
where $\{|q\rangle\}$ are the (unnormalisable) eigenstates of the $\hat{q}$ operator. The output fidelity between operations $V_{\beta}$ and $V_{\gamma}$ is then
\begin{equation}
    \big| \langle \alpha | V_{\beta}^{\dagger} V_{\gamma} | \alpha \rangle  \big|
    = \frac{1}{\sqrt{2\pi}} \Bigg| \int_{-\infty}^{\infty} e^{i(\gamma-\beta)q^3-\frac{1}{2}(q-2\mathrm{Re}[\alpha])^2} dq \Bigg|.
    \label{eq: output fidelity cubic phase gate}
\end{equation}

This integral is difficult to evaluate analytically, but numerically we find that it is a decreasing function of $\Delta_{\gamma} = |\gamma - \beta|$ and of the magnitude of the real part of $\alpha$. To understand why, consider that Eq.~(\ref{eq: output fidelity cubic phase gate}) is the convolution of the Gaussian function $(2\pi)^{-\frac{1}{2}}e^{-\frac{1}{2}q^2}$ (which integrates to $1$) and the oscillatory function $e^{i\Delta_{\gamma}q^3}$ at the point $2\mathrm{Re}[\alpha]$. As $|q|$ (or $\Delta_{\gamma}$) gets larger, the oscillations become more frequent, so the integral over a small locality becomes approximately zero (the positive and negative contributions cancel out). If the Gaussian function is centred in the slowly oscillating region (i.e., if $\mathrm{Re}[\alpha]$ is small), then the contributions will not cancel out much and the magnitude of the integral will be close to $1$. For large $\mathrm{Re}[\alpha]$, the integral will be closer to $0$. As we may expect, the output fidelity is independent of $\mathrm{Im}[\alpha]$, but since we want to bound the output distance over all coherent states in the region, we must choose $\alpha=\pm \mathrm{Re}[\alpha]$. The output trace norm for coherent state inputs is plotted for various values of $\Delta_{\gamma}$ in Fig.~\ref{fig: cubic and kerr}.

If we are guaranteed that, for some $\tau$, $\|(\Psi-\Phi)[| \tau e^{i\phi} \rangle \langle \tau e^{i\phi} |_{\mathrm{coh}}]\|\leq \epsilon_0$, we can numerically upper bound $\Delta_{\gamma}$, and so can construct a function $\epsilon(\epsilon_0,r^2)$ such that $\|(\Psi-\Phi)[| re^{i\phi} \rangle \langle re^{i\phi} |_{\mathrm{coh}}]\|\leq \epsilon(\epsilon_0,r^2)$ for all $r$. Note, however, that we have made the assumption that Eq.~(\ref{eq: output fidelity cubic phase gate}) is a decreasing function of the parameter difference at every point (value of $|\alpha|$), which is suggested by the numerics, but not analytically proven. It would therefore be desirable to find either a more mathematically rigorous proof that this is the case or an analytically solvable bound on Eq.~(\ref{eq: output fidelity cubic phase gate}).

\subsection{Kerr unitaries}\label{sec: kerr}

\begin{figure}[t]
	\centering
	\includegraphics[width=0.9\textwidth]{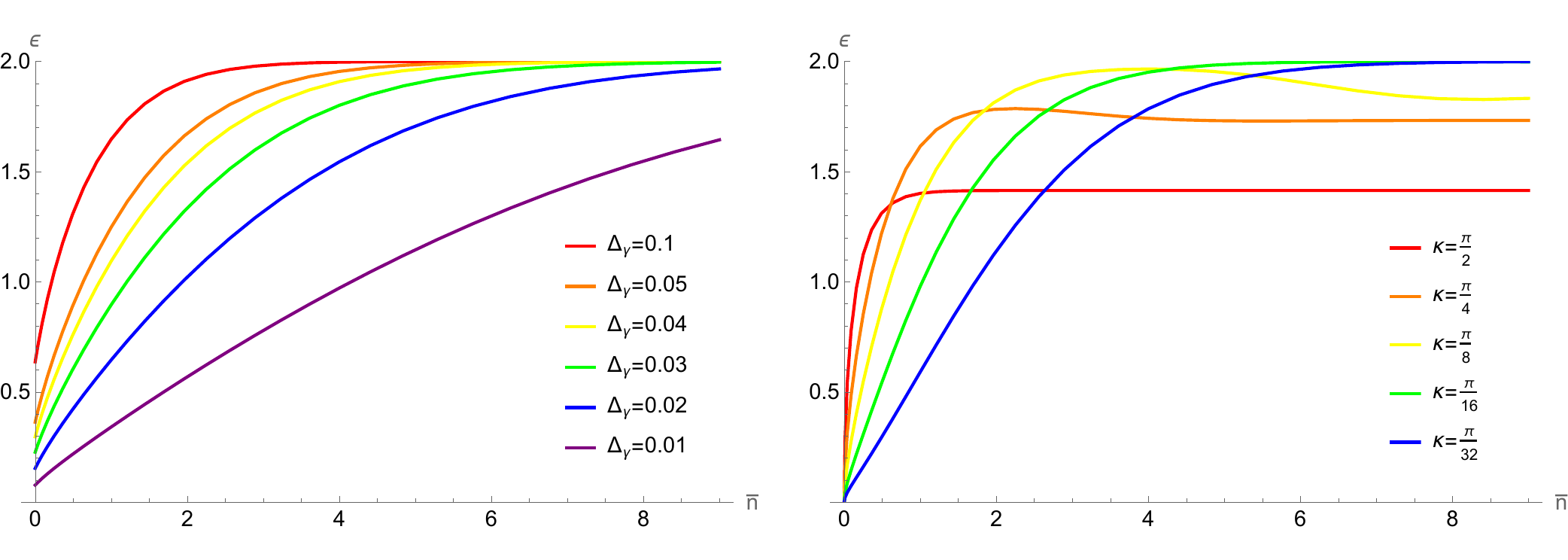}
	\caption{Output trace norm for a coherent state input for a pair of cubic phase unitaries (left) and for a pair of Kerr unitaries (right). The curves are plotted for various values of the parameter difference between the target and learned channels. In the case of cubic phase unitaries, the output trace norm appears to be an increasing function of the parameter difference for all values of $\bar{n}$, whilst for Kerr unitaries, we observe that many of the curves have inflection points and cross each other, so there are points at which a lower value of the parameter difference gives a higher output trace norm.}
	\label{fig: cubic and kerr}
\end{figure}

The Kerr interaction is another example of a physically relevant, non-Gaussian process. A Kerr gate applies the unitary $K_{\kappa}=e^{i\kappa\hat{n}^2}$. Using the Fock basis decomposition of coherent states, we can calculate
\begin{equation}
    \big| \langle re^{i\phi} | K_{\beta}^{\dagger} K_{\gamma} | re^{i\phi} \rangle  \big|
    = e^{-r^2}\Bigg|\sum_{m=0}^{\infty} \frac{r^{2m}}{m!}e^{i(\gamma-\beta)m^2}\Bigg|.
    \label{eq: output fidelity kerr}
\end{equation}
This infinite sum is difficult to evaluate in general, but it can be solved, analytically or numerically, for specific values of $\kappa=|\gamma-\beta|$. In particular, in Fig.~\ref{fig: cubic and kerr}, we have set $\kappa=\frac{\pi}{L}$ for $L=2$, $4$, $8$, $16$, and $32$. We observe that Kerr unitaries are an interesting counterexample to the idea that, for single-parameter unitaries, reducing the parameter difference should reduce the coherent state output distance for every value of $\bar{n}$.

In fact, Fig.~\ref{fig: cubic and kerr} might initially seem quite counter-intuitive, since smaller values of $\kappa$ result in larger asymptotic (in $\bar{n}$) values of the coherent state output distance. This counter-intuitiveness highlights the value of Theorem~\ref{th: eps existence}, as without it, one might suspect that there exist particularly ``weirdly behaved", but physically valid, sets of channels for which it is not always possible to construct a bounding function, $\epsilon(\epsilon_0,r^2)$, of the form we require.

The ``solution" to this counter-intuitiveness can be seen by considering which non-classical probes can perfectly discriminate the target and learned channels. Defining a general, pure probe (with no idler) as $|\Omega\rangle = \sum_{m=0}^{\infty} \omega_m |m\rangle$, with $\sum_{m=0}^{\infty} |\omega_m|^2 =1$, the output fidelity for Kerr unitaries with parameter difference $\kappa$ is
\begin{equation}
    \big| \langle \Omega | K_{\beta}^{\dagger} K_{\gamma} | \Omega \rangle  \big|
    = \Bigg|\sum_{m=0}^{\infty} |\omega_m^2| e^{i\kappa m^2}\Bigg|.
\end{equation}
In order for this sum to be $0$, we need there to exist at least two non-zero components, $\omega_a$ and $\omega_b$, such that $e^{i\kappa a^2}$ and $e^{i\kappa b^2}$ lie in opposite quadrants of the complex plane. In particular, if we have non-zero $\omega_0$ and want to construct a probe that discriminates the two channels perfectly, we need a non-zero $\omega_m$ for some $m \geq \sqrt{\pi\kappa^{-1}}$. Picking, for simplicity, $\kappa=\frac{\pi}{L}$ and restricting to cases in which $L$ is a square number, one can see that the state $|\Omega\rangle = \frac{1}{\sqrt{2}}(|0\rangle+|\sqrt{L}\rangle)$, with average photon number $\frac{L}{2}$, achieves an output fidelity of $0$ (and is the lowest energy state to do so). Thus, a smaller parameter difference (and so a larger $L$) means the minimum energy, single-mode probe that can perfectly discriminate the two unitaries has a higher average photon number, resolving the apparent paradox that closer channels are easier to discriminate using coherent state probes (and further highlighting the importance of understanding out-of-distribution generalisation from classical to non-classical inputs).

\section{Out-of-distribution generalisation for general states}\label{sec: general state}

Now assume we have a concave function $\epsilon(\epsilon_0,r^2)$ that bounds the distance between two channel outputs for any coherent state input. Our goal is to extend the bound to a general state input. Any state $\rho$ can be represented as
\begin{equation}
    \rho = \int P(\alpha) |\alpha\rangle\langle\alpha|_{\mathrm{coh}} d^2\alpha = \int_{0}^{\infty} \int_{0}^{2\pi} P(re^{i\phi}) |re^{i\phi}\rangle\langle re^{i\phi}|_{\mathrm{coh}} r d\phi dr,\label{eq: p representation}
\end{equation}
where we use $d^2\alpha$ to emphasise that this is a double integration over both the real and the imaginary components of $\alpha$. The P-representation takes both positive and negative values, and is only positive-semidefinite for classical states.

We begin by considering a classical state with average photon number $\bar{n}$. I.e., we have a state of the form in Eq.~(\ref{eq: p representation}), but where the P-representation, $P_{\mathrm{class}}(re^{i\phi})$, is constrained to be positive and to obey $\int_{0}^{\infty} r^2 \int_{0}^{2\pi} P_{\mathrm{class}}(re^{i\phi}) r d\phi dr = \bar{n}$. In Appendix~\ref{app: classical}, we apply the convexity of the trace norm, the concavity of $\epsilon(\epsilon_0,r^2)$ (it is worth noting that concavity in $r^2$ is a looser condition than concavity in $r$), and Jensen's inequality to get the upper bound from Corollary~\ref{cor: classical}:
\begin{equation*}
    \|(\Psi-\Phi)[\rho_{\mathrm{class}}]\| \leq \epsilon(\epsilon_0,\bar{n}).
\end{equation*}

\subsection{States with finite negativity}\label{sec: finite negativity}

Now consider the more general case, in which the P-representation can take negative values. We can no longer immediately apply the convexity of the output trace norm. However, we can split $\rho$ into two contributions:
\begin{equation}
    \rho = \rho_{+} - \rho_{-},\quad
    \rho_{+}  = \int_{P(\alpha)>0} P(\alpha) |\alpha\rangle\langle\alpha|_{\mathrm{coh}} d^2\alpha,\quad
    \rho_{-} = \int_{P(\alpha)<0} -P(\alpha) |\alpha\rangle\langle\alpha|_{\mathrm{coh}} d^2\alpha,
\end{equation}
where $\rho_{+}$ and $\rho_{-}$ are given by integrating over the positive and negative domains, respectively, of the P-representation. They represent sub or super-normalised classical states. Defining $\mathcal{N} = \mathrm{Tr}[\rho_{-}]$ (the negativity), we define
\begin{equation}
    \sigma_{+} = \frac{1}{1+\mathcal{N}}\rho_{+},\quad
    \sigma_{-} = \frac{1}{\mathcal{N}}\rho_{-}.
\end{equation}
Except for in the case of infinite negativity, these are now valid, classical quantum states, as they are normalised and have a positive overlap with every coherent state. From the linearity of quantum channels, and using the triangle inequality, we write
\begin{equation}
    \|(\Psi-\Phi)[\rho]\| = \|(1+\mathcal{N})(\Psi-\Phi)[\sigma_{+}] - \mathcal{N}(\Psi-\Phi)[\sigma_{-}]\|
    \leq (1+\mathcal{N})\|(\Psi-\Phi)[\sigma_{+}]\| + \mathcal{N}\|(\Psi-\Phi)[\sigma_{-}]\|.
\end{equation}
We can then bound the trace norm of the outputs of the classical input states $\sigma_{+}$ and $\sigma_{-}$. Note that the average photon numbers of $\sigma_{\pm}$, $\bar{n}_{\pm}$, need not equal $\bar{n}$, the average photon number of $\rho$; we only require $\bar{n} = (1+\mathcal{N})\bar{n}_{+} - \mathcal{N}\bar{n}_{-}$.
Applying Eq.~(\ref{eq: classical bound}),
\begin{equation}
    \|(\Psi-\Phi)[\rho]\| \leq (1+\mathcal{N})\epsilon(\epsilon_0,\bar{n}_{+}) + \mathcal{N}\epsilon(\epsilon_0,\bar{n}_{-}).\label{eq: finite negativity bound pm}
\end{equation}
Hence, from the P-representation of $\rho$, we can bound the distance of the learned channel output from the true channel output, using $\mathcal{N}$ (which can be used as a measure of the non-classicality of the state~\cite{tan_negativity_2020}) and $\bar{n}_{\pm}$. Applying the concavity of function $\epsilon$,
\begin{equation}
    \begin{split}
        (1+\mathcal{N})\epsilon(\epsilon_0,\bar{n}_{+}) + \mathcal{N}\epsilon(\epsilon_0,\bar{n}_{-})
        &= (1+2\mathcal{N})\left(\frac{1+\mathcal{N}}{1+2\mathcal{N}}\epsilon(\epsilon_0,\bar{n}_{+}) + \frac{\mathcal{N}}{1+2\mathcal{N}}\epsilon(\epsilon_0,\bar{n}_{-})\right)\\
        &\leq (1+2\mathcal{N}) \epsilon\left(\epsilon_0, \frac{(1+\mathcal{N})\bar{n}_{+} + \mathcal{N}\bar{n}_{-}}{1+2\mathcal{N}}\right)
        = \mu \epsilon\left(\epsilon_0, \frac{\nu}{\mu}\right),\label{eq: finite bound with mu}
    \end{split}
\end{equation}
where we have defined the quantities $\mu = 1+2\mathcal{N}$ and $\nu = (1+\mathcal{N})\bar{n}_{+} + \mathcal{N}\bar{n}_{-}$.
Eq.~(\ref{eq: finite negativity bound pm}) provides a tighter bound, but Eq.~(\ref{eq: finite bound with mu}) only requires knowledge of two quantities, rather than three, and will prove useful in the next subsection. Together, Eqs.~(\ref{eq: finite negativity bound pm}) and (\ref{eq: finite bound with mu}) give Corollary~\ref{cor: finite negativity}.

\subsection{States with infinite negativity}\label{sec: infinite negativity}

In general, $\mathcal{N}$ can be infinite. Examples include Fock states and squeezed vacuum states. In this case, the bound in Eq.~(\ref{eq: finite negativity bound pm}) becomes infinite, and hence trivial. The intuition we use to get around this is that every state with infinite negativity is $\delta$-close to a state with finite (but potentially large) negativity. Suppose we have a state $\rho$ with $\mathcal{N}(\rho) = \infty$. Then, for any small $\delta$, we can find another state, $\sigma$, with finite negativity, such that $\|\rho-\sigma\|\leq \delta$. By the data processing inequality,
\begin{equation}
    \|(\Psi-\Phi)[\rho]\| \leq \|(\Psi-\Phi)[\sigma]\| + 2\delta.
\end{equation}
We can calculate $\|(\Psi-\Phi)[\sigma]\|$ using Eq.~(\ref{eq: finite bound with mu}). As $\epsilon_0\to 0$, $\|(\Psi-\Phi)[\sigma]\|\to 0$, but $\delta$ will remain fixed, so we will need to find another state with smaller $\delta$ but higher negativity.

Suppose we have a parametrised sequence of states $\{\sigma_s\}$, for $s\geq 0$, with distance $\delta_s$ from $\rho$ and negativity $\mathcal{N}_s$, so that $\delta_s$ is an increasing function of $s$ that starts at $0$ for $s=0$ and $\mathcal{N}_s$ is a decreasing function of $s$ that is finite for any $s\neq 0$. The average photon number for $\sigma_s$, $\bar{n}_s$, may differ from $\bar{n}$. Then we can decrease $s$ as $\epsilon_0\to 0$, so that we eventually converge to $0$. For all $s$,
\begin{equation}
    \|(\Psi-\Phi)[\rho]\| \leq 
    \mu_s \epsilon\left(\epsilon_0, \frac{\nu_s}{\mu_s}\right) +2\delta_s.\label{eq: convolution only}
\end{equation}

The remaining task is therefore to find an appropriate sequence of states close to $\rho$. Taking the convolution of the P-representation of any state, $P(\alpha)[\rho]$, with $\frac{1}{s\pi}e^{-\frac{1}{s}|\alpha|^2}$ results in a new, valid state (for $s>0$). In fact, if we set $s=\frac{1}{2}$ or $s=1$, we get the W and Q-representations respectively of the original state. Since the Q-representation is always non-negative, this suggests the sequence of states with P-representations given by
\begin{equation}
    P_s(\alpha)[\rho] = \rho_s = \frac{1}{s\pi}e^{-\frac{1}{s}|\alpha|^2} \star P(\alpha)[\rho]
\end{equation}
could be useful for our purposes. For convenience, we define $\mathcal{C}_s$ as the channel enacted by this convolution (equivalent to a Gaussian additive noise channel).
In Appendix~\ref{app: sigma distance}, we show that, for this sequence, $\delta_s$ is upper bounded by
\begin{equation}
    \delta_s
    \leq \delta_s^{(\mathrm{UB})} = 2\sqrt{s (1+ 2 \bar{n})}.\label{eq: sigma distance}
\end{equation}
We know that for the Fock states, any non-zero value of $s$ has a finite negativity. However, this is not true in general: for squeezed vacuums, the negativity remains infinite up to some (squeezing dependent) threshold value, at which point it becomes $0$. This threshold value is the non-classical depth for a squeezed vacuum state.

To make our sequence of states universal, we can therefore combine an energy truncation with application of the channel $\mathcal{C}_s$. That is, we define $\sigma_{s,M} = \mathcal{C}_s[\rho^{(M)}]$, with P-representation $P_{s,M}$, where $\rho^{(M)}$ is the (normalised) truncation of $\rho$ to an $M$-dimensional qudit state (i.e., applying a hard energy cut-off of $M-1$ photons and renormalising). This adds an extra distance from the true state, $\delta_M = \|\rho-\rho^{(M)} \|$. We are free to choose any relationship between parameters $s$ and $M$ (so as to combine them into a single parameter) as long as $s\to 0$ when $M\to\infty$ (some later bounds will take $s\sim M^{-1}$).
In Appendix~\ref{app: sigma distance}, we show that for pure states, $\delta_M=\sqrt{1-\eta_M}$, where $\eta_M$ is the probability that a photon counting measurement on $\rho$ would have an outcome less than $M$, whilst in general, $\delta_M$ is bounded by
\begin{equation}
    \delta_M \leq 2\sqrt{\frac{\bar{n}}{M}}.
    \label{eq: deltaM bound}
\end{equation}

Finally, we can write
\begin{equation}
    \|(\Psi-\Phi)[\rho]\| \leq \mu_{s,M} \epsilon\left(\epsilon_0, \frac{\nu_{s,M}}{\mu_{s,M}}\right)  + 2\delta_{s} + 2\delta_M.
    \label{eq: bound infinite negativity full}
\end{equation}
In Appendices~\ref{app: mu UB} and \ref{app: sigma properties}, we show that the term $\mu_{s,M} \epsilon\left(\epsilon_0, \frac{\nu_{s,M}}{\mu_{s,M}}\right)$ can be upper bounded as
\begin{equation}
    \mu_{s,M} \epsilon\left(\epsilon_0, \frac{\nu_{s,M}}{\mu_{s,M}}\right) \leq
    \mu_{s,M}^{(\mathrm{UB})}\epsilon \left(\epsilon_0,\frac{s(1-s)(M+1)}{1-2s}\right),
    \label{eq: bound on mu epsilon}
\end{equation}
where $\mu_{s,M}^{(\mathrm{UB})}$ has an analytical expression, based on the number state decomposition of $\rho$, given in Eqs.~(\ref{eq: mu UB full}) and (\ref{eq: mu elements def}).
Crucially, Eq.~(\ref{eq: bound on mu epsilon}) means that, for an appropriate scaling of $s$ with $M$ ($s\lesssim
 M^{-1}$), we can fix the argument of $\epsilon$ to a constant, eliminating our dependence on the specific form of function $\epsilon$ entirely.

On the other hand, if we have less information about the state (i.e., we only know the average energy and nothing about the form of the state), we can write a looser but even more general bound. In Appendix~\ref{app: ultimate bound}, we show that
\begin{equation*}
    \|(\Psi-\Phi)[\rho]\| \leq \frac{2(1-s)^{M}M}{s^{M-1}(1-2s)} \epsilon\left(\epsilon_0, \frac{s(1-s)(M+1)}{1-2s} \right)  + 4\sqrt{s(1+ 2\bar{n})} + 4\sqrt{\frac{\bar{n}}{M}},
\end{equation*}
per Theorem~\ref{th: general states}. The condition, given in Theorem~\ref{th: general states}, that $\frac{(1-s)(1-2s)}{s(M-1)}>1$ comes from how we have bounded $\mu_{s,M}$ over all states. Per Corollary~\ref{cor: energy scaling}, we can choose particular relationships between $s$, $M$, and $\bar{n}$, to find the worst case scaling of the error.
Whilst this bound may be extremely loose (and exponential in $M$, so that even a fairly low $M$ results in an error bound much larger than $\epsilon_0$), it shows that out-of-distribution generalisation is always possible, for any input state and with almost no prior knowledge about the input state (except for $\bar{n}$).

The physical meaning of the $M=1$ case can be a little confusing, since it means we have truncated to a $1$-dimensional state. In fact, what this means is we have set $\sigma_{0,1}=|0\rangle\langle0|$, i.e., to the vacuum, and $\sigma_{s,1}$, for $s>0$, are thermal states. Eq.~(\ref{eq: generic UB}) still holds but is unnecessarily loose, since the vacuum is classical ($\mathcal{N}=0$), so we can write
\begin{equation}
    \|(\Psi-\Phi)[\rho]\| \leq \epsilon(\epsilon_0, 0 ) + 4\sqrt{\bar{n}}.
    \label{eq: tighter bound for M=1}
\end{equation}

We may be interested in how small we require $\epsilon(\epsilon_0,r^2)$ to be in order to obtain a non-trivial bound (i.e., a value less than $2$). When using Eq.~(\ref{eq: bound on mu epsilon}), this is highly state dependent, however the parameter region in which Eq.~(\ref{eq: generic UB}) is non-trivial is an important question, both for understanding the usefulness of the most general bound and because of the physical implications. An exact characterisation of this region depends on the form of the function $\epsilon$, and would require optimisation over both $s$ and $M$. However, one interesting case is obtained by setting $\frac{s(1-s)(M+1)}{1-2s}=\bar{n}$ (thus fixing $s$ as a function of $M$). In this case, we have a non-trivial bound if and only if
\begin{equation}
    \begin{split}
        \epsilon(\epsilon_0,\bar{n})
        < \frac{ 2\big(\sqrt{(M+1)^2+4\bar{n}^2} - 2\bar{n}\big) \big(M+1+2\bar{n}-\sqrt{(M+1)^2+4\bar{n}^2}\big)^{M-1}}{M\big(M+1-2\bar{n}+\sqrt{(M+1)^2+4\bar{n}^2}\big)^{M}}\\
        \times\left( 1 - 2\sqrt{\frac{\bar{n}}{M}} - \sqrt{\frac{2(2\bar{n}+1)\big(M+1+2\bar{n}-\sqrt{(M+1)^2+4\bar{n}^2}\big)}{M+1}} \right).
    \end{split}
    \label{eq: triviality condition fixed to nbar}
\end{equation}

\begin{figure}[t]
	\centering
	\includegraphics[width=1\textwidth]{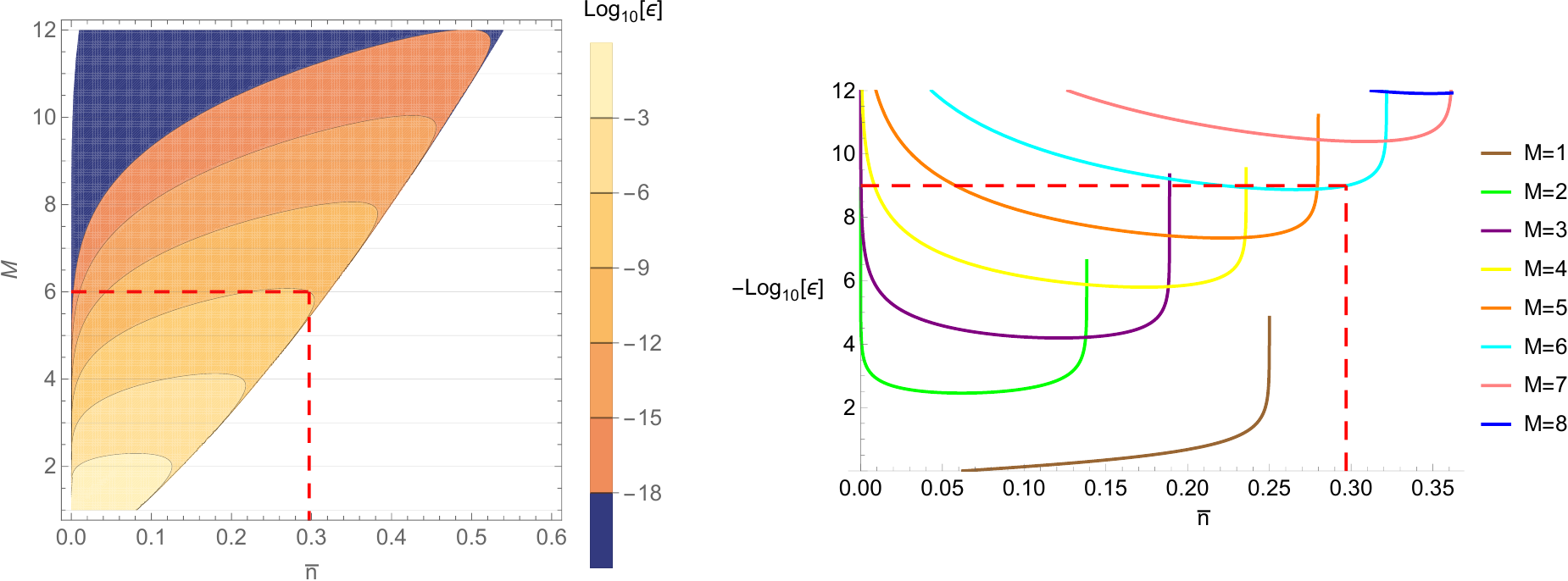}
	\caption{The minimum value of $\epsilon(\epsilon_0,\bar{n})$ for which the bound from Theorem~\ref{th: general states} is non-trivial, when we fix $\frac{s(1-s)(M+1)}{1-2s}=\bar{n}$ (given on a logarithmic scale). For a given value of $\bar{n}$, we can choose any integer value for $M$ (indicated by the horizontal lines on the left plot). Then, the plot on the left can be interpreted as follows: if $\epsilon(\epsilon_0,\bar{n})$ (the bound on the output trace norm for classical inputs) is less than the value indicated by the colour, then the bound will be non-trivial. The white region (in the bottom right) does not admit a non-trivial bound (for this choice of relationship between $s$ and $M$), whilst the navy region (in the top left) has an extremely small value for the required $\epsilon(\epsilon_0,\bar{n})$. The plot on the right shows the same thing from a slightly different perspective: each curve is for a fixed value of $M$. As an example, by following the red, dashed line, we can see that for any input with average photon number $\bar{n} \leq 0.297$, we always have a non-trivial bound if $\epsilon(\epsilon_0,\bar{n})\leq 10^{-9}$.}
	\label{fig: non-trivial region}
\end{figure}

In Fig.~\ref{fig: non-trivial region}, we plot the required value of $\epsilon(\epsilon_0,\bar{n})$ in terms of $\bar{n}$ and $M$. We note the caveat that this is for a particular relationship between $s$ and $M$, so Eq.~(\ref{eq: generic UB}) can be made non-trivial in a broader region by optimally tuning $s$ and $M$. In general, we want $M$ to be as small as possible, since Eq.~(\ref{eq: generic UB}) is exponential in $M$.
The plot on the right uses the tighter bound for the $M=1$ case, coming from Eq.~(\ref{eq: tighter bound for M=1}). Although the $M=1$ curve is lower than the $M=2$, $M=3$, and $M=4$ curves, we should be careful not to interpret this as meaning that setting $M=1$ is always the best choice. The curves only show the region in which the bounds are non-trivial, not the comparative values of the bounds for each parameter choice; for small values of $\epsilon(\epsilon_0,\bar{n})$, the term coming from $\delta_M$ (i.e., the error due to approximating $\rho$ with the vacuum state) will dominate. It is also not surprising that the $1$-dimensional case, for which we know $\mathcal{N}=0$, since $\sigma$ is always the vacuum, sometimes gives a tighter bound than bounding over all $d$-dimensional states, for $d>1$.

Fig.~\ref{fig: non-trivial region} also has an interesting interpretation in terms of channel discrimination. Eq.~(\ref{eq: generic UB}) can be interpreted as a bound on the maximum quantum advantage for channel discrimination tasks in the following sense: if we set $\bar{n}=\frac{s(1-s)(M+1)}{1-2s}$ then we can directly compare the (upper bound on the) best single-shot discrimination probability using any finite energy probe (proportional to the output trace distance) to the best single-shot discrimination probability using a classical probe of the same energy (proportional to $\epsilon(\epsilon_0,\bar{n})$).
Since Fig.~\ref{fig: non-trivial region} depicts the values of $\epsilon(\epsilon_0,\bar{n})$ for which Theorem~\ref{th: general states} provides a non-trivial ($<2$) bound on the output trace norm, it can be interpreted as the region in which we can meaningfully bound quantum advantage in channel discrimination. We can say, for instance, that (per the red, dashed line on Fig.~\ref{fig: non-trivial region}) any pair of channels for which the maximum output trace norm using a classical probe of average photon number $0.297$ is less than $\sim 10^{-9}$ (corresponding to a discrimination success probability of $\sim \frac{1}{2}+2.5\times 10^{-10}$) cannot be perfectly discriminated by using any non-classical probe of the same energy.
We revisit this interpretation in Section~\ref{sec: metrology}, whilst a more extensive discussion of when the bounds are non-trivial and whether they can be saturated is provided in Appendix~\ref{app: triviality}.

\section{Examples: Out-of-distribution generalisation for general states}\label{sec: general examples}

We will consider three different types of non-classical input state, to demonstrate how the various bounds can be applied: single-photon-added thermal states (SPATSs), Fock states, and one-mode squeezed vacuums. SPATSs have finite negativity. Fock states have infinite negativity for any non-zero energy, but their negativity becomes finite with convolution alone (without the need for an energy truncation). Squeezed vacuums require both truncation and convolution to reach finite negativity.

\subsection{Single-photon-added thermal states}\label{sec: SPAT}

SPATSs have the P-representation~\cite{kiesel_experimental_2008}
\begin{equation}
    P_{\mathrm{SPAT}}(r e^{i\phi}) =
    \frac{1+q}{\pi q^3} \bigg(r^2 -\frac{q}{1+q}\bigg)e^{-\frac{r^2}{q}},
    \label{eq: P SPAT}
\end{equation}
where $q > 0$ is the average photon number of the thermal state before the photon addition (a conditional process, involving post-selection). $q$ is connected to the average photon number of the SPATS by $\bar{n}=1+2q$. From Eq.~(\ref{eq: P SPAT}), we can see that the P-representation has no angular dependence and starts negative for small $r$ before becoming positive for $r^2 > \frac{q}{1+q} = \frac{\bar{n}-1}{\bar{n}+1}$. We can explicitly calculate $\mathcal{N}$ and $\bar{n}_{\pm}$:
\begin{equation}
    \mathcal{N} = e^{-\frac{1}{1+q}}\bigg(1+\frac{1}{q}\bigg) - 1,
    \quad \bar{n}_{-} = 1 + 2q - \frac{1}{1+ \Big( 1-e^{\frac{1}{1+q}} \Big) q},
    \quad \bar{n}_{+} = 1+2q-\frac{1}{1+q}.
\end{equation}
We can now use the bound in Eq.~(\ref{eq: finite negativity bound pm}), or can simplify our calculations by instead using the bound in Eq.~(\ref{eq: finite bound with mu}), with
\begin{equation}
    \mu = 2e^{-\frac{1}{1+q}} \frac{1+q}{q} - 1,
    \quad \frac{\nu}{\mu} = 1+2q - \frac{2}{2+ 2q-e^{\frac{1}{1+q}}q}
    < 1+2q.
\end{equation}
The bound becomes
\begin{equation}
    \|(\Psi-\Phi)[\rho_{\mathrm{SPAT}}]\| \leq
    \left(2e^{-\frac{1}{1+q}}\frac{1+q}{q} - 1\right) \epsilon\left(\epsilon_0, 1+2q - \frac{2}{2+ 2q-e^{\frac{1}{1+q}}q} \right).
    \label{eq: bound SPAT}
\end{equation}
Since $\bar{n}_{-}< \frac{\nu}{\mu} <\bar{n}_{+}$, Eq.~(\ref{eq: bound SPAT}) is no more than a factor of $\frac{\mu}{\mu-1}=2-e^{\frac{1}{1+q}}\frac{q}{1+q}$ looser than the bound from Eq.~(\ref{eq: finite negativity bound pm}).

The quantity $\frac{\nu}{\mu}$ increases approximately linearly with $\bar{n}$ (or $k$). Since the function $\epsilon$ is concave, $\epsilon(\epsilon_0,\frac{\nu}{\mu})$ increases sublinearly with the energy of the SPATS. On the other hand, the negativity diverges for small $\bar{n}$; this is to be expected, as the limiting case of $\bar{n}\to 1$ ($q\to 0$) is the Fock state $\left|1\middle>\middle<1\right|$. The overall bound on the closeness of the two channel outputs for an input SPATS is therefore larger for lower energies (but higher non-classicalities). It should be noted that the bound being larger does not necessarily imply the channel outputs are further away; the bound may simply be looser for such inputs. As briefly discussed in Appendix~\ref{app: SPAT}, it may be possible to apply the technique from Section~\ref{sec: infinite negativity} of using a sequence of states with lower negativities in order to tighten the bounds somewhat.

\subsection{Fock states}\label{sec: Fock}

The Fock state $|m \rangle \langle m|$ has the P-representation
\begin{equation}
    P_{s,\mathrm{Fock}}(re^{i\phi})
    = \frac{(-1)^m}{\pi}\frac{(1-s)^m}{s^{m+1}}e^{-\frac{1}{s} r^2}\mathrm{L}_m\left[\frac{r^2}{s(1-s)}\right].\label{eq: Ps Fock on-diag}
\end{equation}
In Appendices~\ref{app: sigma distance} and \ref{app: sigma properties}, we show that the values of the relevant quantities are
\begin{equation}
    \mu^{(\mathrm{UB})}_s =
    2\frac{(1-s)^{m+1}}{s^{m}(1-2s)},
    \quad \nu^{(\mathrm{UB})}_s =
    4\frac{(1-s)^{m+2}}{s^{m-1}(1-2s)^2},
    \quad \delta_s=2\sqrt{1-\frac{G[m,m,0,s]}{(1+s)^{2m+1}}},
    \quad G[m,m,0,s] = \sum_{k=0}^m \binom{m}{k}^2 s^{2k},
\end{equation}
where the function $G$ is a polynomial defined in Appendix~\ref{app: sigma distance} (Eq.~(\ref{eq: polynomial})), so that the bound in Eq.~(\ref{eq: convolution only}) becomes
\begin{equation}
    \|(\Psi-\Phi)[|m \rangle \langle m|]\| \leq
    2\frac{(1-s)^{m+1}}{s^{m}(1-2s)} \epsilon\left(\epsilon_0, \frac{2s(1-s)}{1-2s}\right) +4\sqrt{1-\frac{G[m,m,0,s]}{(1+s)^{2m+1}}}.
\end{equation}
The first term is exponential in the energy of the Fock state, since it is proportional to $(\frac{1-s}{s})^m$, whilst the second term is sublinear in $m$. The $s$-dependence is less straightforward, and depends on the form of $\epsilon$, but the multiplicative factor quickly becomes large for small $s$ and $m>0$.

\subsection{One-mode squeezed vacuums}\label{sec: 1msv}

A one-mode squeezed vacuum with squeezing parameter $r=\mathrm{arctanh}(\lambda)$ (and average photon number $\bar{n} = \sinh(r)^2 = \frac{\lambda^2}{1-\lambda^2}$) has density matrix~\cite{weedbrook_gaussian_2012}
\begin{equation}
    \rho_{\mathrm{sq}}
    = \sqrt{1-\lambda^2} \sum_{p,q=0}^{\infty} \frac{\lambda^{p+q}\sqrt{(2p)!(2q)!}}{2^{p+q}p!q!}
    |2p \rangle \langle 2q|.
    \label{eq: 1msv density}
\end{equation}
$\lambda$ lies in the range $(-1,1)$, but, for convenience, we assume positive values (similar results can be derived for $\lambda<0$). Note that we should also restrict to odd $M$ (even-dimensional truncated systems).
In Appendix~\ref{app: squeezed vacuum}, we show that
\begin{equation}
    \delta_s = 2\sqrt{1-\bigg(1+s^2+2s\frac{1+\lambda^2}{1-\lambda^2}\bigg)^{-\frac{1}{2}}}.
    \label{eq: deltas 1msv}
\end{equation}
If $s>\frac{\lambda}{1+\lambda}$, then $\mathcal{N}=0$. Otherwise, since squeezed vacuums are pure, $\delta_M$ is bounded by $\delta_M=\sqrt{1-\eta_M}$,
where $\eta_M$ is the probability that a photon counting measurement on $\rho_{\mathrm{sq}}$ gives a result of $M-1$ or less. $\eta_M$ obeys
\begin{equation}
    \eta_M = \sqrt{1-\lambda^2}\sum_{p=0}^{\frac{M-1}{2}} \frac{\lambda^{2p}(2p)!}{4^p (p!)^2}
    = 1- \frac{\beta\big[\lambda^2;\frac{M+1}{2},\frac{1}{2}\big]\Gamma[\frac{M}{2}+1]}{\sqrt{\pi}(\frac{M-1}{2})!}
    \geq 1 - \frac{\lambda^2}{M(1-\lambda^2)},
    \label{eq: eta 1msv}
\end{equation}
where $\beta$ is the incomplete beta function. The exact value is easy to calculate numerically for a specific value of $M$, whilst the bound can be used for a simpler analytic expression. We find that $\mu$ is upper bounded by
\begin{align}
    &\mu^{(\mathrm{UB})}_{s,M}
    = \frac{2(1-s)\sqrt{1-\lambda^2}}{\eta_M(1-2s)} \bigg( \frac{4(y^{\frac{M+1}{2}}_1 - 1)}{\pi\sqrt{1+x}(y_1 - 1)}
    + \frac{y^{\frac{M+1}{2}}_2 - 1}{y_2 - 1} \bigg),
    \label{eq: mu bound 1msv}\\
    & y_1 = \frac{\lambda (1-s) (1+x)}{s(1-2s)},
    \quad y_2 = \frac{\lambda (1-s) x}{s(1-2s)},
    \quad x=\frac{(1-2s)(1-s)\lambda}{s}.
    \label{eq: x y defs}
\end{align}
The approximate scaling of the negativity with $s$, $M$, and $\lambda$ (for $s\ll\frac{\lambda}{1+\lambda}$) is $\mathcal{O}[\lambda^{M-1} s^{1-M}]$. 

We can then construct a piecewise bound, with
\begin{equation}
    \|(\Psi-\Phi)[\rho_{\mathrm{sq}}]\| \leq
    \epsilon\left(\epsilon_0, \frac{\lambda^2}{1-\lambda^2}+s\right) +4\sqrt{1-\bigg(1+s^2+2s\frac{1+\lambda^2}{1-\lambda^2}\bigg)^{-\frac{1}{2}}}
    \quad \mathrm{for~} s\geq \frac{\lambda}{1+\lambda},
    \label{eq: 1msv piecewise}
\end{equation}
and Eq.~(\ref{eq: bound infinite negativity full}), substituting in Eqs.~(\ref{eq: deltas 1msv}), (\ref{eq: eta 1msv}), and (\ref{eq: mu bound 1msv}), otherwise. This bound is discontinuous in $s$, so exact optimisation over $s$ and $M$ may be tricky; often, it may be easier to set $s=\frac{\lambda}{1+\lambda}=\sqrt{\bar{n}(\bar{n}+1)}-\bar{n}$ and use Eq.~(\ref{eq: 1msv piecewise}).

\subsection{Tightness of the bounds for specific examples}\label{sec: tightness}

By looking numerically at some specific examples of channels and input states, we can investigate whether and when the bounds can be tight. If there exist examples of channels and input states that come close to saturating the various bounds, it would suggest that those bounds cannot be greatly improved whilst maintaining full generality. Using the input states studied previously in this section as examples, we performed a brief numerical investigation of this type, the full details of which are presented in Appendix~\ref{app: triviality}.

\begin{figure}[t]
	\centering
	\includegraphics[width=0.9\textwidth]{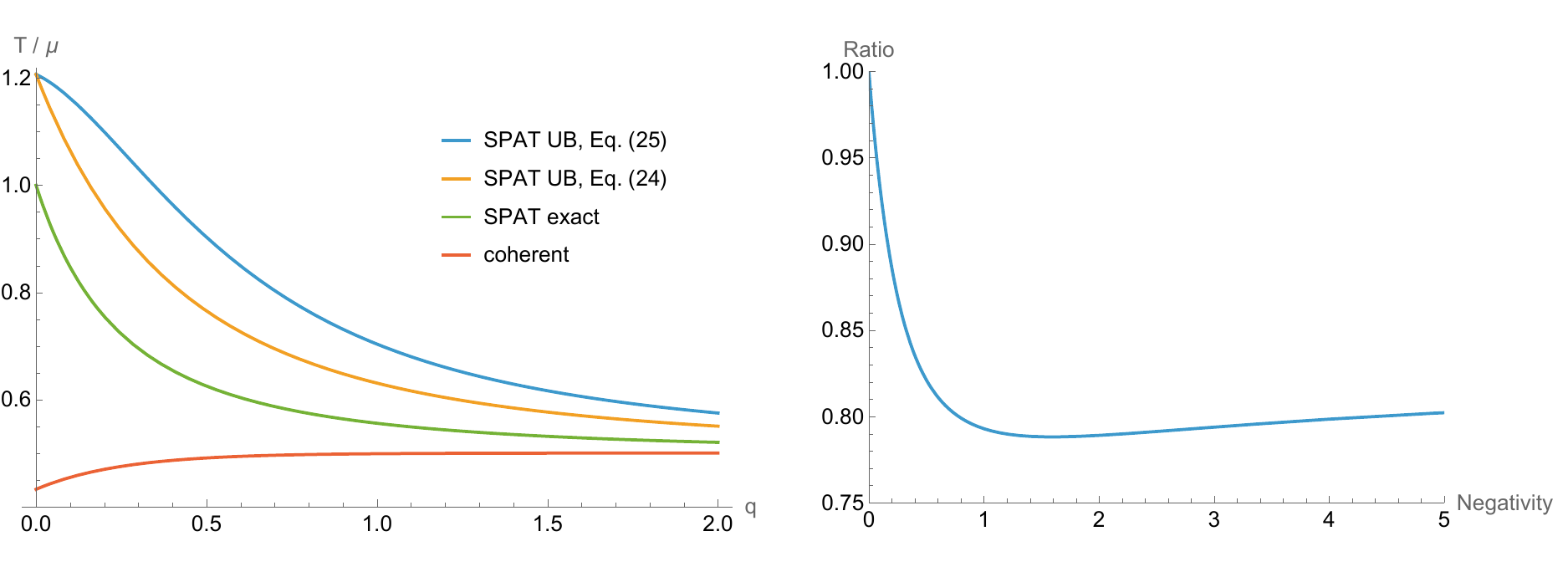}
	\caption{The plot on the left shows the output trace norm, $T$, up to a factor of $\gamma$, for a pair of parity channels. The green line shows the exact value for SPATSs, whilst the orange and blue lines show the bounds coming from Eqs.~(\ref{eq: finite negativity bound pm}) and (\ref{eq: finite bound with mu}) respectively. The red line shows the output trace norm for a coherent state of the same energy. Note that the x-axis tracks $q$, the average photon number of the thermal state before the photon addition, which is connected to the average photon number of the SPATS by $\bar{n}=1+2q$, so that $q=0$ corresponds to $\bar{n}=1$.
    The graph on the right plots the negativity of SPATSs against the ratio between the exact output trace norm and the upper bound from Eq.~(\ref{eq: finite negativity bound pm}).}
	\label{fig: SPAT}
\end{figure}

Using SPATS inputs, we were able to find a class of channels, which we dubbed parity channels, for which the upper bound, found by using Eq.~(\ref{eq: finite negativity bound pm}) and the quantities from Section~\ref{sec: SPAT}, is close to the exact output trace norm. We depict this in Fig.~\ref{fig: SPAT}, in which we plot the output trace norm and the bounds, up to a multiplicative factor of $\gamma$ (a parameter of the target and learned channels). We also plot the ratio between the true value and the bound from Eq.~(\ref{eq: finite negativity bound pm}) as a function of the negativity of the corresponding SPATS. We observe that the ratio never drops below $\sim0.788$, suggesting that the bound from Corollary~\ref{cor: finite negativity} can be fairly tight.

On the other hand, using one-mode squeezed vacuum inputs and focusing on phase rotation channels, we find that the upper bound from Section~\ref{sec: 1msv} is much larger than the exact output trace norm. This does not necessarily mean that the bound for the case of infinite negativity is loose in general; we could simply have picked a poor example for the channel class. Indeed, we might suspect that a pair of non-Gaussian channels could have a much larger output trace norm, with the same behaviour for coherent states, than phase rotation channels (which are a particularly simple class of Gaussian channels).

\subsection{When to apply the various bounds}

We have presented a variety of bounds for different kinds of input states, $\rho$. The more information we have about $\rho$, the tighter our bounds can be, as demonstrated previously in this section, where we have calculated state-specific bounds. As a rule, explicit calculation of, or state-specific bounds on, the various quantities ($\mathcal{N}$, $\delta_s$, $\delta_M$, etc.) results in tighter bounds, whilst more general bounds on the quantities give looser bounds (see Appendix~\ref{app: triviality} for more discussion of the tightness of the bounds). In Table~\ref{tab: table of bounds}, we present a brief summary of when and how the various bounds should be used.

\begin{table}[ht]
    \centering
    \renewcommand{\arraystretch}{1.2}
    \begin{tabular}{|p{35mm}|p{31mm}|p{41mm}|p{31mm}|p{31mm}|}
        \hline
        \textbf{Type of state} &\textbf{Examples} &\bm{$\mathcal{N}$} \textbf{and} \bm{$\bar{n}_{\pm}$} \textbf{or} \bm{$\mu$} \textbf{and} \bm{$\nu$} &\bm{$\delta_s$} &\bm{$\delta_M$} \\
        \hline
        Classical &Mixture of Gaussians &$\mathcal{N}=0$ &$\delta_s=0$ &$\delta_M=0$\\
        \hline
        Finite negativity &SPATSs &Calculate $\mathcal{N}$ and $\bar{n}_{\pm}$ (analytically or numerically). &$\delta_s=0$ &$\delta_M=0$\\
        \hline
        Infinite negativity, finite after any convolution &Fock states &Calculate analytically or use Eqs.~(\ref{eq: mu UB full}), (\ref{eq: mu elements def}), and (\ref{eq: nu mu ratio UB}). &Calculate analytically or use Eq.~(\ref{eq: sigma distance}). &$\delta_M=0$\\
        \hline
        Infinite negativity, still infinite after small convolutions &One-mode squeezed vacuums &Calculate analytically or use Eqs.~(\ref{eq: mu UB full}), (\ref{eq: mu elements def}), and (\ref{eq: nu mu ratio UB}). For larger $s$, $\mathcal{N}$ may be $0$. &Calculate analytically or use Eq.~(\ref{eq: sigma distance}). &Calculate analytically or use Eq.~(\ref{eq: deltaM bound}). For larger $s$, we may not need to truncate.\\
        \hline
        General/unknown states &Arbitrary\newline P-representation &Use Eq.~(\ref{eq: mu ultimate UB}). &Use Eq.~(\ref{eq: sigma distance}). &Use Eq.~(\ref{eq: deltaM bound}).\\
        \hline
    \end{tabular}
    \renewcommand{\arraystretch}{1}
    \caption{A summary of the various bounds and when each should be applied.}
    \label{tab: table of bounds}
\end{table}

\section{Quantum process tomography, metrology, and machine learning interpretations of channel learning and comparison with existing works}\label{sec: metrology}

The language and framing of our theorems is more typical of a ``learning" framework than of metrology or process tomography, however this same problem is an important task in all three fields. Our results have general applicability, and so we set out here the connections between the various frameworks. However, this is not intended to be an exhaustive review of quantum machine learning, CV channel tomography, or quantum metrology.

First, let us describe the task of finding a bound of the form $\|(\Psi-\Phi)[| r e^{i\phi} \rangle \langle r e^{i\phi} |_{\mathrm{coh}}]\|\leq \epsilon_0$ from a learning perspective.
$\Phi$ could take the form of a physical device or process that we can choose optimal parameters for, but it could also simply represent our state of knowledge about the target. E.g., it could be a quantum process matrix or an equation, stored on a classical computer, that we can use to calculate the output state for a given input.
The goal is to learn such a channel using a finite number, $N$, of sample states of the form $\Psi[| r e^{i\phi} \rangle \langle r e^{i\phi} |_{\mathrm{coh}}]$ (where $r\leq \tau$).
A learning algorithm might need multiple samples for each value of $re^{i\phi}$ as well as channel outputs for many different values of $re^{i\phi}$.
On a quantum device, an algorithm of this type could work by using samples of both $\Psi[| r e^{i\phi} \rangle \langle r e^{i\phi} |_{\mathrm{coh}}]$ and $\Phi[| r e^{i\phi} \rangle \langle r e^{i\phi} |_{\mathrm{coh}}]$ (at the same points) to assess and bound $\|(\Psi-\Phi)[| r e^{i\phi} \rangle \langle r e^{i\phi} |_{\mathrm{coh}}]\|$, over the region $r\leq \tau$, for a particular $\Phi$, and then evolving $\Phi$ to a new channel $\Phi'$, with a smaller output distance (using further samples from the target channel at each update step). This would be a form of quantum machine learning.
If, instead, we only have a classical computer, we might perform tomography on all of our samples of $\Psi[| r e^{i\phi} \rangle \langle r e^{i\phi} |_{\mathrm{coh}}]$ and then classically reconstruct $\Psi$, by finding the learned channel, $\Phi$, that best reproduces the measurement results amongst all channels in a certain class.
These two scenarios are depicted in parts (a) and (b) of Fig.~\ref{fig: three_steps}.
In Appendix~\ref{app: measurements}, we take a brief look at the sample efficiency of possible measurements.

Working in the learning framework can provide us with useful tools for formulating guarantees on the output trace distance. If the ``cost function" that we minimise is the maximum value of the output distance, evaluated over the training region, then we are immediately working in the formalism of our theorems. If the inputs are sampled randomly from the training region, we can apply in-distribution generalisation techniques~\cite{arunachalam_survey_2017,huang_power_2021,banchi_generalization_2021,caro_generalization_2022} to bound (with high probability) the difference between the observed and true cost functions. Alternatively, we could choose the sampled inputs according to some fixed scheme that guarantees nowhere in the training region is far from a sampled point; bounds over the region would then follow from the data processing inequality.
In Ref.~\cite{volkoff_universal_2021}, it is shown that certain cost functions based on Bell state or coherent state inputs are ``faithful", for the problem of learning unitaries, in the sense that they only reach $0$ when the learned and target unitaries coincide. This is in line with our Lemma~\ref{th: eps convergence}.

In the language of quantum process tomography, characterising a quantum process involves learning the quantum process matrix that relates the elements of the input density matrix to the elements of the output density matrix, in some basis. However, for a completely arbitrary CV channel, this matrix has infinite elements. A number of works have therefore developed the theory of coherent state quantum process tomography (csQPT)~\cite{lobino_complete_2008,rahimi-keshari_quantum_2011,anis_maximum-likelihood_2012}. We will look in a little more detail at two foundational works.

The idea (as in this work) is to exploit the P-representation of Fock states in order to construct the process matrix. In Ref.~\cite{lobino_complete_2008}, the process matrix is constructed by characterising the outputs for coherent state inputs ($\Psi[| \alpha \rangle \langle \alpha |_{\mathrm{coh}}]$) and directly substituting them into the P-representations of the Fock state elements. The authors deal with the singularity of the P-representation using the Klauder approximation, which works similarly to the convolution with a Gaussian that we apply in Section~\ref{sec: infinite negativity}. Ref.~\cite{rahimi-keshari_quantum_2011} instead differentiates the Fock basis elements of $\Psi[| \alpha \rangle \langle \alpha |_{\mathrm{coh}}]$ with regard to (the real and imaginary parts of) $\alpha$. In both cases, a Fock basis truncation must be applied to the input.

In both cases, the goal was to write the output of the channel as a function of its inputs. Here, instead, we assume we have such a function and want to rigorously bound its error. Indeed, the function could have been obtained by one of these methods.
Hence, although a Fock decomposition of the input we want to know about may tighten our bounds, we do not need one in order to apply our theorems. Although the method in Section~\ref{sec: infinite negativity} also uses a Fock basis truncation, we also show how this truncation can be taken to infinity as our learning error, $\epsilon_0$, approaches $0$.

Both papers discuss the errors in their output states, however they do so in the context of the additional errors introduced by their approximations/truncations. That is, they discuss the additional error introduced by using a finite Fock cut-off and (in the case of Ref.~\cite{lobino_complete_2008}) Klauder approximation, but assuming the channel is perfectly understood over the sampled input states. Hence, they do not account for the fact that any measurements of the output states will themselves have non-zero errors; here we examine and bound how our imperfect understanding of the channel, even for the set of input states we do test, propagates as we look at inputs outside of that set. Since our work compares the input-output relations of the target and learned channels, it automatically accounts for all sources of error.

Further, in Ref.~\cite{lobino_complete_2008}, interpolation is applied between the input states, to understand the behaviour for states that lie in-distribution but that were not previously sampled. Whilst this may be physically justified in many cases, a rigorous treatment of errors should bound the worst case scenario; our previous discussion of the learning framework gives examples of how this can be accomplished (in-distribution generalisation bounds or uniform sampling).

Our work could be applied to understand the error in quantum process tomography using, e.g., the method from Ref.~\cite{rahimi-keshari_quantum_2011}. In this case, many of the bounds could be tightened by combining the error analysis from Ref.~\cite{rahimi-keshari_quantum_2011} with our own work, since the learned channel takes a specific form. This could be an interesting topic for future study.

Another physically relevant framework is to see channel learning as a problem of parameter estimation. In many cases, the channel we want to learn about is not some completely arbitrary transformation. Rather, we may know it belongs to some parametrised class of channels and want to learn the (potentially many) parameters, which could represent some important information about a physical process. Examples could include the loss of a medium or the strength of some particular non-linear interaction. Quantum metrology then involves probing the channels in order to learn the parameter values. Errors in parameter estimation of $\mathcal{O}[N^{-\frac{1}{2}}]$ can be achieved using coherent states, and better error scalings are possible using clever strategies.

For the important class of Gaussian channels, protocols have been proposed that use csQPT to perform parameter estimation~\cite{wang_efficient_2013,teo_highly_2021}. One-mode Gaussian channels are fully characterised by a limited number of parameters and it is possible to learn all of them to within a small uncertainty.

The difficulty, from our perspective, is understanding how errors in the parameters translate to errors in output distance for arbitrary inputs. A small change in displacement is likely to have a drastically different effect on an arbitrary state than a similar magnitude change in a squeezing parameter. This is especially true if we consider non-Gaussian effects. Thus, a condition on the variance of whichever parameter we want to learn (often the goal in parameter estimation) does not automatically translate to a useful bound on the output distance of our target and learned channels. In Section~\ref{sec: coherent examples}, we illustrate via examples how knowledge about the output distance can be converted to bounds on the parameter difference and used to construct $\epsilon(\epsilon_0,r^2)$.
This works both ways: for many channel classes, we may be able to convert a bound on the difference in parameters into a bound on the coherent state output distance and then apply our bounds for general states. This is the case regardless of how we estimated the parameters (i.e., even if we learned them using a different limited set of inputs to the low energy coherent states, such as a small number of non-coherent Gaussian probes~\cite{bina_continuous-variable_2018,candeloro_quantum_2021}), so our results are not limited to a specific model of quantum metrology.

We can connect parameter estimation to the idea of quantum machine learning. If we have an optical circuit that we want to imitate our target channel as closely as possible, we would do this by tuning the parameters of our circuit so that the outputs are as close to the target outputs as possible. We can view this as estimating the optimal parameters.

The precise form of the bound for low energy coherent state inputs that we use as our basic ingredient may not be the format in which the error of a sensing protocol is presented elsewhere. However, we chose this form for its generality, as any similar bound on the output distance can be rewritten in this way.

Finally, we note that our results are also of interest from a purely channel discrimination perspective (i.e., separately from both learning and quantum metrology).
The output trace distance, $\|(\Psi-\Phi)[\rho]\|$, corresponds to the optimal probability of successfully deciding whether an unknown channel is $\Psi$ or $\Phi$ by inputting $\rho$ and measuring the output. We are often interested in the success probability optimised over both probes and measurements, but in the CV setting, we must generally impose an energy constraint on the input state.
Our results can therefore be interpreted as bounds on the probability of successfully discriminating between two channels, given any probe of a certain energy. One difference between this setting and ours is that, typically, in channel discrimination tasks, we know the two channels that we want to choose between. Here, we give bounds for all pairs of channels constrained only by a certain maximum success probability with low energy coherent state probes.

As discussed at the end of Section~\ref{sec: infinite negativity}, our results can therefore be used to bound quantum advantage in channel discrimination tasks. Per Fig.~\ref{fig: non-trivial region}, if the maximum discrimination probability using a coherent state probe is close enough to $\frac{1}{2}$, then no non-classical probe is able to discriminate with a success probability of $1$ (we could draw a similar region for a different maximum success probability for non-classical probes, e.g., $0.9$).
This is not a particularly strong statement, as it may be intuitive that there exists some limit to the benefit of non-classicality, but it is non-trivial that such a limit should exist for every possible pair of channels that are very close for classical inputs, even if we pick extremely non-Gaussian channels and input states that are tailored to show as much quantum advantage as possible.

Further, as a counterpoint to the intuition that there should exist a limit to the benefit of non-classicality, we note that when using non-classical states for quantum metrology, and using the quantum Fisher information (QFI) as our metric, no such limit exists. At any finite energy, there exist ``infinite precision" probe states that result in an output with infinite QFI, even for a Gaussian (phase rotation) encoding channel (strictly speaking, infinite QFI is obtained as the limit of a sequence of physically achievable, fixed energy probes with finite and increasing QFI)~\cite{moore_secure_2024}. As such, for metrology, there is no upper bound on the benefit of non-classicality (at least in terms of QFI), so it is non-trivial that such a bound should exist for channel discrimination.
We also note that the topic of ultimate bounds for channel discrimination has largely been broached in the more limited cases of DV or teleportation covariant channels, rather than the extremely general case of arbitrary CV channels~\cite{pirandola_fundamental_2019,zhuang_ultimate_2020}.

In many channel discrimination settings, we are also allowed an idler system, i.e., a system that does not pass through the unknown channel, but that may be entangled with the probe. In such contexts, we are often interested in the diamond norm, which is the output trace norm maximised over all probes, including those entangled with idler systems. In CV systems, we generally need to apply an energy constraint, arriving at the energy-constrained diamond norm~\cite{winter_energy-constrained_2017,pirandola_fundamental_2017,shirokov_energy-constrained_2018}. If our results were extended to multi-mode channels, it would be possible to also bound the energy-constrained diamond norm in terms of the output distance for classical states. This would provide a fully general bound on the quantum advantage of every channel discrimination task, and would be an interesting extension of this work.

\section{Conclusion}\label{sec: conclusion}

When learning the action of a quantum channel, it is not possible to test every possible input state. Instead, we can aim to find a learned channel that reproduces the input-output relations for simple, low energy, classical probes. We want to be sure that this learned channel also reproduces the input-output relations for higher energy and non-classical inputs, and to understand how an error in learning the channel over low energy coherent states propagates when considering inputs outside of the sampled distribution.

In this paper, we have shown that a bound on the distance of between the outputs of two channels for low energy coherent state inputs can always give rise to a bound on the output distance for higher energy coherent state inputs, as long as the initial bound is tight enough. Such a bound, in turn, gives rise to a bound on the output distance for all input states. This bound converges to $0$ as the output distance for coherent state inputs converges to $0$.
This is a useful result -- with implications for subjects such as quantum machine learning and quantum metrology, amongst others -- as it shows that it is sufficient to probe a process with experimentally simple, classical probes.
We have given general statements that hold for all classes of channel and for all input states, as well as looking at specific, useful examples of channel classes and types of input state.

One avenue for future research is to extend this result from one-mode channels to multi-mode channels.
This is likely to be a simple extension, since all multimode CV states also have a P-representation (i.e., a quasiprobability distribution over multimode coherent states), however it would require recalculating the distances $\delta_s$ and $\delta_M$ and reformulating the bound on $\mu_{s,M}$ for this more general setting. The energy truncation would also need to be handled with care, since there are different ways of applying an energy truncation in the multimode case (local or global). This research direction would be interesting, as it would allow one to formulate bounds that account for idler modes, and so bound the diamond norm as well as the output trace norm.

Another option would be to look at other specific classes of channels or input states. We have looked at some examples in this work, but bounds for specific, physically relevant channels or states could allow our results to be applied more broadly.

Many of the existing bounds could also be tightened. From numerically studying some examples (see Appendix~\ref{app: triviality}), it appears that at least Eqs.~(\ref{eq: generic UB}) and (\ref{eq: mu bound 1msv}) can be significantly tightened. It is also possible that there is a more optimal sequence of states that we could use in the infinite negativity case (rather than $\sigma_{s,M} = \mathcal{C}_s[\rho^{(M)}]$).

Finally, it would be useful to bound the worst case scaling of the function $\epsilon$ with $r^2$; in this work we have only demonstrated that there exists an $\epsilon$ that approaches $0$ as $\epsilon_0$ approaches $0$.

\smallskip
\begin{acknowledgments}
The authors acknowledge financial support from:
CLUSTEC, which has received funding from the European Union's Horizon Europe programme, under grant agreement No.~101080173 (J.L.P.);
NGI Enrichers, which has received funding from the European Union’s Horizon Europe Research and Innovation Programme, under grant agreement 101070125 (J.L.P.);
U.S. Department of Energy, Office of Science, National Quantum Information Science Research Centers, Superconducting Quantum Materials and Systems Center (SQMS), under the Contract No. DE-AC02-07CH11359 (J.L.P. and Q.Z.);
the European Union’s Horizon Europe research and innovation program
under EPIQUE Project GA No. 101135288 (L.B.). 
Q.Z. also acknowledges support from NSF (CCF-2240641, OMA-2326746, 2350153), ONR (N00014-23-1-2296), AFOSR MURI FA9550-24-1-0349 and DARPA (MeasQUIT HR0011-24-9-0362, HR00112490453, D24AC00153-02). Part of the work was carried out during J.L.P.'s research visit to USC.
J.L.P. also thanks Matteo Paris and Sean Moore for helpful discussions.
\end{acknowledgments}

\bigskip

\appendix

\section{Measurements of the outputs}\label{app: measurements}

Let us consider how we can assess $\|(\Psi-\Phi)[| r e^{i\phi} \rangle \langle r e^{i\phi} |_{\mathrm{coh}}]\|$ for a particular value of $re^{i\phi}$. One option is to carry out tomography on our copies of $\Psi[| r e^{i\phi} \rangle \langle r e^{i\phi} |_{\mathrm{coh}}]$ (up to some trace norm error $\eta$) and then assess the output distance classically.
Another option, if we are carrying out the learning algorithm using a quantum device capable of producing copies of $\Phi[| r e^{i\phi} \rangle \langle r e^{i\phi} |_{\mathrm{coh}}]$, is to directly measure $\|(\Psi-\Phi)[| r e^{i\phi} \rangle \langle r e^{i\phi} |_{\mathrm{coh}}]\|$, without explicitly obtaining classical descriptions of the output states.

There exist a variety of methods for performing tomography on CV states, some of which are specific to certain types of state. Some more general methods involve the reconstruction of the Husimi Q-function or the Wigner function by applying homodyne, heterodyne, or other measurements, after displacing, squeezing, or otherwise operating on the state~\cite{lvovsky_continuous-variable_2009,landon-cardinal_quantitative_2018,kalash_wigner_2023}. If we know the output states have a property called reflection symmetry, we can estimate their characteristic function at several points~\cite{wu_efficient_2024}. If we know the outputs are Gaussian states, we need only accurately learn their first and second moments in order to fully characterise them. In Ref.~\cite{mele_learning_2024}, it is shown that we need only learn these moments up to $\mathcal{O}[\eta^2]$, and explicit bounds on the trace distance between Gaussian states are given, based on the difference in first and second moments and the average energies of the states.

The downside of tomography based methods is that they can be extremely inefficient for CV states. Even for pure states, one requires $S=\mathcal{O}[E \eta^{-2}]$ copies for tomography of a non-Gaussian output, and at least $S=\mathcal{O}[E^2 \eta^{-2}]$ copies (at most $S=\mathcal{O}[E^2 \eta^{-3}]$ copies) are required for mixed states~\cite{mele_learning_2024}. Here, $E$ is the average energy of the output that we want to characterise. Crucially, this is not necessarily the same as $r^2$, the energy of the input coherent state, and could, depending on the target channel, be much higher (and also may not be known a priori). Note that the error in estimation, $\eta$, must be added to the assessed value of $\|(\Psi-\Phi)[| r e^{i\phi} \rangle \langle r e^{i\phi} |_{\mathrm{coh}}]\|$, when formulating the bound.

For pure output states, a direct assessment of the output distance can be accomplished using the SWAP test. The SWAP test is a two-outcome measurement on a pair of states, $\psi_1$ and $\psi_2$, that outputs $0$ with probability $\frac{1}{2}+\frac{1}{2}\mathrm{Tr}[\psi_1 \psi_2]$. If the outputs, $\Psi[| r e^{i\phi} \rangle \langle r e^{i\phi} |_{\mathrm{coh}}]$ and $\Phi[| r e^{i\phi} \rangle \langle r e^{i\phi} |_{\mathrm{coh}}]$, are pure, this immediately tells us their squared fidelity, up to an error of $\mathcal{O}[\sqrt{S^{-1}}]$. Applying the Fuchs-van de Graaf relation, we get $S=\mathcal{O}[\eta^2]$. Though this error scaling may seem similar to the tomography expressions, a crucial difference is that it is entirely independent of the energy of the output state.

The SWAP test can be enacted using a controlled beamsplitter~\cite{nguyen_experimental_2021,pietikainen_controlled_2022}. However, this is a non-Gaussian component, and so may be difficult to implement. In Ref.~\cite{volkoff_ancilla-free_2022}, it is shown that the SWAP test can be implemented in the CV setting using only linear optics and photon counting. Thus, for unitary $\Psi$ and $\Phi$, we can assess the output distance using just a single beamsplitter and photon counters (although the finite threshold of any real photon counter will introduce errors).

For mixed state outputs, the SWAP test only lets us learn the overlap of $\Psi[| r e^{i\phi} \rangle \langle r e^{i\phi} |_{\mathrm{coh}}]$ and $\Phi[| r e^{i\phi} \rangle \langle r e^{i\phi} |_{\mathrm{coh}}]$. This would not give a bound on the output distance that converges to $0$ when the target and learned channels coincide. One alternative is to use the implementation of the Helstrom measurement from Ref.~\cite{lloyd_quantum_2020}. This method is based on the state exponentiation algorithm~\cite{lloyd_quantum_2014} and phase estimation, and it uses copies of (unknown) states $\psi_1$ and $\psi_2$ to apply the optimal discriminator between $\psi_1$ and $\psi_2$ to an arbitrary state. In Ref.~\cite{banchi_statistical_2024}, it was shown that enacting the Helstrom measurement on a single state with a failure probability of $\zeta$ requires $S=\mathcal{O}[-\log(\zeta)\zeta^{-3}]$ copies of each state. Since the optimal measurement should correctly discriminate between the states with a probability linear in the trace distance (in fact, this is precisely its operational meaning), we could estimate the output distance by applying the Helstrom measurement to multiple copies of the output states and finding the error probability. To estimate it to error probability $\mathcal{O}[\zeta]$ (so that the two contributors to the error are equal), we would carry out the protocol $\sim \zeta^{-2}$ times, and so the total number of copies required is $S=\mathcal{O}[-\log(\zeta)\zeta^{-5}]$. At first glance, this is quite costly, but the complete lack of dependence on the output states themselves (including the lack of dependence on $E$) could make this algorithm more efficient than tomography, in some circumstances. In terms of implementation, however, this would require many uses of a controlled beamsplitter, and so would not be practical for near-term applications.

\section{Out-of-distribution generalisation for coherent states when $\epsilon_0=0$}\label{app: oodg for coherent exact}

We want to show that if two channels (i.e., the target channel and the learned channel) have exactly the same output for low energy coherent state inputs, then they must also have the same output for any coherent state input. More precisely, suppose $\|(\Psi-\Phi)[| r e^{i\phi} \rangle \langle r e^{i\phi} |_{\mathrm{coh}}]\| = 0$ for $r\leq \tau$. Then, this holds for $r>\tau$ too. Whilst intuitive, this is worth proving, since it guarantees that if we learn a channel exactly over a small area of the phase space, we have also learned it over all of phase space. 

If $\|(\Psi-\Phi)[| r e^{i\phi} \rangle \langle r e^{i\phi} |_{\mathrm{coh}}]\| = 0$ for $r\leq \tau$, then
\begin{equation}
    e^{-r^2}\Bigg\| \sum_{m,n=0}^{\infty} \frac{r^{m+n}e^{i(m-n)\phi}}{\sqrt{m!n!}} (\Psi-\Phi) \big[ | m \rangle \langle n | \big] \Bigg\|
    = 0
    \quad \forall~r\leq\tau.
\end{equation}
Choosing any (infinite-dimensional) basis, every element of $\sum_{m,n=0}^{\infty} \frac{r^{m+n}e^{i(m-n)\phi}}{\sqrt{m!n!}} (\Psi-\Phi) \big[ | m \rangle \langle n | \big]$ must equal $0$, for any $r\leq\tau$.
In particular, we choose the Fock basis.
Then, the condition can only be fulfilled if every element of $(\Psi-\Phi) \big[| m \rangle \langle n | \big]$ is identically $0$, so $\|(\Psi-\Phi)[| r e^{i\phi} \rangle \langle r e^{i\phi} |_{\mathrm{coh}}]\| = 0$ for all $r$ (including $r>\tau$).
This is also in line with what we expect from~\cite{rahimi-keshari_quantum_2011}, since it is shown there that the quantum process matrix of a channel, $\Psi$, is completely determined by all of the (infinite) derivatives of $\|\Psi[| r e^{i\phi} \rangle \langle r e^{i\phi} |_{\mathrm{coh}}]\|$ assessed at the origin. Since they must all be the same for both $\Psi$ and $\Phi$ over a finite region that includes the origin, the processes are identical everywhere.

We revisit the case of $\epsilon_0 > 0$ in Appendix~\ref{app: oodg for coherent bound}.

\section{Gaussian channels}\label{app: gaussian}

The output fidelity for Gaussian channels takes the form given in Eq.~(\ref{eq: output fidelity Gaussian gen form}). By bounding $x$, $y$, and $z$ individually, we can bound the output fidelity for any $r$. We have a constraint on the output distance for $r\leq \tau$, which translates into a lower bound on the fidelity via a Fuchs-van de Graaf inequality. Specifically, for all $r\leq \tau$ (and replacing $F$ with $F^2$ for pure states),
\begin{equation}
    F\Big[ \mathcal{G}_1 \big[ | r e^{i\phi} \rangle \langle r e^{i\phi} |_{\mathrm{coh}} \big],\mathcal{G}_2 \big[ | r e^{i\phi} \rangle \langle r e^{i\phi} |_{\mathrm{coh}} \big] \Big] \geq \frac{2-\epsilon_0}{2}.
\end{equation}

At $r=0$, the output fidelity is $x$, so the lower bound on $x$ follows directly. Now consider $r=\pm\tau$, for which we have
\begin{equation}
    xe^{-y\tau^2 \mp z\tau} \geq \Big(\frac{2-\epsilon_0}{2}\Big)^2.
\end{equation}
If we want to maximise $y$ and $|z|$, we should make $x$ as large as possible, i.e., $1$. To bound $y$, we set $z=0$ and get
\begin{equation}
    e^{-y\tau^2} \geq \Big(\frac{2-\epsilon_0}{2}\Big)^2,
\end{equation}
and so recover our bound on $y$. Finally, we bound $|z|$ by comparing $F^2$ at $r=0$ and at $r=\pm \tau$. Concavity can be confirmed by differentiating Eq.~(\ref{eq: gaussian epsilon}) twice with regard to $r^2$.

Applying single-mode squeezing to $|\alpha\rangle_{\mathrm{coh}}$, the first and second moments become
\begin{equation}
    V =
    \begin{pmatrix}
        e^{2s} &0\\
        0 &e^{-2s}
    \end{pmatrix},
    \quad x =
    \begin{pmatrix}
        2e^{s}\alpha_R\\
        2e^{-s}\alpha_I
    \end{pmatrix},
\end{equation}
so the output fidelity between a target and learned squeezing unitary is $F^2_{\mathrm{sq}} = \mathrm{sech}(s_T-s_L)\exp\big[2|\alpha|^2 (\mathrm{sech}(s_T-s_L)-1)\big]$. If we are again given $\|(\Psi-\Phi)[| \tau e^{i\phi} \rangle \langle \tau e^{i\phi} |_{\mathrm{coh}}]\|\leq \epsilon_0$ for some $\tau$, we get the condition $\mathrm{sech}(s_T-s_L)\leq \frac{1}{2\tau^2}W_0\big[e^{2\tau^2}\tau^2(2-\frac{1}{2}\epsilon_0^2)\big]$, where $W_0$ is the Lambert W function. The right hand side goes to $1$ as $\epsilon_0\to 0$, as expected. We therefore recover Eq.~(\ref{eq: squeezing epsilon}).

\section{Out-of-distribution generalisation for classical states}\label{app: classical}

From the convexity of the trace norm, we can write
\begin{equation}
    \|(\Psi-\Phi)[\rho_{\mathrm{class}}]\| \leq \int_{0}^{\infty} \epsilon(\epsilon_0,r^2) \int_{0}^{2\pi} P_{\mathrm{class}}(re^{i\phi}) r d\phi dr = \int_{0}^{\infty} \epsilon(\epsilon_0,r^2) p_{\mathrm{class}}(r) dr,\label{eq: classical trace norm}
\end{equation}
where we have defined $p(r) = \int_{0}^{2\pi} P(re^{i\phi}) r d\phi$. Using this definition, $\int_{0}^{\infty} p(r) dr = 1$ for any normalised P-representation, and our energy constraint takes the form $\int_{0}^{\infty} r^2 p(r) dr \leq \bar{n}$.
Now we apply Jensen's inequality for concave functions. The right hand side of Eq.~(\ref{eq: classical trace norm}) can be expressed as the expectation value of $\epsilon(\epsilon_0,r^2)$ over the classical probability density function $p_{\mathrm{class}}(r)$. For any valid probability density function
\begin{equation}
    \mathbb{E}_{p_{\mathrm{class}}}[\epsilon(\epsilon_0,r^2)]
    \leq \epsilon(\epsilon_0,\mathbb{E}_{p_{\mathrm{class}}}[r^2])
    \leq \epsilon(\epsilon_0,\mathbb{E}_{p_{\mathrm{class}}}[\bar{n}]).
\end{equation}
Note that since $\bar{n}$ is the expectation value of $r^2$ and not of $r$, we only require concavity in $r^2$.

\section{Sufficiency of upper bounding $\mu$}\label{app: mu UB}

Recall that the quantity in Eq.~(\ref{eq: bound infinite negativity full}) that we want to bound is $\mu\epsilon(\frac{\nu}{\mu})$, where $\epsilon(x)$ is concave in $x$ and where we have temporarily dropped the dependence on $\epsilon_0$. Recall that $\mu$ and $\nu$ are defined by
\begin{equation}
    \mu = 1 + 2\mathcal{N},\quad
    \nu = (1+\mathcal{N})\bar{n}_{+} + \mathcal{N}\bar{n}_{-}.
\end{equation}
The negativity of a state $\rho$ is defined as the negative volume of the P-representation, i.e.,
\begin{equation}
    \mathcal{N} = -\int_{P(\alpha)[\rho]<0} P(\alpha)[\rho] d^2 \alpha,
\end{equation}
and $\bar{n}_{\pm}$ are defined by
\begin{equation}
    \bar{n}_{+} = \frac{1}{1+\mathcal{N}}\int_{P(\alpha)>0} P(\alpha)|\alpha|^2 d^2 \alpha,\quad
    \bar{n}_{-} = -\frac{1}{\mathcal{N}}\int_{P(\alpha)<0} P(\alpha)|\alpha|^2 d^2 \alpha.
\end{equation}
Hence, $\mu$ and $\nu$ can be calculated as
\begin{equation}
    \mu = \int |P(\alpha)| d^2\alpha,\quad
    \nu = \int |P(\alpha)| |\alpha|^2 d^2\alpha,
\end{equation}
where we integrate over both the negative and the positive domains of the P-distribution, but take the absolute value.

At first glance, upper bounding $\mu\epsilon(\frac{\nu}{\mu})$ requires us to find both an upper and a lower bound on $\mu$ (and an upper bound on $\nu$), since it is used both as a multiplicative factor and as the denominator, inside the function $\epsilon$. However, by applying the concavity, we can show that it suffices to find an upper bound on $\mu$.

Taking the partial derivative of $\mu\epsilon(\frac{\nu}{\mu})$ with regard to $\mu$, we get
\begin{equation}
    \frac{\partial}{\partial \mu}\left[ \mu\epsilon\left(\frac{\nu}{\mu}\right) \right]
    = \epsilon\left(\frac{\nu}{\mu}\right) - \frac{\nu}{\mu}\epsilon'\left(\frac{\nu}{\mu}\right),
\end{equation}
where $\epsilon'$ is the partial derivative of $\epsilon$ with regard to $\mu$. Recalling that $\epsilon'(\frac{\nu}{\mu})$ is the gradient of $\epsilon$ at $\frac{\nu}{\mu}$ and applying the concavity of $\epsilon$, we see that $\frac{\partial}{\partial \mu}\left[ \mu\epsilon(\frac{\nu}{\mu}) \right]\geq 0$ for any (valid) value of $\mu$. Thus, an overestimate of $\mu$ always results in an overestimate of $\mu\epsilon(\frac{\nu}{\mu})$, and so it suffices to use an upper bound of $\mu$, both inside and outside $\epsilon$, when bounding $\mu\epsilon(\frac{\nu}{\mu})$.

\section{Bounding the output distance for $\sigma_{s,M}$}\label{app: sigma properties}

The P-representation, $P_{s,M}$, of $\sigma_{s,M}$ (the state obtained by first truncating $\rho$ to a maximum photon number of $M-1$ and then convolving the resulting P-representation with a Gaussian function of width $s^{-1}$), can be written as
\begin{equation}
    P_{s,M}(\alpha)[\rho] = P_{s,M}(re^{i\phi})[\rho] = 
    \sum_{m,n=0}^{M-1} \langle m|\rho^{(M)}|n\rangle P_s(re^{i\phi})[|m\rangle\langle n|],
\end{equation}
where $|m\rangle$ and $|n\rangle$ are Fock states.
We specify $\langle m|\rho^{(M)}|n\rangle$ rather than $\langle m|\rho|n\rangle$ to emphasise that $\rho^{(M)}$ is normalised.
The contribution to $P_s(re^{i\phi})$ of the terms $|m \rangle \langle n|$ and $|n \rangle \langle m|$, where $m>n$, is
(\cite{wuensche_remarks_1998}, Eq.~(2.15))
\begin{equation}
    \begin{split}
        P_s(re^{i\phi})\left[\frac{1}{2}(e^{i\theta}|m \rangle \langle n| + e^{-i\theta}|n \rangle \langle m|)\right] &= P^{(m,n)}_{s,\theta}(re^{i\phi})\\
        &= \cos(\theta-(m-n)\phi)\frac{(-1)^n}{\pi}\sqrt{\frac{n!}{m!}}\frac{(1-s)^n}{s^{m+1}}e^{-\frac{1}{s}r^2}r^{m-n}\mathrm{L}^{m-n}_n\left[\frac{r^2}{s(1-s)}\right],\label{eq: Ps Fock off-diag}
    \end{split}
\end{equation}
where $\mathrm{L}_x^y$ is a generalised Laguerre polynomial. We include $\theta$ for completeness, but will find it has no effect on our calculations and so will generally drop it from the subscript. For Fock states, $|m \rangle \langle m|$, we get Eq.~(\ref{eq: Ps Fock on-diag}). Eq.~(\ref{eq: Ps Fock on-diag}) has no angular dependence, whilst the angular component of Eq.~(\ref{eq: Ps Fock off-diag}) is a cosine function of period $\frac{2\pi}{m-n}$. We will upper bound $\mu$ and $\nu$ by finding and summing the contributions from each term $P^{(m,n)}_s$; this is just an upper bound and is not tight unless the different contributions to the total P-representations never cancel each other out (which is not generally the case). Noting that the P-representation contributions of the on and off-diagonal elements of $\rho^{(M)}$ have different properties, it is helpful to separate their contributions to $\mu$ and $\nu$.

Starting with the off-diagonals, and using Eq.~(\ref{eq: Ps Fock off-diag}), the exact contributions to $\mu$ and $\nu$ are
\begin{align}
    &\mu[P^{(m,n)}_s]
    = \frac{1}{\pi} \frac{(1-s)^n}{s^{m+1}} \sqrt{\frac{n!}{m!}} \int_{0}^{2\pi} |\cos(\theta - (m-n)\phi)| d\phi \int_{0}^{\infty} e^{-\frac{1}{s}r^2}r^{m-n+1} \left|\mathrm{L}^{m-n}_n\left[\frac{r^2}{s(1-s)} \right]\right| dr,\\
    &\nu[P^{(m,n)}_s]
    = \frac{1}{\pi} \frac{(1-s)^n}{s^{m+1}} \sqrt{\frac{n!}{m!}} \int_{0}^{2\pi} |\cos(\theta - (m-n)\phi)| d\phi \int_{0}^{\infty} e^{-\frac{1}{s}r^2}r^{m-n+3} \left|\mathrm{L}^{m-n}_n\left[\frac{r^2}{s(1-s)} \right]\right| dr.
\end{align}
The integral over $\phi$ is always equal to $4$. Then, applying a bound (by Szeg\"{o}) on the magnitude of Laguerre polynomials,
\begin{align}
    &\mu[P^{(m,n)}_s]
    \leq \frac{4}{\pi} \frac{(1-s)^n}{s^{m+1}} \frac{(m-n+1)_n}{\sqrt{m!n!}} \int_{0}^{\infty} r^{m-n+1} e^{\frac{r^2}{s}\left(\frac{1}{2(1-s)}-1\right)} dr,\\
    &\nu[P^{(m,n)}_s]
    \leq \frac{4}{\pi} \frac{(1-s)^n}{s^{m+1}} \frac{(m-n+1)_n}{\sqrt{m!n!}} \int_{0}^{\infty} r^{m-n+3} e^{\frac{r^2}{s}\left(\frac{1}{2(1-s)}-1\right)} dr,
\end{align}
where $(x)_n$ is the Pochhammer symbol. These integrals converge for $s<\frac{1}{2}$. Evaluating them~\cite{supplemental}, we get
\begin{align}
    &\mu[P^{(m,n)}_s]
    \leq \frac{2}{\pi} \frac{(1-s)^n}{s^{m+1}} \frac{(m-n+1)_n}{\sqrt{m!n!}} \left(\frac{2s(1-s)}{1-2s}\right)^{1+\frac{m-n}{2}} \Gamma\left[1+\frac{m-n}{2}\right],\\
    &\nu[P^{(m,n)}_s]
    \leq \frac{2}{\pi} \frac{(1-s)^n}{s^{m+1}} \frac{(m-n+1)_n}{\sqrt{m!n!}} \left(\frac{2s(1-s)}{1-2s}\right)^{2+\frac{m-n}{2}} \Gamma\left[2+\frac{m-n}{2}\right].
\end{align}
After some simplification, we end up with the inequalities:
\begin{align}
    &\mu[P^{(m,n)}_s]
    \leq \frac{2^{2+\frac{m-n}{2}} (1-s)^{1+\frac{m+n}{2}}}{\pi s^{\frac{m+n}{2}}(1-2s)^{1+\frac{m-n}{2}}} \frac{(m-n+1)_n}{\sqrt{m!n!}} \Gamma\left[1+\frac{m-n}{2}\right],\label{eq: mu mn}\\
    &\nu[P^{(m,n)}_s]
    \leq \frac{2^{3+\frac{m-n}{2}} (1-s)^{2+\frac{m+n}{2}}}{\pi s^{\frac{m+n}{2}-1}(1-2s)^{2+\frac{m-n}{2}}} \frac{(m-n+1)_n}{\sqrt{m!n!}} \Gamma\left[2+\frac{m-n}{2}\right].\label{eq: nu mn}
\end{align}
We now use Eq.~(\ref{eq: Ps Fock on-diag}) to find the contributions from the on-diagonals:
\begin{align}
    &\mu[P^{(m,m)}_s]
    = \frac{1}{\pi} \frac{(1-s)^m}{s^{m+1}} \int_{0}^{2\pi} 1 d\phi \int_{0}^{\infty} e^{-\frac{1}{s}r^2} r \left|\mathrm{L}_m\left[\frac{r^2}{s(1-s)} \right]\right| dr,\\
    &\nu[P^{(m,m)}_s]
    = \frac{1}{\pi} \frac{(1-s)^m}{s^{m+1}} \int_{0}^{2\pi} 1 d\phi \int_{0}^{\infty} e^{-\frac{1}{s}r^2} r^{3} \left|\mathrm{L}_m\left[\frac{r^2}{s(1-s)} \right]\right| dr.
\end{align}
Again using the bound on the magnitude of the Laguerre polynomials,
\begin{align}
    &\mu[P^{(m,m)}_s]
    \leq 2\frac{(1-s)^m}{s^{m+1}} \int_{0}^{\infty} r e^{\frac{r^2}{s}\left(\frac{1}{2(1-s)}-1\right)} dr
    = 2\frac{(1-s)^{m+1}}{s^{m}(1-2s)},\label{eq: mu mm}\\
    &\nu[P^{(m,m)}_s]
    \leq 2\frac{(1-s)^m}{s^{m+1}} \int_{0}^{\infty} r^{3} e^{\frac{r^2}{s}\left(\frac{1}{2(1-s)}-1\right)} dr
    = 4\frac{(1-s)^{m+2}}{s^{m-1}(1-2s)^2}.\label{eq: nu mm}
\end{align}
We can therefore upper bound $\mu[P_{s,M}] = \mu_{s,M}$, based on the number state decomposition of $\rho^{(M)}$, as
\begin{align}
    &\mu_{s,M}^{(\mathrm{UB})}
    = \sum_{m,n=0}^{M-1} \big|\langle m|\rho^{(M)}|n \rangle\big| \mu_{s,m,n}
    \geq \mu_{s,M},
    \label{eq: mu UB full}\\
    & \mu_{s,m,m} = \frac{2(1-s)^{m+1}}{s^{m}(1-2s)},
    \quad \mu_{s,m,n\neq m} =
    \frac{2^{2+\frac{|m-n|}{2}} (1-s)^{1+\frac{m+n}{2}}}{\pi s^{\frac{m+n}{2}}(1-2s)^{1+\frac{|m-n|}{2}}} \frac{(|m-n|+1)_{\mathrm{min}[m,n]}}{\sqrt{m!n!}} \Gamma\left[1+\frac{|m-n|}{2}\right].
    \label{eq: mu elements def}
\end{align}

By comparing Eqs.~(\ref{eq: mu mn}) and (\ref{eq: nu mn}) and Eqs.~(\ref{eq: mu mm}) and (\ref{eq: nu mm}), we see that, in both the on and off-diagonal cases, the ratio between (the upper bounds on) the contributions to $\nu$ and $\mu$ can be expressed (for $m \geq n$) as
\begin{equation}
    \frac{\nu^{(\mathrm{UB})}[P^{(m,n)}_s]}{\mu^{(\mathrm{UB})}[P^{(m,n)}_s]}
    = \frac{s(1-s)}{1-2s}(2+m-n).\label{eq: nu mu ratio}
\end{equation}
We now have two options: we can explicitly write upper bounds on $\mu$ and $\nu$, by combining Eqs.~(\ref{eq: mu mn}), (\ref{eq: nu mn}), (\ref{eq: mu mm}), and (\ref{eq: nu mm}) with a number state description of $\rho$ or we can upper bound the ratio between $\nu^{(\mathrm{UB})}$ and $\mu^{(\mathrm{UB})}$ (though perhaps loosely) as
\begin{equation}
    \frac{\nu^{(\mathrm{UB})}[P_{s,M}]}{\mu^{(\mathrm{UB})}[P_{s,M}]}
    \leq \frac{s(1-s)}{1-2s}(M+1).\label{eq: nu mu ratio UB}
\end{equation}
Eq.~(\ref{eq: nu mu ratio UB}) is less than $\sim\frac{M}{2}$ times Eq.~(\ref{eq: nu mu ratio}) and $\epsilon$ is concave, so we choose this method, and so arrive at Eq.~(\ref{eq: bound on mu epsilon}).

\section{Distance of $\sigma_{s,M}$ from $\rho^{(M)}$}\label{app: sigma distance}

Our goal is to determine how far each state in our parametrised sequence, $\{\sigma_{s,M}\}$, is from the truncated state, $\rho^{(M)}$. Recall that each state $\sigma_{s,M}$ is obtained from $\rho$ by first truncating it to an $M$-dimensional representation, $\rho^{(M)}$, and then convolving the P-representation of the resulting state, $P_{M}(\alpha)$, with $\frac{1}{s\pi}e^{-\frac{1}{s}|\alpha|^2}$ (equivalently, applying channel $\mathcal{C}_s$), to obtain state $\sigma_{s,M}$, with P-representation $P_{s,M}(\alpha)$. It is clear that as $M\to\infty$ and $s\to 0$, $\sigma_{s,M}$ converges to $\rho$, but we are interested in the rate of convergence.
We split $\|\rho-\sigma_{s,M} \|$ into separate contributions from the truncation and from the convolution, $\delta_M$ and $\delta_s$.

Some care is required in handling the energy truncation from $\rho$ to $\rho^{(M)}$. We define
\begin{equation}
    \rho^{(M)} = \eta_M^{-1} \sum_{m,n=0}^{M-1} \langle m|\rho|n \rangle |m\rangle\langle n |,
    \quad \eta_M = \sum_{m=0}^{M-1} \langle m|\rho|m \rangle.
\end{equation}
$\eta_M$ is the probability that a photon counting measurement on $\rho$ would give a result of $M-1$ or less. We can calculate the fidelity between $\rho^{(M)}$ and $\rho_{s,M}$ for an arbitrary pure state, $|\Gamma\rangle$, by expressing $|\Gamma\rangle$ as $\sum_{m=0}^{\infty} \gamma_{m} |m\rangle$ for some complex parameters $\{\gamma_m\}$ such that $\sum_{m=0}^{\infty} |\gamma_{m}|^2 = 1$. Then,
\begin{equation}
    F\big(|\Gamma\rangle\langle\Gamma|, |\Gamma\rangle\langle\Gamma|^{(M)}\big)^2
    = \eta_M^{-1}\sum_{l,p=0}^{\infty} \sum_{m,n=0}^{M-1} \gamma_l^{*} \gamma_m \gamma_n^{*} \gamma_p \langle l|m\rangle\langle n|p\rangle
    = \eta_M^{-1}\sum_{m,n=0}^{M-1} |\gamma_m|^2 |\gamma_n|^2 = \eta_M,
\end{equation}
since $\sum_{m=0}^M |\gamma_m|=\eta_M$. 
Then, from the Fuchs-van de Graaf inequality, $\delta_M = 2\sqrt{1-\eta_M}$. $\eta_M$ can be lower bounded, using the average photon number of the state, as $\eta_M\geq 1-\frac{\bar{n}}{M}$, so that, for pure states, $\delta_M = 2\sqrt{1-\eta_M} \leq 2\sqrt{\frac{\bar{n}}{M}}$.

For mixed states, using the joint concavity of quantum fidelity, $F(\rho,\rho^{(M)})\geq \sum p_i \sqrt{\eta_{M,i}} \geq \sum p_i \eta_{M,i}=\eta_M$, where $\eta_{M,i}$ are the values of $\eta_M$ calculated on each of the individual states $|\Gamma_i\rangle$ in the decomposition of $\rho$ and $p_i$ are their respective probabilities (replacing the sum with an integral and the probabilities with probability densities in the case of a continuous distribution). Hence, $\delta_M  \leq 2\sqrt{1-\eta_M^2}$.
Alternatively, we can use the convexity of the trace norm, the concavity of the square root function, and Jensen's inequality to get $\delta_M \leq \frac{2}{\sqrt{M}}\sum p_i\sqrt{\bar{n}_i} \leq \frac{2}{\sqrt{M}}\sqrt{\sum p_i\bar{n}_i}=2\sqrt{\frac{\bar{n}}{M}}$, where $\bar{n}_i$ are the average photon numbers of the states $|\Gamma_i\rangle$, recovering the same bound as in the pure case. The bound we choose depends on which is tighter, in a particular case.

As for $\|\rho-\rho^{(M)} \|$, we bound $\delta_s$ by calculating the fidelity between $\rho^{(M)}$ and $\sigma_{s,M}$ and applying a Fuchs-van de Graaf inequality. The overlap between the states, $\mathrm{Tr}[\rho^{(M)} \sigma_{s,M}]$ is given by
\begin{equation}
   \mathrm{Tr}[\rho^{(M)} \sigma_{s,M}] = \pi \int P_s(\alpha)[\rho^{(M)}] Q(\alpha)[\rho^{(M)}] d^2\alpha
    = \pi \int \left(\frac{1}{s\pi}e^{-\frac{1}{s}|\alpha|^2} \star P(\alpha)[\rho^{(M)}]\right) \left(\frac{1}{\pi}e^{-|\alpha|^2} \star P(\alpha)[\rho^{(M)}]\right) d^2\alpha,
\end{equation}
where $Q(\alpha)$ is the Husimi Q-representation. Since $F(A,B)^2 \geq \mathrm{Tr}[A B]$, we can use this overlap to bound the fidelity (sometimes defined as the square of the quantity that we call fidelity here). Some care is required here, since our lower bound on the squared fidelity, based on the overlap, is not tight and does not converge to $1$ as $s\to 0$ unless the state is pure.

First, we assess $\mathrm{Tr}[\left|\psi\middle>\middle<\psi\right| \mathcal{C}_s[\left|\psi\middle>\middle<\psi\right|]]$ for an arbitrary, pure state $|\psi\rangle$ (lying in the Hilbert space of $\rho^{(M)}$). Expressing $|\psi\rangle$ as $\sum_{m=0}^{M-1} a_{m} |m\rangle$ for some $\{a_m\}$ such that $\sum_{m=0}^{M-1} a_{m}a^{*}_{m} = 1$, the P-representation of $\mathcal{C}_s[ \left|\psi\middle>\middle<\psi\right| ]$ is
\begin{equation}
    P_s\left[ \left|\psi\middle>\middle<\psi\right| \right] = \sum_{m,n=0}^{M-1} |a_{m} a^{*}_{n}| P^{(m,n)}_{s,\theta(m,n)},\quad
    \theta(m,n) = \arg[a_{\max[m,n]} a^{*}_{\min[m,n]}],\label{eq: P rep pure}
\end{equation}
where the expressions for $P^{(m,n)}_s$ are given by Eqs.~(\ref{eq: Ps Fock off-diag}) and (\ref{eq: Ps Fock on-diag}) and where we have again included the subscript $\theta$ for completeness. The linearity of the trace means we can decompose $\mathrm{Tr}[\left|\psi\middle>\middle<\psi\right| \mathcal{C}_s[\left|\psi\middle>\middle<\psi\right|]]$ into a sum of contributions:
\begin{equation}
    \begin{split}
        \mathrm{Tr}[\left|\psi\middle>\middle<\psi\right| \mathcal{C}_s[\left|\psi\middle>\middle<\psi\right|]]
        &= \pi \int P_s(\alpha)[\left|\psi\middle>\middle<\psi\right|] Q(\alpha)[\left|\psi\middle>\middle<\psi\right|] d^2\alpha\\
        &= \sum_{m_1,n_1=0} \sum_{m_2,n_2=0} |a_{m_1} a^{*}_{n_1} a_{m_2} a^{*}_{n_2}| \gamma_s(m_1,n_1,\theta_1,m_2,n_2,\theta_2),
    \end{split}\label{eq: pure overlap}
\end{equation}
where each contribution $\gamma_s(m_1,n_1,\theta_1,m_2,n_2,\theta_2)$ is defined by
\begin{align}
    & \gamma_s(m_1,n_1,\theta_1,m_2,n_2,\theta_2) = \pi \int_{0}^{\infty} \int_{0}^{2\pi} P^{(m_1,n_1)}_{s,\theta_1}(re^{i\phi}) Q^{(m_2,n_2)}_{\theta_2}(re^{i\phi}) r d\phi dr,\\
    &Q^{(m,n)}_{\theta}(re^{i\phi}) = \lim_{s\to 1} P^{(m,n)}_{s,\theta}(re^{i\phi})
    = \cos(\theta-(m-n)\phi)\frac{e^{-r^2}r^{m+n}}{\pi\sqrt{m!n!}},\\
    &Q^{(m,m)}(re^{i\phi}) = \lim_{s\to 1} P^{(m,m)}_s(re^{i\phi})
    = \frac{e^{-r^2}r^{2m}}{\pi m!}.
\end{align}

From the angular dependence of the P and Q-representations, it is immediate that $\gamma_s(m_1,n_1,\theta_1,m_2,n_2,\theta_2)$ is only non-zero if $|m_1 - n_1| = |m_2 - n_2|$. This is because, for $|m_1 - n_1| \neq |m_2 - n_2|$,
\begin{equation}
    \int_{0}^{2\pi} \cos(\theta_1-(m_1-n_1)\phi)\cos(\theta_2-(m_2-n_2)\phi) d\phi
    = 0.
\end{equation}
Note too that $\gamma_s(m_1,n_1,\theta_1,m_2,n_2,\theta_2)$ is symmetric under swapping $m$ and $n$, so we can set $m \geq n$.

Starting by looking at the off-diagonals, and setting $\Delta=m-n$ and $m>n$,
\begin{equation}
    \begin{split}
        \gamma_s(m_1,m_1-\Delta,\theta_1,m_2,m_2-\Delta,\theta_2)
        =& \sqrt{\frac{(m_1-\Delta)!}{m_1!m_2!(m_2-\Delta)!}} \frac{(s-1)^{m_1-\Delta}}{\pi s^{m_1+1}}
        \int_{0}^{2\pi} \cos(\theta_1-\Delta\phi)\cos(\theta_2-\Delta\phi) d\phi\\
        &\times \int_{0}^{\infty} e^{-\frac{1+s}{s}r^2}r^{2m_2+1} \mathrm{L}^{\Delta}_{m_1-\Delta}\left[\frac{r^2}{s(1-s)}\right] dr.
    \end{split}
\end{equation}
Then, by carrying out the integration over $\phi$ and explicitly expanding the Laguerre polynomial
\begin{equation}
    \begin{split}
        \gamma_s(m_1,m_1-\Delta,\theta_1,m_2,m_2-\Delta,\theta_2)
        =& \cos(\theta_1-\theta_2) \sqrt{\frac{(m_1-\Delta)!}{m_1!m_2!(m_2-\Delta)!}} \frac{(s-1)^{m_1-\Delta}}{s^{m_1+1}}\\
        & \times \sum_{i=0}^{m_1-\Delta} \frac{(-1)^i}{s^{i}(1-s)^{i} i!} \binom{m_1}{\Delta+i} \int_{0}^{\infty} e^{-\frac{1+s}{s}r^2}r^{2(m_2+i)+1} dr.
    \end{split}
\end{equation}
Next, we calculate that for any non-negative integer $x$,
\begin{equation}
    \int_{0}^{\infty} e^{-\frac{1+s}{s}r^2}r^{2x+1} dr = \frac{x!}{2}\left( \frac{s}{1+s} \right)^{1+x},\label{eq: r integration}
\end{equation}
where, per convention, $0! = 1$~\cite{supplemental}. Using Eq.~(\ref{eq: r integration}), we get
\begin{equation}
    \begin{split}
        \gamma_s(m_1,m_1-\Delta,\theta_1,m_2,m_2-\Delta,\theta_2)
        =& \cos(\theta_1-\theta_2) \frac{s^{m_2-m_1}}{2}\sqrt{\frac{(m_1-\Delta)!}{m_1!m_2!(m_2-\Delta)!}} \frac{(s-1)^{m_1-\Delta}}{(1+s)^{m_2+1}}
        \sum_{i=0}^{m_1-\Delta} \frac{(m_2+i)!}{(s^2-1)^{i} i!} \binom{m_1}{\Delta+i}\\
        =& \cos(\theta_1-\theta_2) \frac{s^{m_2-m_1}}{2} \sqrt{\binom{m_1}{\Delta}\binom{m_2}{\Delta}} \frac{{}_{2}F_{1}(\Delta-m_1,m_2+1,\Delta+1,(1-s^2)^{-1})}{(1+s)^{m_2+1}(s-1)^{\Delta-m_1}},\label{eq: gamma off-diag}
    \end{split}
\end{equation}
where ${}_{2}F_{1}(a,b,c,z)$ is the hypergeometric function. Applying Pfaff's transformation rule for hypergeometric functions,
\begin{equation}
    \gamma_s(m_1,m_1-\Delta,\theta_1,m_2,m_2-\Delta,\theta_2)
    = \frac{\cos(\theta_1-\theta_2)}{2} \sqrt{\binom{m_1}{\Delta}\binom{m_2}{\Delta}} \frac{{}_{2}F_{1}(\Delta-m_1,\Delta-m_2,\Delta+1,s^{-2})}{s^{2\Delta-m_1-m_2}(1+s)^{m_1+m_2+1-\Delta}}.\label{eq: gamma off-diag 2}
\end{equation}
This rearrangement is useful because it highlights the fact that $\gamma_s$ is symmetric under swapping $m_1$ and $m_2$ (and $
\theta_1$ and $\theta_2$), but also because ${}_{2}F_{1}(\Delta-m_1,\Delta-m_2,\Delta+1,s^{-2}){}_{2}F_{1}(\Delta-m_1,\Delta-m_2,\Delta+1,s^{-2})$ can be expressed as a sum over only positive terms (rather than a sum of terms with alternating signs). Since $\gamma_s$ is symmetric, we set $m_1 \leq m_2$. Then, since $\Delta-m_1=-n_1$ and $\Delta-m_2=-n_2$ are both negative (we will substitute back and forth between $m$ and $n$ depending on what is easiest),
\begin{equation}
    \begin{split}
        s^{m_1+m_2-2\Delta}{}_{2}F_{1}(\Delta-m_1,\Delta-m_2,\Delta+1,s^{-2})
        &= s^{n_1+n_2} \sum_{k=0}^{n_1} (-1)^{k} \binom{n_1}{k} \frac{(-n_2)_{k}}{(\Delta+1)_{k}}s^{-2k}\\
        &= s^{n_2-n_1} \sum_{k=0}^{n_1} \binom{n_1}{k} \frac{n_2!}{(n_2-k)!} \frac{\Delta!}{(\Delta+1)!} s^{2(n_1-k)}.
    \end{split}
\end{equation}
We define the polynomial
\begin{equation}
    G[m_1,m_2,\Delta,s]
    = \sum_{k=0}^{n_1} \binom{n_1}{k} \frac{n_2!}{(n_2-k)!} \frac{\Delta!}{(\Delta+k)!} s^{2(n_1-k)},\label{eq: polynomial}
\end{equation}
where we assume $m_2 \geq m_1 \geq \Delta$. Crucially, this is a polynomial in $s$ that only has positive coefficients, so we are guaranteed that it will evaluate to a finite, positive value for any value of $s$. Finally, we can rewrite Eq.~(\ref{eq: gamma off-diag 2}) as
\begin{equation}
    \gamma_s(m_1,m_1-\Delta,\theta_1,m_2,m_2-\Delta,\theta_2)
    = \frac{\cos(\theta_1-\theta_2)}{2} \sqrt{\binom{m_1}{\Delta}\binom{m_2}{\Delta}} \frac{s^{m_2-m_1} G[m_1,m_2,\Delta,s]}{(1+s)^{m_1+m_2+1-\Delta}}.\label{eq: gamma off-diag final}
\end{equation}

For on-diagonals ($m=n$), we have
\begin{equation}
    \begin{split}
        \gamma_s(m_1,m_1,m_2,m_2)
        &= \frac{1}{m_2!} \frac{(s-1)^{m_1}}{\pi s^{m_1+1}}
        \int_{0}^{2\pi} 1 d\phi
        \int_{0}^{\infty} e^{-\frac{1+s}{s}r^2}r^{2m_2+1} \mathrm{L}_{m_1}\left[\frac{r^2}{s(1-s)}\right] dr\\
        &= \frac{2}{m_2!} \frac{(s-1)^{m_1}}{s^{m_1+1}}
        \sum_{i=0}^{m_1} \binom{m_1}{i} \frac{(-1)^i}{s^{i}(1-s)^{i} i!}
        \int_{0}^{\infty} e^{-\frac{1+s}{s}r^2}r^{2(m_2+i)+1} dr\\
        &= \frac{s^{m_2-m_1}}{m_2!} \frac{(s-1)^{m_1}}{(1+s)^{m_2+1}}
        \sum_{i=0}^{m_1} \binom{m_1}{i} \frac{(m_2+i)!}{(s^2-1)^{i} i!}\\
        &= s^{m_2-m_1} \frac{(s-1)^{m_1}}{(1+s)^{m_2+1}}
        {}_{2}F_{1}(-m_1,m_2+1,1,(1-s^2)^{-1})\\
        &= s^{m_1+m_2} \frac{{}_{2}F_{1}(-m_1,-m_2,1,s^{-2})}{(1+s)^{m_1+m_2+1}},\label{eq: gamma on-diag}
    \end{split}
\end{equation}
where the only difference from the off-diagonal case is in the prefactor, coming from the integration of the angular part. We have dropped the $\theta$-dependence because, per Eq.~(\ref{eq: P rep pure}), for $m=n$, $\theta=\arg[a_m a^*_m]=0$. Again, the expression is symmetric under interchange of $m_1$ and $m_2$, so we again set $m_2 \geq m_1$. Then,
\begin{equation}
    \gamma_s(m_1,m_1,m_2,m_2)
    = s^{m_2-m_1} \sum_{k=0}^{m_1} \binom{m_1}{k} \frac{m_2!}{(m_2-k)!k!} s^{2(m_1-k)}
    = \frac{s^{m_2-m_1} G[m_1,m_2,0,s]}{(1+s)^{m_1+m_2+1}}.\label{eq: gamma on-diag final}
\end{equation}

Recall that we are only interested in the fidelity for small $s$. We will therefore construct a small-$s$ approximation of the squared fidelity. From Eqs.~(\ref{eq: gamma off-diag final}) and (\ref{eq: gamma on-diag final}), we can see that the gamma functions go to $0$ as $s\to 0$ except in the case of $m_1 = m_2$, due to the prefactor of $s^{m_2-m_1}$. For $m_1 = m_2$, Eqs.~(\ref{eq: gamma off-diag final}) and (\ref{eq: gamma on-diag final}) become
\begin{align}
    &\gamma_s(m,n,\theta,m,n,\theta)
    = \frac{1}{2} \binom{m}{n} \frac{G[m,m,m-n,s]}{(1+s)^{m+n+1}},\label{eq: gamma off-diag mm}\\
    &\gamma_s(m,m,m,m)
    = \frac{G[m,m,0,s]}{(1+s)^{2m+1}}.\label{eq: gamma on-diag mm}
\end{align}
As an aside, Eq.~(\ref{eq: gamma on-diag mm}) is exactly the squared fidelity if our input, $\rho$, is a Fock state, and so can be used directly in this case (as we do in Section~\ref{sec: Fock}).
From Eq.~(\ref{eq: polynomial}), we can see that only the $k = n_1$ term survives as $s \to 0$, so Eqs.(\ref{eq: gamma off-diag mm}) and (\ref{eq: gamma on-diag mm}) go to $\frac{1}{2}$ and $1$ respectively as $s \to 0$. Assessing Eq.~(\ref{eq: pure overlap}) for $s=0$ (and again dropping unnecessary $\theta$-dependence), we therefore get
\begin{equation}
    \mathrm{Tr}[\left|\psi\middle>\middle<\psi\right| \mathcal{C}_0[\left|\psi\middle>\middle<\psi\right|]]
    = \sum_{m=0} |a^{4}_{m}| \gamma_0(m,m,m,m) + 2\sum_{\substack{m,n=0,\\m\neq n}} |a^{2}_{m} a^{2}_{n}| \gamma_0(m,n,m,n)
    = \sum_{m,n=0} |a^{2}_{m} a^{2}_{n}| = 1,
\end{equation}
where the factor of $2$ comes from summing both $\gamma_0(m,n,m,n)$ and $\gamma_0(m,n,n,m)$.

To lower bound the fidelity for non-zero (but small) $s$, we take the lowest ordered terms of $\gamma_s(m,m,m,m)$ and $\gamma_s(m,n,m,n)$, i.e., we take the $s^0$ terms of Eqs.~(\ref{eq: gamma off-diag mm}) and (\ref{eq: gamma on-diag mm}) and neglect all other terms. This is valid because, as we will now show, the sum over all terms of the form $\gamma_s(m,n,m + q,n + q)$ (for $q > 0$) is positive. I.e.,
\begin{equation}
    \sum_{q=1}^{M-2} \sum_{m,n=0}^{M-1-q}
    (2-\delta_{\mathrm{kron}}(m,n))
    |a_{m+q} a_{n+q} a_{m} a_{n}| \gamma_s(m,n,\theta_1,m + q,n + q,\theta_2)
    \geq 0,\label{eq: positive condition}
\end{equation}
where $\delta_{\mathrm{kron}}$ is the Kronecker delta and the prefactor of $(2-\delta_{\mathrm{kron}}(m,n))$ is so that we count both $\gamma_s(m,n,m + q,n + q)$ and $\gamma_s(n,m,m + q,n + q)$ for $m\neq n$. For completeness, we should also sum over $\gamma_s(m + q,n + q,m,n)$ and $\gamma_s(m + q,n + q,n,m)$, but since this only results in a prefactor of $2$ on all terms, we can neglect this. We rewrite Eq.~(\ref{eq: gamma off-diag final}) (for $q>0$ and $m \neq n$) as
\begin{equation}
    \begin{split}
        \gamma_s(m,n,\theta_1,m+q,n+q,\theta_2)
        &= \frac{\cos(\theta_1-\theta_2)}{2} \frac{\sqrt{m!n!(m+q)!(n+q)!}}{(1+s)^{m+n+q+1}}
        \sum_{j=0}^{\min[m,n]} \frac{s^{q+2j}}{j!(q+j)!(n-j)!(m-j)!}\\
        &= \frac{1}{4 |a_{m+q} a_{n+q} a_{m} a_{n}|} 
        \sum_{j=0}^{\min[m,n]} \psi_{q,j,m}\psi^{*}_{q,j,n} + \psi^{*}_{q,j,m}\psi_{q,j,n},
    \end{split}
\end{equation}
where we define
\begin{equation}
    \psi_{q,j,r} =
    \frac{a_{r} a^*_{r+q} s^{j+\frac{q}{2}} \sqrt{r!(r+q)!}}{(1+s)^{r+\frac{q+1}{2}} (r+q)! \sqrt{j!(q+j)!}}
\end{equation}
and we use the fact that
\begin{equation}
    a_{m} a^*_{m+q} a^*_{n} a_{n+q} + a^*_{m} a_{m+q} a_{n} a^*_{n+q} = 2|a_{m+q} a_{n+q} a_{m} a_{n}|\cos\left(\arg[a_m a^{*}_n]-\arg[a_{m+q} a^{*}_{n+q}]\right).
\end{equation}
Similarly, Eq.~(\ref{eq: gamma on-diag final}) becomes
\begin{equation}
    \gamma_s(m,m,m+q,m+q)
    = \frac{1}{|a_{m+q}^2 a_{m}^2|} 
    \sum_{j=0}^{m} \psi_{q,j,m}\psi^{*}_{q,j,m}
\end{equation}
Thus, the condition from Eq.~(\ref{eq: positive condition}), which we want to prove, becomes
\begin{equation}
    \sum_{q=1}^{M-2} \sum_{m,n=0}^{M-1-q} \sum_{j=0}^{\min[m,n]} \psi_{q,j,m}\psi^{*}_{q,j,n}
    = \sum_{q=1}^{M-2} \sum_{j=0}^{M-1-q} \sum_{m,n=j}^{M-1-q} \psi_{q,j,m}\psi^{*}_{q,j,n}
    \geq 0.\label{eq: positive condition 2}
\end{equation}
Defining $\psi_{q,j}$ as the $1$ by $M-j-q$ vector with entries $\psi_{q,j,r}$ for $r$ ranging from $j$ to $M-1-q$, we note that the left hand side of Eq.~(\ref{eq: positive condition 2}) is the sum over $q$ and $j$ of the sum of all entries of the matrices $\psi_{q,j}\psi^{\dagger}_{q,j}$. Summing all entries of a positive semi-definite matrix gives a number that is $\geq 0$, and since $\psi^{\dagger}_{q,j}\psi_{q,j} \geq 0$, $\psi_{q,j}\psi^{\dagger}_{q,j}$ is positive semi-definite.

Hence, we can validly lower bound the fidelity by summing only the 
$s^0$ terms in Eqs.~(\ref{eq: gamma off-diag mm}) and (\ref{eq: gamma on-diag mm}). Specifically,
\begin{equation}
    \mathrm{Tr}[\left|\psi\middle>\middle<\psi\right| \mathcal{C}_s[\left|\psi\middle>\middle<\psi\right|]]
    \geq \sum_{m,n=0}^{M-1} \frac{|a^{2}_{m} a^{2}_{n}|}{(1+s)^{m+n+1}},
\end{equation}
where we only use the first term in $G[m,m,n,0]$. This is a convex function in $s$, so for pure states, we can write
\begin{equation}
    F^2(\left|\psi\middle>\middle<\psi\right|, \mathcal{C}_s[\left|\psi\middle>\middle<\psi\right|])
    \geq 1 - s \sum_{m,n=0}^{M-1} |a^{2}_{m} a^{2}_{n}|(m+n+1)
    = 1 - s \bigg(1+ 2\sum_{m=0}^{M-1} |a^{2}_{m}|m\bigg),
\end{equation}
where we have lower bounded it using the first order Taylor expansion around $s=0$. Using the definition of the average energy
\begin{equation}
    F^2(\left|\psi\middle>\middle<\psi\right|, \mathcal{C}_s[\left|\psi\middle>\middle<\psi\right|])
    \geq 1 - s (1+ 2 \bar{n}(\left|\psi\middle>\middle<\psi\right|)),
\end{equation}
and from the Fuchs-van de Graaf relations,
\begin{equation}
    \delta_s(\left|\psi\middle>\middle<\psi\right|)
    \leq 2\sqrt{s (1+ 2 \bar{n}(\left|\psi\middle>\middle<\psi\right|))}.
\end{equation}
Using the convexity of the trace norm and Jensen's inequality, we arrive at Eq.~(\ref{eq: sigma distance}) from the main text.

Finally, we note that we are free to decompose $\|\rho-\sigma_{s,M}\|$ in a different way, namely
\begin{equation}
    \begin{split}
        \|\rho-\sigma_{s,M}\|
        &= \|\rho-\mathcal{C}_s[\rho^{(M)}]\|
        = \|\rho-\mathcal{C}_s[\rho]+\mathcal{C}_s[\rho]-\mathcal{C}_s[\rho^{(M)}]\|\\
        &\leq \|\rho-\mathcal{C}_s[\rho]\| + \|\mathcal{C}_s[\rho]-\mathcal{C}_s[\rho^{(M)}]\|
        \leq \|\rho-\mathcal{C}_s[\rho]\| + \|\rho-\rho^{(M)}\|
        = \|\rho-\mathcal{C}_s[\rho]\| + \delta_M,
    \end{split}
    \label{eq: alternative delta split}
\end{equation}
where, on the second line, we have used the data processing inequality. Where convenient, we can therefore calculate $\|\rho-\mathcal{C}_s[\rho]\|$ instead of $\|\rho^{(M)}-\mathcal{C}_s[\rho^{(M)}]\|$. We will call the quantity $\delta_s$ in both cases, to avoid unnecessary extra notation.

\section{Bounding the output distance over all states}\label{app: ultimate bound}

To upper bound $\mu^{(\mathrm{UB})}$ over all states $\rho^{(M)}$, we use the fact $\rho^{(M)}$ is a positive semi-definite matrix, and hence:
\begin{equation}
    |\langle m|\rho^{(M)}|n \rangle| \leq \frac{\langle m|\rho^{(M)}|m \rangle + \langle n|\rho^{(M)}|n \rangle}{2}.
\end{equation}
Upper bounding the upper bound on $\mu$ from Eq.~(\ref{eq: mu UB full}), we get
\begin{equation}
    \mu_{s,M}^{(\mathrm{UB})}
    \leq \sum_{m,n=0}^{M-1} \frac{\langle m|\rho^{(M)}|m \rangle + \langle n|\rho^{(M)}|n \rangle}{2} \mu_{s,m,n}
    = \sum_{m=0}^{M-1} \langle m|\rho^{(M)}|m \rangle \sum_{n=0}^{M-1} \mu_{s,m,n}
    \leq \max_{m} \bigg[\sum_{n=0}^{M-1} \mu_{s,m,n}\bigg].\label{eq: UB on mu UB}
\end{equation}
For small $s$, we will show that this maximum is achieved by setting $m=M-1$. We must note that we are taking an upper bound on an upper bound here, so the resulting bound on $\mu$ may be very loose. However, a loose bound is to be expected, since our goal is to make an extremely general statement that holds even without having any description of the input state at all, other than knowing its average photon number. We could improve this bound somewhat by using the average photon number (i.e., by accounting for the fact $\sum_m \langle m|\rho^{(M)}|m \rangle m = \bar{n}$), but this would result in more complicated expressions.

We can upper bound the row sum $\sum_{n=0}^{M-1} \mu_{s,m,n}$ by bounding the ratio between neighbouring terms. We start by calculating
\begin{equation}
    \frac{\mu_{s,m,n}}{\mu_{s,m,n-1}} = \sqrt{\frac{(1-s)(1-2s)}{2s}} \frac{m-n+1}{\sqrt{n}}\frac{\Gamma[\frac{m-n+2}{2}]}{\Gamma[\frac{m-n+3}{2}]}
    > \sqrt{\frac{(1-s)(1-2s)}{s}} \frac{m-n+1}{\sqrt{n(m-n+3)}}
    \quad \mathrm{for~} m>n,
    \label{eq: m>n ratio}
\end{equation}
where we bound the ratio between gamma functions using Gautschi's inequality. Similarly,
\begin{equation}
    \frac{\mu_{s,m,n+1}}{\mu_{s,m,n}} = \sqrt{\frac{2(1-s)}{s(1-2s)}} \frac{\sqrt{n+1}}{n-m+1} \frac{\Gamma[\frac{n-m+3}{2}]}{\Gamma[\frac{n-m+2}{2}]}
    > \sqrt{\frac{1-s}{s(1-2s)}} \sqrt{\frac{n+1}{n-m+1}}
    \quad \mathrm{for~} m<n.
    \label{eq: m<n ratio}
\end{equation}
Next, we calculate
\begin{equation}
    \frac{\mu_{s,m,m}}{\mu_{s,m,m-1}}
    = \sqrt{\frac{\pi}{2}} \sqrt{\frac{(1-s)(1-2s)}{sm}},
    \quad \frac{\mu_{s,m,m+1}}{\mu_{s,m,m}}
    = \sqrt{\frac{2}{\pi}} \sqrt{\frac{(1-s)(m+1)}{s(1-2s)}},
    \label{eq: m=n ratios}
\end{equation}
where we have used $\Gamma[\frac{3}{2}]=\frac{\sqrt{\pi}}{2}$. Upper and lower bounding the $n$-dependent terms (for $m \neq 0$),
\begin{equation}
    \frac{\mu_{s,m,n}}{\mu_{s,m,n-1}}
    > \sqrt{\frac{(1-s)(1-2s)}{s m}}
    \quad \mathrm{for~} m \geq n,
    \quad \frac{\mu_{s,m,n+1}}{\mu_{s,m,n}}
    > \sqrt{\frac{1-s}{s(1-2s)}}
    \quad \mathrm{for~} m \leq n.
    \label{eq: both ratios}
\end{equation}
From Eq.~(\ref{eq: both ratios}), we can see that, so long as $s$ decreases at least with $M^{-1}$ (more specifically, as long as $\frac{(1-s)(1-2s)}{s(M-1)}>1$), each $\mu_{s,m,n-1} < \mu_{s,m,n}$ (for both $m\geq n$ and $m<n$). Since $\mu_{s,m,n}=\mu_{s,n,m}$, this means $\sum_{n=0}^{M-1} \mu_{s,m-1,n}<\sum_{n=0}^{M-1} \mu_{s,m,n}$. Hence, the $m=M-1$ row has the largest row sum, and thus (using Eq.~(\ref{eq: mu elements def}))
\begin{equation}
    \mu_{s,M}^{(\mathrm{UB})} \leq \sum_{n=0}^{M-1} \mu_{s,M-1,n}
    < \mu_{s,M-1,M-1}\sum_{k=0}^{M-1} \left(\frac{s (M-1)}{(1-s)(1-2s)}\right)^{\frac{k}{2}}
    < \frac{2(1-s)^{M}M}{s^{M-1}(1-2s)},
    \label{eq: mu ultimate UB}
\end{equation}
where we could have used the formula for the sum of a geometric sequence to obtain a slightly tighter bound.

Substituting Eq.~(\ref{eq: mu ultimate UB}) into Eq.~(\ref{eq: bound infinite negativity full}), we recover Eq.~(\ref{eq: generic UB}) from the main text, i.e.,
\begin{equation*}
    \|(\Psi-\Phi)[\rho]\| \leq \frac{2(1-s)^{M}M}{s^{M-1}(1-2s)} \epsilon\left(\epsilon_0, \frac{s(1-s)(M+1)}{1-2s} \right)  + 4\sqrt{s(1+ 2\bar{n})}  + 4\sqrt{\frac{\bar{n}}{M}}.
\end{equation*}
Finally, we consider one particular relationship between $s$ and $M$. To study the scaling with the input energy, we will set $\frac{s(1-s)(M+1)}{1-2s}$ to a constant with regard to $\bar{n}$, allowing us to eliminate any dependence on the specific form of $\epsilon$.
For this, we need $s$ to be linear in $M^{-1}$. If we choose $s=\frac{1}{\kappa (M+3)}$, for some $\kappa > 1$, then the argument to $\epsilon$ is upper bounded by $\frac{s(1-s)(M+1)}{1-2s}<\frac{1}{\kappa}$ for all $\kappa > 1$. Consequently, for any $\kappa > 1$,
\begin{equation}
    \|(\Psi-\Phi)[\rho]\| \leq  2(M+3)^{M}\kappa^{M-1} \epsilon\left(\epsilon_0, \frac{1}{\kappa} \right)  + 4\sqrt{\frac{1+2\bar{n}}{\kappa (M+3)}} + 4\sqrt{\frac{\bar{n}}{M}}.
    \label{eq: s M relation UB}
\end{equation}
This can also hold for $\kappa <1$ as long as $M$ is not very small (and with some care required with the condition $\frac{(1-s)(1-2s)}{s(M-1)}>1$). In terms of the leading order terms, this can be expressed, per Corollary~\ref{cor: energy scaling}, as
\begin{equation*}
    \|(\Psi-\Phi)[\rho]\| \leq 
    \mathcal{O}[e^{M\log(M\kappa)}\kappa^{-1} \epsilon (\epsilon_0, \kappa^{-1} ) ]
    + \mathcal{O}[\bar{n}^{\frac{1}{2}} M^{-\frac{1}{2}} ]
\end{equation*}
Per the discussion after Corollary~\ref{cor: energy scaling}, we can retrieve an upper bound on the scaling with $\bar{n}$ by setting $M\sim \bar{n}$. Some caution is needed here, as strictly $\bar{n}$ is a continuous variable and $M$ is discrete, but we could get around this by using the ceiling function (and we are focused on the scaling, so the difference is not very important).

\section{Out-of-distribution generalisation for coherent states when $\epsilon_0 > 0$}\label{app: oodg for coherent bound}

We now use some of the techniques from the previous appendices to revisit out-of-distribution generalisation for coherent states. The aim now is to explicitly show that we can always (i.e., regardless of the class of the target channel) construct a concave function $\epsilon(\epsilon_0,r^2)$ that bounds the output distance for coherent states and that has the properties listed in Theorem~\ref{th: eps existence}.

Recall that coherent states have the number state decomposition
\begin{equation}
    | r e^{i\phi} \rangle \langle r e^{i\phi} |_{\mathrm{coh}}
    = e^{-r^2}\sum_{m,n=0}^{\infty} \frac{r^{m+n}e^{i(m-n)\phi}}{\sqrt{m!n!}} | m \rangle \langle n |.
\end{equation}
We use the same method as in Section~\ref{sec: infinite negativity}, replacing $| r e^{i\phi} \rangle \langle r e^{i\phi} |_{\mathrm{coh}}$ with $\mathcal{C}_s[| r e^{i\phi} \rangle \langle r e^{i\phi} |_{\mathrm{coh}}]$. Per Eq.~(\ref{eq: sigma distance}), $| r e^{i\phi} \rangle \langle r e^{i\phi} |_{\mathrm{coh}}$ has a distance from $\mathcal{C}_s[| r e^{i\phi} \rangle \langle r e^{i\phi} |_{\mathrm{coh}}]$ of no more than $2\sqrt{s(1+2r^2)}$. We then upper bound $\|(\Psi-\Phi)[\mathcal{C}_s[| r e^{i\phi} \rangle \langle r e^{i\phi} |_{\mathrm{coh}}]]\|$ by bounding $\|(\Psi-\Phi)[ \mathcal{C}_s[| m \rangle \langle m |] ]\|$ and $\|(\Psi-\Phi)[ \frac{1}{2}\mathcal{C}_s[e^{i\theta} | m \rangle \langle n | + e^{-i\theta} | n \rangle \langle m |] ]\|$ for every $m$, $n$, and $\theta$.

Note that we have previously upper bounded these same quantities (in Section~\ref{sec: Fock} and Appendix~\ref{app: sigma properties}), however here our starting point is different. Previously, we assumed we had a known, concave function $\epsilon(\epsilon_0,r^2)$, but now finding such a function is the goal. Instead, we only assume that $\|(\Psi-\Phi)[| r e^{i\phi} \rangle \langle r e^{i\phi} |_{\mathrm{coh}}]\| \leq \epsilon_0$ for $r^2\leq\tau^2$. Then, we can apply the step function ($\epsilon_{\mathrm{step}}$ from Eq.~(\ref{eq: step function})) that assigns $\epsilon_0$ for $r \leq \tau$ and $2$ for $r>\tau$.

Recall that the P-representation of $| m \rangle \langle m |$ is given by Eq.~(\ref{eq: Ps Fock on-diag}). $\|(\Psi-\Phi)[ \mathcal{C}_s[| m \rangle \langle m |] ]\|$ can therefore be upper bounded by
\begin{align}
    &\|(\Psi-\Phi)[ \mathcal{C}_s[| m \rangle \langle m |] ]\|
    \leq \epsilon_0 \mu_{\tau}\Big[P^{(m,m)}_s\Big]
    + 2 \Big(\mu\Big[P^{(m,m)}_s\Big] - \mu_{\tau}\Big[P^{(m,m)}_s\Big]\Big),\\
    &\mu_{\tau}\Big[P^{(m,m)}_s\Big] = \int_{0}^{2\pi} \int_{0}^{\tau} \Big|P^{(m,m)}_s (re^{i\phi})\Big| r d\phi dr,
    \quad \mu\Big[P^{(m,m)}_s\Big] = \int_{0}^{2\pi} \int_{0}^{\infty} \Big|P^{(m,m)}_s (re^{i\phi})\Big| r d\phi dr,
\end{align}
where we have applied the same technique as in Section~\ref{sec: infinite negativity} of separating the negative and positive parts of the P-representation, but applying the step function instead of $\epsilon(\epsilon_0,r^2)$. Since all of the mass of $P^{(m,m)}_s$ concentrates around the origin as $s\to 0$ (since the decay exponent becomes larger as $s$ decreases), the second term approaches $0$ as $s$ does. On the other hand, $\mu_{\tau}\Big[P^{(m,m)}_s\Big]$ approaches $\infty$. Following the methods used in Eq.~(\ref{eq: mu mm}), we get (for $s < \frac{1}{2}$)
\begin{equation}
    \mu_{\tau}\Big[P^{(m,m)}_s\Big]
    \leq 2\Big(1-e^{-\tau^2\frac{1-2s}{2s(1-s)}}\Big)\frac{(1-s)^{m+1}}{s^{m}(1-2s)},
\end{equation}
so we can write the upper bound
\begin{equation}
    \big\|(\Psi-\Phi)[ \mathcal{C}_s[| m \rangle \langle m |] ]\big\|
    \leq 2\frac{(1-s)^{m+1}}{s^{m}(1-2s)} \Big(\epsilon_0 + (2 - \epsilon_0)e^{-\tau^2\frac{1-2s}{2s(1-s)}}\Big).
    \label{eq: on-diag bound for coh}
\end{equation}
Noting that Eq.~(\ref{eq: on-diag bound for coh}) is trivial for large $m$, we provide the non-trivial bound $\big\|(\Psi-\Phi)[ \mathcal{C}_s[| m \rangle \langle m |] ]\big\| \leq \xi_{\tau,s}^{(m,m)}$, where
\begin{equation}
    \xi_{\tau,s}^{(m,m)}
    = \min\bigg[2\frac{(1-s)^{m+1}}{s^{m}(1-2s)} \Big(\epsilon_0 + (2 - \epsilon_0)e^{-\tau^2\frac{1-2s}{2s(1-s)}}\Big),2\bigg].
    \label{eq: xi mm}
\end{equation}

Following a similar approach for the off-diagonals, we recall that the P-representation of $\frac{1}{2}\mathcal{C}_s[e^{i\theta} | m \rangle \langle n | + e^{-i\theta} | n \rangle \langle m |]$ is given by Eq.~(\ref{eq: Ps Fock off-diag}). Proceeding similarly to the on-diagonal case, we must find $\mu_{\tau}\Big[P^{(m,n)}_s\Big]$. Following Eq.~(\ref{eq: mu mn}),
\begin{equation}
    \mu_\tau \Big[ P^{(m,n)}_s \Big]
    \leq \frac{2^{2+\frac{m-n}{2}} (1-s)^{1+\frac{m+n}{2}}}{\pi s^{\frac{m+n}{2}}(1-2s)^{1+\frac{m-n}{2}}} \frac{(m-n+1)_n}{\sqrt{m!n!}}
    \left( \Gamma\left[1+\frac{m-n}{2}\right] - \Gamma\left[1+\frac{m-n}{2},\frac{\tau^2 (1-2s)}{2s(1-s)}\right] \right),
\end{equation}
where $\Gamma[a,b]$ is the incomplete gamma function~\cite{supplemental}. Hence, $\frac{1}{2}\big\|(\Psi-\Phi)[ \mathcal{C}_s[e^{i\theta} | m \rangle \langle n | + e^{-i\theta} | n \rangle \langle m |] ]\big\| \leq \xi_{\tau,s}^{(m,n)}$, where
\begin{equation}
    \xi_{\tau,s}^{(m,n)}
    = \min\Bigg[\frac{2^{2+\frac{m-n}{2}} (1-s)^{1+\frac{m+n}{2}}}{\pi s^{\frac{m+n}{2}}(1-2s)^{1+\frac{m-n}{2}}} \frac{(m-n+1)_n}{\sqrt{m!n!}}
    \bigg( \epsilon_0\Gamma\left[1+\frac{m-n}{2}\right]\\
    + (2-\epsilon_0)\Gamma\left[1+\frac{m-n}{2},\frac{\tau^2 (1-2s)}{2s(1-s)}\right] \bigg),2\Bigg].
    \label{eq: xi mn}
\end{equation}

Finally, putting these various elements together, we get
\begin{equation}
    \|(\Psi-\Phi)[| r e^{i\phi} \rangle \langle r e^{i\phi} |_{\mathrm{coh}}]\|
    \leq \epsilon(\epsilon_0,r^2)
    = \min\bigg[ \inf_{0<s<\frac{1}{2}}\bigg\{ e^{-r^2}\sum_{m,n=0}^{\infty} \frac{r^{m+n}}{\sqrt{m!n!}}\xi_{\tau,s}^{(m,n)} + 4\sqrt{s(1+2r^2)} \bigg\}, 2\bigg].
    \label{eq: coherent state bound}
\end{equation}
From Eqs.~(\ref{eq: xi mm}) and (\ref{eq: xi mn}), we see that decreasing $\epsilon_0$ (or increasing $\tau$) for fixed $s$ decreases $\xi_{\tau,s}^{(m,n)}$. For sufficiently small $\epsilon_0$, we can choose $s$ such that Eq.~(\ref{eq: coherent state bound}) is non-trivial for any $r$. Eq.~(\ref{eq: coherent state bound}) is not known to be concave, but we can always make it concave by taking its upper concave hull. Nonetheless, it is not a practically useful bound, as we require $\epsilon_0$ to be extremely small for it to be non-trivial for large $r$. However, it is sufficient to prove that a concave bounding function $\epsilon(\epsilon_0,r^2)$ can always be constructed, and hence that out-of-distribution generalisation is always possible.

\section{Tightening the bound for SPATSs}\label{app: SPAT}

For low energy SPATSs, the negativity is large, so the bound on the distance between the output states, given by Eq.~(\ref{eq: bound SPAT}), may be loose. To tighten it, we may consider using the same technique as for states with infinite negativity, i.e., we can bound the output distance for the input $\sigma_{s,\mathrm{SPAT}}$, where $\sigma_{s,\mathrm{SPAT}}$ is given by applying the Gaussian additive noise channel $\mathcal{C}_s$ to $\rho_{\mathrm{SPAT}}$. The distance between $\rho_{\mathrm{SPAT}}$ and $\sigma_{s,\mathrm{SPAT}}$ is given by Eq.~(\ref{eq: sigma distance}). Parametrising with $q$ instead of $\bar{n}$, we get $\delta_s \leq 2\sqrt{s (3+4q)}$.

Using the P-representation of $\rho_{\mathrm{SPAT}}$ from Eq.~(\ref{eq: P SPAT}), we find the P-representation of $\sigma_{s,\mathrm{SPAT}}$ is
\begin{equation}
    \bigg(\frac{1}{s\pi}e^{-\frac{1}{s}|\alpha|^2}\bigg) \star P_{\mathrm{SPAT}}(r e^{i\phi}) =
    \frac{1+q}{\pi (q+s)^3} \bigg(r^2 -\frac{(q+s)(1-s)}{1+q}\bigg)e^{-\frac{r^2}{q+s}}.
\end{equation}
Integrating separately over the negative ($r^2<\frac{(q+s)(1-s)}{1+q}$) and positive regions, we find the values of $\mu_s$ and $\frac{\nu_s}{\mu_s}$ for $\sigma_{s,\mathrm{SPAT}}$ are
\begin{equation}
    \mu_s = 2e^{-\frac{1-s}{1+q}}\frac{1+q}{q+s} - 1,
    \quad \frac{\nu}{\mu} =
    1+2q+s - \frac{2(1-s)^2}{2+2q-e^{-\frac{1-s}{1+q}}(q+s)}
    < 1+2q+s.
\end{equation}
We can then write the bound
\begin{equation}
    \|(\Psi-\Phi)[\rho_{\mathrm{SPAT}}]\| \leq
    \left(2e^{-\frac{1-s}{1+q}}\frac{1+q}{q+s} - 1\right) \epsilon\left(\epsilon_0, 1+2q+s - \frac{2(1-s)^2}{2+2q-e^{-\frac{1-s}{1+q}}(q+s)}\right) + 4\sqrt{s (3+4q)},
\end{equation}
which reduces to Eq.~(\ref{eq: bound SPAT}) in the limit of $s\to 0$. Since the argument to concave function $\epsilon$ is approximately linear in $s$, a small change in $s$ may not increase $\epsilon(\epsilon_0,\frac{\nu}{\mu})$ by much. The 
$\delta_s$ term is also sublinear in $s$. On the other hand, for small $q$, a small change in $s$ can result in a significant change in the multiplicative factor $\mu_s$. Hence, we may obtain tighter bounds by choosing $s>0$.

\section{Application to squeezed vacuums}\label{app: squeezed vacuum}

Recall, from Eq.~(\ref{eq: alternative delta split}), that we are free to calculate $\delta_s$ as $\| \rho_{\mathrm{sq}} - \mathcal{C}_s [\rho_{\mathrm{sq}}] \|$, rather than $\| \rho^{(M)}_{\mathrm{sq}} - \mathcal{C}_s [\rho^{(M)}_{\mathrm{sq}}] \|$. Since squeezed vacuums and the channel $\mathcal{C}_s$ are both Gaussian, we can calculate the fidelity between $\rho_{\mathrm{sq}}$ and $\mathcal{C}_s [\rho_{\mathrm{sq}}]$~\cite{banchi_quantum_2015}:
\begin{equation}
    F(\rho_{\mathrm{sq}},\mathcal{C}_s [\rho_{\mathrm{sq}}])
    = \bigg(1+s^2+2s\frac{1+\lambda^2}{1-\lambda^2}\bigg)^{-\frac{1}{4}}.
\end{equation}
We therefore retrieve Eq.~(\ref{eq: deltas 1msv}) from the main text.

Using Eqs.~(\ref{eq: mu UB full}), (\ref{eq: mu elements def}), and (\ref{eq: 1msv density}), we can calculate $\mu_{s,M}$ by calculating the contributions from each element of $\rho_{\mathrm{sq}}$.
For any off-diagonal element with $p > q$,
\begin{equation}
    |\langle 2p|\rho|2q \rangle| \mu_{s,2p,2q}
    = \frac{4\sqrt{1-\lambda^2}(1-s)}{\pi (1-2s)^{p-q+1}}
    \frac{(1-s)^{p+q}\lambda^{p+q}}{s^{p+q}}
    \frac{(1+2p-2q)_{2q}(p-q)!}{2^{2q}p!q!}
\end{equation}
where we have separated out terms based on their dependence on $p$ and $q$. For on-diagonal elements,
\begin{equation}
    |\langle 2p|\rho|2p \rangle| \mu_{s,2p,2p}
    = \frac{2\sqrt{1-\lambda^2}(1-s)}{(1-2s)}
    \frac{(1-s)^{2p}\lambda^{2p}}{s^{2p}}
    \frac{(2p)!}{2^{2p}(p!)^2}.
\end{equation}
Note that if and only if $\frac{(1-s)\lambda}{s}<1$, the sum $\sum_{q}|\langle 2(p_0+q)|\rho|2q \rangle| \mu_{s,2(p_0+q),2q}$ for $q$ ranging from $0$ to $\infty$ is convergent. This is as expected, since $\frac{(1-s)\lambda}{s}<1$ is precisely the condition for the negativity of a squeezed vacuum to be finite (in fact, zero). However, we want to be able to take $s$ to $0$ as $\epsilon_0\to 0$, so we are interested in the divergent region. Instead, we set $x=\frac{(1-2s)(1-s)\lambda}{s}$, so that
\begin{equation}
    \begin{split}
        \sum_{q=0}^{p-1} |\langle 2p|\rho|2q \rangle| \mu_{s,2p,2q}
        &= \frac{4\sqrt{1-\lambda^2}}{\pi}
        \frac{\lambda^{p}(1-s)^{p+1}}{s^{p}(1-2s)^{p+1}}
        \sum_{q=0}^{p-1} 
        \frac{x^q (1+2p-2q)_{2q}(p-q)!}{2^{2q}p!q!}\\
        &= \frac{4\sqrt{1-\lambda^2}}{\pi}
        \frac{\lambda^{p}(1-s)^{p+1}}{s^{p}(1-2s)^{p+1}}
        \bigg( (1+x)^{p-\frac{1}{2}} -
        \frac{x^p {}_{2}F_{1}\big( \frac{1}{2},1,1+p,-x \big) (2p)!}{2^{2p}(p!)^2} \bigg).
    \end{split}
\end{equation}
Multiplying by two to account for $p<q$ and adding the on-diagonal contribution,
\begin{equation}
    \begin{split}
        |\langle 2p|\rho|2p \rangle| \mu_{s,2p,2p} + 2 \sum_{q=0}^{p-1} |\langle 2p|\rho|2q \rangle| \mu_{s,2p,2q}
        = \frac{8\sqrt{1-\lambda^2}}{\pi}
        &\frac{\lambda^{p}(1-s)^{p+1}}{s^{p}(1-2s)^{p+1}}
        \bigg( (1+x)^{p-\frac{1}{2}}\\
        &+ \frac{x^p (2p)!}{2^{2p}(p!)^2} \Big(\frac{\pi}{4} - {}_{2}F_{1}\Big( \frac{1}{2},1,1+p,-x \Big)\Big) \bigg).
    \end{split}
\end{equation}
Applying a Pfaff transformation to the hypergeometric function so that its final argument is $\frac{x}{1+x}$ and then writing it in power series form, we can easily verify that it is always positive. Hence, we can upper bound this expression with
\begin{equation}
    \begin{split}
        |\langle 2p|\rho|2p \rangle| \mu_{s,2p,2p} + 2 \sum_{q=0}^{p-1} |\langle 2p|\rho|2q \rangle| \mu_{s,2p,2q}
        \leq \frac{2(1-s)\sqrt{1-\lambda^2}}{(1-2s)} \bigg( \frac{4}{\pi\sqrt{1+x}}&
        \left(\frac{\lambda (1-s) (1+x)}{s(1-2s)}\right)^{p}\\
        &+ \frac{(2p)!}{2^{2p}(p!)^2} \left(\frac{\lambda (1-s) x}{s(1-2s)}\right)^{p} \bigg).
    \end{split}
\end{equation}
The first term is a geometric sequence in $p$, so is simple to sum. For the second term, $\frac{(2p)!}{2^{2p}(p!)^2} \leq 1$ (this can be verified by taking the ratio between this expression for $p$ and $p+1$ and seeing it is a decreasing function of $p$). Hence, we can upper bound the second term by replacing $\frac{(2p)!}{2^{2p}(p!)^2}$ with its maximum value of $1$. Then, both terms are geometric sequences, so we can recover Eq.~(\ref{eq: mu bound 1msv}) by summing over $p$ (note that the maximum value of $p$ is $\frac{M-1}{2}$, not $M-1$) and then normalising by dividing by $\eta_M$ (since we recall that $\rho^{(M)}$ is normalised).

\section{Analytical and numerical investigation of the tightness of the bounds for non-classical states}\label{app: triviality}

The trace norm of a difference between quantum states is constrained to lie in the range $[0,2]$, so an upper bound $\geq 2$ is trivial. Theorems~\ref{th: eps existence} and \ref{th: general states}, taken together, show that for any fixed input state, $\rho$, we can non-trivially bound the output distance, provided $\epsilon_0$ is small enough. In a learning setting, we are able to make $\epsilon_0$ arbitrarily small by taking more samples, i.e., by increasing $N$. If we fixed a specific learning model and class of channels, it might also be possible to understand the minimum $N$ required to non-trivially bound the output distance for a particular $\rho$, but this would be very problem specific.

On the other hand, for fixed $\epsilon_0$, we might wonder for which set of states we are able to non-trivially bound the output distance. In the finite negativity scenario of Corollary~\ref{cor: finite negativity}, it is simple to check whether the right hand side of Eq.~(\ref{eq: bound finite negativity}) is $\geq 2$. However, the region of non-triviality for the more general, infinite negativity scenario of Theorem~\ref{th: general states} is harder to characterise, since there are two free parameters to tune.
For Eq.~(\ref{eq: generic UB}) to be non-trivial, we require there to exist $0\leq s<\frac{1}{2}$ and integer $M\geq \bar{n}$ (and the condition $\frac{(1-s)(1-2s)}{s(M-1)}>1$ to be met) such that the bound in Eq.~(\ref{eq: generic UB}) is $<2$.

There are two approaches we can take. If we know the form of $\epsilon(\epsilon_0,r^2)$, we can investigate the region in which a non-trivial bound is possible, in terms of $\bar{n}$ and $\epsilon_0$. Alternatively, we can eliminate the dependence on the function $\epsilon$ by setting $\frac{s(1-s)(M+1)}{1-2s}=\bar{n}$. We must set $s=\frac{M+1+2\bar{n}-\sqrt{(M+1)^2+4\bar{n}^2}}{2(M+1)}$, so that the condition becomes (per Eq.~(\ref{eq: triviality condition fixed to nbar})
\begin{equation*}
    \begin{split}
        \epsilon(\epsilon_0,\bar{n})
        < \frac{ 2\big(\sqrt{(M+1)^2+4\bar{n}^2} - 2\bar{n}\big) \big(M+1+2\bar{n}-\sqrt{(M+1)^2+4\bar{n}^2}\big)^{M-1}}{M\big(M+1-2\bar{n}+\sqrt{(M+1)^2+4\bar{n}^2}\big)^{M}}\\
        \times\left( 1 - 2\sqrt{\frac{\bar{n}}{M}} - \sqrt{\frac{2(2\bar{n}+1)\big(M+1+2\bar{n}-\sqrt{(M+1)^2+4\bar{n}^2}\big)}{M+1}} \right).
    \end{split}
\end{equation*}
Note that we do still require $\frac{(1-s)(1-2s)}{s(M-1)}>1$, but this condition is met for all $M$, for this choice of relationship between $s$ and $M$, as long as $\bar{n}< 1$, which is the case for Fig.~\ref{fig: non-trivial region}; for larger $\bar{n}$ we would need to take care that this still holds.
The region in which the bound is non-trivial is then defined by $\bar{n}$ and $\epsilon(\epsilon_0,\bar{n})$, where $\epsilon(\epsilon_0,\bar{n})$ is the maximum output trace norm for a classical probe of average photon number $\bar{n}$. Whilst the function $\epsilon$ still appears in this inequality, we can treat $\epsilon(\epsilon_0,\bar{n})$ as a single variable, and so carry out a comparison between the classical and non-classical cases. From Eq.~(\ref{eq: triviality condition fixed to nbar}), we observe that a necessary condition for there to exist a positive $\epsilon(\epsilon_0,\bar{n})$ satisfying the condition is
\begin{equation}
    1 - \sqrt{\frac{2(2\bar{n}+1)\big(M+1+2\bar{n}-\sqrt{(M+1)^2+4\bar{n}^2}\big)}{M+1}} - 2\sqrt{\frac{\bar{n}}{M}}>0.
    \label{eq: necessary triviality condition}
\end{equation}

The minimal value of ($\log_{10}$ of) $\epsilon(\epsilon_0,\bar{n})$ required to satisfy Eq.~(\ref{eq: triviality condition fixed to nbar}) is plotted in Fig.~\ref{fig: non-trivial region}. In theory, we could solve for $M$, and so only have a condition in $\bar{n}$ and $\epsilon(\epsilon_0,\bar{n})$, but this is difficult to do analytically. Note that in Fig.~\ref{fig: non-trivial region}, $M$ is treated as a continuous variable, but in practice we must constrain it to take integer values. We can see that the largest minimal value of $\epsilon(\epsilon_0,\bar{n})$ is generally obtained by choosing an $M$ as close as possible to the curve defined by Eq.~(\ref{eq: necessary triviality condition}), i.e., the smallest value of $M$ that fulfils the condition, although we can see from the plotted contours that this may not be exactly true in all cases. We could also substitute in a particular expression for $\epsilon(\epsilon_0,\bar{n})$, in order to investigate the dependence on $\epsilon_0$, for a particular channel model, although in theory we could get a slightly tighter bounds by tuning $s$ and $M$ in Eq.~(\ref{eq: generic UB}) directly.

Per Section~\ref{sec: tightness}, we now look numerically at some specific examples of channels and input states, in order to investigate whether and when the bounds can be tight.
First, we will consider whether Theorem~\ref{th: general states} and Corollaries~\ref{cor: classical} and \ref{cor: finite negativity} can be tight for cases in which $\epsilon(\epsilon_0,r^2)$ is known and tight. I.e., we will fix a target channel, $\Psi$, and a learned channel, $\Phi$, for which we can calculate the output distance for coherent states, and will investigate whether our bounds on the extension to non-classical states can be tight.
Trivially, Corollary~\ref{cor: classical} can be saturated by choosing a pure coherent state, so we will start by focusing on Corollary~\ref{cor: finite negativity}, i.e., checking whether Eq.~(\ref{eq: finite negativity bound pm}) can be tight.

Since SPATSs are relatively simple states, with finite P-representations, that we have already studied (in Section~\ref{sec: SPAT}), we will use them as an example. Eq.~(\ref{eq: finite negativity bound pm}) makes use of the triangle inequality, summing separate contributions from the negative and positive supports of the input state's P-representation, so we might expect it to be tight when the action of the channels is very different on these two domains. Consequently, we choose what we call the parity channel, which is defined by the Kraus operators $|\psi_E\rangle\langle 2m|$ and $|\psi_O\rangle\langle 2m+1|$, for $m$ from $0$ to $\infty$. In other words, the channel performs a photon number counting measurement and returns state $|\psi_E\rangle$ if the result is even and $|\psi_O\rangle$ if the result is odd. We choose that for both the target channel and the learned channel, $|\psi_E\rangle = |0\rangle$, but that the outputs of each channel for odd photon numbers, $|\psi^{\Psi}_O\rangle$ and $|\psi^{\Phi}_O\rangle$, differ by $\| |\psi^{\Psi}_O\rangle \langle\psi^{\Psi}_O| - |\psi^{\Phi}_O\rangle \langle\psi^{\Phi}_O| \| = \gamma$. We can immediately see that the Fock state $|1\rangle$ would be an optimal probe for discriminating between channels $\Psi$ and $\Phi$, so we may expect that low energy, high negativity SPATSs will have a large output distance.

For a coherent state input, the output distance for channels $\Psi_{\mathrm{par}}$ and $\Phi_{\mathrm{par}}$ is
\begin{equation}
    \|(\Psi_{\mathrm{par}}-\Phi_{\mathrm{par}})[|re^{i\phi}\rangle \langle re^{i\phi}|]\|
    = \gamma e^{-r^2}\sum_{m=0}^{\infty} \frac{r^{4m+2}}{(2m+1)!}
    = \gamma e^{-r^2}\sinh(r^2).
\end{equation}
This starts at $0$ for $r^2=0$ and tends towards $\frac{\gamma}{2}$. SPATSs have the Fock basis decomposition
\begin{equation}
    \rho_{\mathrm{SPAT}}(q) = \frac{1}{1+q}\sum_{m=0}^{\infty} \frac{(m+1) q^{m}}{(1+q)^{m+1}} |m+1\rangle\langle m+1|,
    \label{eq: eps parity}
\end{equation}
so we can calculate the output distance exactly as
\begin{equation}
    \|(\Psi_{\mathrm{par}}-\Phi_{\mathrm{par}})[\rho_{\mathrm{SPAT}}(q)]\|
    = \frac{\gamma}{1+q}\sum_{m=0}^{\infty} \frac{(2m+1) q^{2m}}{(1+q)^{2m+1}}
    = \frac{\gamma}{2}\bigg(1+\frac{1}{(1+2q)^2}\bigg).
    \label{eq: parity distance SPAT}
\end{equation}

Substituting our expression for $\|(\Psi_{\mathrm{par}}-\Phi_{\mathrm{par}})[|re^{i\phi}\rangle \langle re^{i\phi}|]\|$ into Eqs.~(\ref{eq: finite negativity bound pm}) and (\ref{eq: finite bound with mu}) and using the quantities derived in Section~\ref{sec: SPAT}, we can compare the bounds from Corollary~\ref{cor: finite negativity} with the exact value of the output distance, from Eq.~(\ref{eq: parity distance SPAT}). This comparison is carried out in Fig.~\ref{fig: SPAT}, in which we plot the output trace norm and the bounds, up to a multiplicative factor of $\gamma$. We can see that the bound from Eq.~(\ref{eq: finite negativity bound pm}) is not much larger than the true value (although any such assessment is subjective), and the two converge as $q$ becomes large. This latter observation is not surprising, as the negativity tends to $0$ as $q$ tends to $\infty$. To illustrate this, we also plot the ratio between the true value and the bound from Eq.~(\ref{eq: finite negativity bound pm}) as a function of the negativity of the corresponding SPATS. Since the ratio never drops below $\sim0.788$, this suggests that our bounds can be fairly tight. However, we note that we are only looking here at a particular channel model and that SPATSs can only have low energy and high negativity or high energy and low negativity, i.e., we have not looked at states with both high energy and high negativity, to see if the bounds can still be tight. It remains an open question whether it is possible, for every $\bar{n}$ and $\mathcal{N}$, to find a triple of $\Psi$, $\Phi$, and $\rho_{\mathrm{in}}$ such that $\rho_{\mathrm{in}}$ has average photon number $\bar{n}$ and negativity $\mathcal{N}$ and Eq.~(\ref{eq: finite negativity bound pm}) is tight.

\begin{figure}[t]
	\centering
	\includegraphics[width=0.6\textwidth]{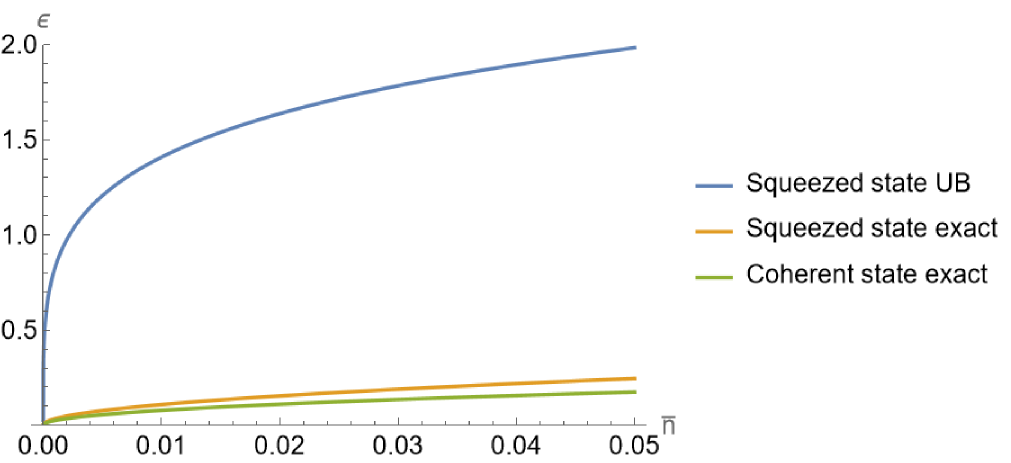}
	\caption{The output trace norm for a pair of phase rotation channels. The parameter difference between the target and learned channels is $\frac{\pi}{8}$. We plot the exact value for a coherent state input, the exact value for a one-mode squeezed vacuum, and the bound for a one-mode squeezed vacuum, coming from Eq.~(\ref{eq: 1msv piecewise}).}
	\label{fig: squeezed phase}
\end{figure}

Next, we look numerically at an example for the case of input states with infinite negativity. Specifically, we focus on the phase rotation channel (previously discussed in Section~\ref{sec: coherent examples}) for one-mode squeezed vacuum inputs. Recall, from Eq.~(\ref{eq: PR coherent distance}), that the output trace norm for coherent state inputs is $\|(\Psi_{\mathrm{PR}}-\Phi_{\mathrm{PR}})[|re^{i\phi}\rangle \langle re^{i\phi}|]\| = 2\sqrt{1-e^{-2r^2(1-\cos(\theta))}}$,
where $\theta$ is the difference in the applied phases.

Since squeezed vacuums are Gaussian states and phase rotations are Gaussian unitaries, we can calculate the exact output fidelity for a coherent state input using the formula for the fidelity of a Gaussian state~\cite{banchi_quantum_2015}. Specifically, we want to find the fidelity between two zero mean Gaussian states with covariance matrices
\begin{equation}
    V_{\Psi} = \begin{pmatrix}
        e^{-2q} &0\\
        0 &e^{2q}
    \end{pmatrix},
    \quad V_{\Phi} = \begin{pmatrix}
        e^{-2q} \cos^2(\theta) + e^{2q} \sin^2(\theta) &\sin(2\theta) \sinh(2q)\\
        \sin(2\theta) \sinh(2q) &e^{2q} \cos^2(\theta) + e^{-2q} \sin^2(\theta)
    \end{pmatrix},
\end{equation}
where $q=\mathrm{arctanh}\Big(\sqrt{\frac{\bar{n}}{\bar{n}+1}}\Big)$, and the resulting fidelity is
\begin{equation}
    F_{\mathrm{out}} = \big(2\bar{n}(\bar{n}+1)(1-\cos(2\theta))+1\big)^{-\frac{1}{4}}.
\end{equation}
Using the Fuchs-van de Graaf relation, $\|(\Psi_{\mathrm{PR}}-\Phi_{\mathrm{PR}})[\rho_{\mathrm{sq}}]\| = 2\sqrt{1-\big(2\bar{n}(\bar{n}+1)(1-\cos(2\theta))+1\big)^{-\frac{1}{2}}}$.

When applying the bound from Section~\ref{sec: 1msv}, we must minimise over parameters $s$ and $M$. Numerically (see the Supplemental Material~\cite{supplemental}), we find that the bound from Eq.~(\ref{eq: 1msv piecewise}) (i.e., setting $s=\frac{\lambda}{1+\lambda}$) is often not much worse than the bound we would obtain from a full minimisation using Eq.~(\ref{eq: generic UB}), unless $\theta$ is very small. As such, for our comparison, we set $s=\sqrt{\bar{n}(\bar{n}+1)}-\bar{n}$, so that Eq.~(\ref{eq: 1msv piecewise}) becomes
\begin{equation}
    \|(\Psi_{\mathrm{PR}}-\Phi_{\mathrm{PR}})[\rho_{\mathrm{sq}}]\| \leq
    \epsilon_{\mathrm{PR}}\Big(\epsilon_0, \sqrt{\bar{n}(\bar{n}+1)} \Big) +4\sqrt{1-\frac{\bar{n}+1+\sqrt{\bar{n}(\bar{n}+1)}}{(\bar{n}+1)\sqrt{4\bar{n}+1+4\sqrt{\bar{n}(\bar{n}+1)}}}}.
\end{equation}
We then plot this bound, for $\theta=\frac{\pi}{8}$, in Fig.~\ref{fig: squeezed phase}.

From Fig.~\ref{fig: squeezed phase}, we can see that the upper bound is very loose. There could be a few reasons for this. Firstly, we are looking here at a very simple class of Gaussian channels, but our bound holds over all channels that have the same output distance for coherent state inputs. There could exist non-Gaussian channels with the same coherent state output distance but a much higher output distance for squeezed states. We have focused on the phase rotation channels because, due to their simplicity, we can calculate the exact output distance for squeezed states.

\begin{figure}[t]
	\centering
	\includegraphics[width=0.9\textwidth]{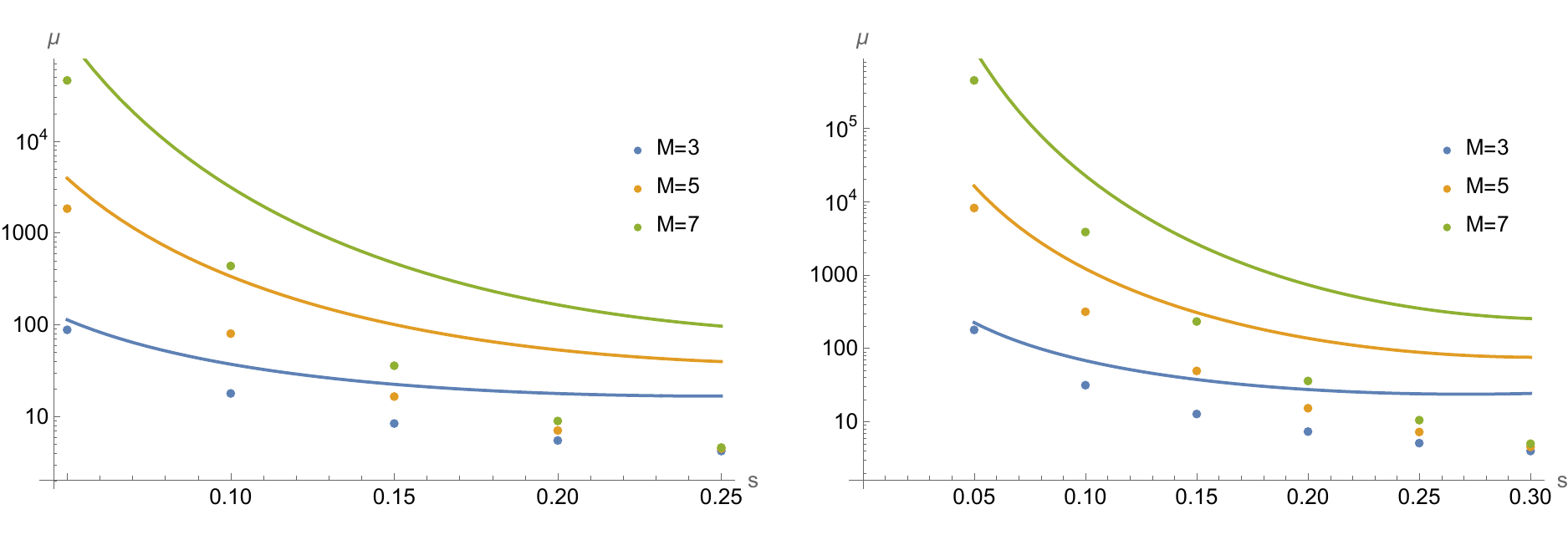}
	\caption{A comparison of the true value of $\mu_{s,M}$ for a (truncated) squeezed state, $\rho_{\mathrm{sq}}^{(M)}$, and the upper bound on $\mu_{s,M}$ from Eq.~(\ref{eq: mu bound 1msv}). The true values are given as points, whilst the upper bounds are the continuous lines. The plot on the left is for a squeezed state with average photon number $\frac{1}{4}$ and the plot on the right is for a squeezed state with average photon number $\frac{1}{10}$.}
	\label{fig: squeezed negativities}
\end{figure}

Alternatively, our estimation of $\mu_{s,M}$, using Eq.~(\ref{eq: mu bound 1msv}), could be too large. If so, a tighter upper bound on $\mu_{s,M}$ could allow us to decrease $s$, and so $\delta_s$, without making $\mu_{s,M} \epsilon\left(\epsilon_0, \frac{\nu_{s,M}}{\mu_{s,M}}\right)$ large. In Fig.~\ref{fig: squeezed negativities}, demonstrates that this is indeed the case. The true values of $\mu_{s,M}$ were calculated by constructing the P-representation, using Eqs.~(\ref{eq: Ps Fock on-diag}), (\ref{eq: Ps Fock off-diag}), and (\ref{eq: 1msv density}), and then numerically integrating the absolute value of it, for various values of $s$. The downside of this method is that we do not get an analytical formula, just a numerical value. However, this is enough to demonstrate that the gap between $\mu_{s,M}$ and $\mu_{s,M}^{\mathrm{UB}}$ can be quite extreme. This suggests a way the bounds for squeezed states could be improved.

\end{document}